\documentclass[journal,10pt,twocolumn,twoside]{IEEEtran}
\usepackage{longtable}
\usepackage{mathtools}
\usepackage{empheq}
\mathtoolsset{showonlyrefs}
\usepackage[acronym]{glossaries}
\usepackage{acronym}
\usepackage{algpseudocode, algorithm}
\usepackage{indentfirst}
\usepackage{amsthm}
\usepackage{amsfonts}
\usepackage{multirow}
\usepackage{color}
\usepackage{MnSymbol}
\usepackage{subfig}
\usepackage{fmtcount} 
\usepackage{enumitem}
\usepackage[open,openlevel=1]{bookmark}
\usepackage{breakurl}

\usepackage{xcolor}

\ifCLASSOPTIONcompsoc
   \usepackage[nocompress]{cite}
\else
   \usepackage{cite}
\fi
\ifCLASSINFOpdf
   \usepackage[pdftex]{graphicx}
   \DeclareGraphicsExtensions{.pdf,.jpeg,.png}
\else
   \usepackage[dvips]{graphicx}
   \DeclareGraphicsExtensions{.eps}
\fi
\usepackage{url}

\begin{document}
%
\title{A Survey of Multi-Access Edge Computing in 5G and Beyond: Fundamentals, Technology Integration, and State-of-the-Art}

%
%
%
%

\author{Quoc-Viet Pham, Fang Fang, Vu Nguyen Ha, Md. Jalil Piran, Mai Le, \\ Long Bao Le, Won-Joo Hwang, and Zhiguo Ding
\IEEEcompsocitemizethanks{Q.-V. Pham is with High Safety Vehicle Key Technology Research Center and Mai Le is with Department of Information and Communications System, 197, Inje-ro, Gimhae-si, Gyeongsangnam-do 50834, Korea. W.-J. Hwang is with School of Biomedical Convergence Engineering	Pusan National, 49, Busandaehak-ro, Mulgeum-eup, Yangsan-si, Gyeongsangnam-do 50612, Korea. F. Fang and Z. Ding are with School of Electrical and Electronic Engineering, The University of Manchester, M13 9PL, UK. Vu Nguyen Ha is with Ecole Polytechnique de Montreal, Montreal, Quebec, Canada. Md. Jalil Piran is with Department of Computer Science and Engineering, Sejong University, 05006 Seoul, South Korea. Long Bao Le is with the Institut National de la Recherche Scientifique, University of Quebec, Montreal, QC H5A 1K6, Canada. E-mail: \{vietpq90@gmail.com, fang.fang@manchester.ac.uk, vu.ha-nguyen@polymtl.ca, piran@sejong.ac.kr, maile2108@gmail.com, zhiguo.ding@manchester.ac.uk, long.le@emt.inrs.ca, wjhwang@pusan.ac.kr\}.}
\thanks{This work was supported by the National Research Foundation of Korea (NRF) grant funded by the Korea government (MSIT) (No.NRF-2019R1C1C1006143). Quoc-Viet Pham is the corresponding author.}
}

\markboth{IEEE Communications Surveys and Tutorials}{Q.-V. PHAM \MakeLowercase{\textit{et al.}}: A Survey of MEC in 5G and Beyond: Fundamentals, Technology Integration, and State-of-the-Art}


\IEEEcompsoctitleabstractindextext{%
\begin{abstract}
Driven by the emergence of new compute-intensive applications and the vision of the Internet of Things (IoT), it is foreseen that the emerging 5G network will face an unprecedented increase in traffic volume and computation demands. However, end users mostly have limited storage capacities and finite processing capabilities, thus how to run compute-intensive applications on resource-constrained users has recently become a natural concern. Mobile edge computing (MEC), a key technology in the emerging fifth generation (5G) network, can optimize mobile resources by hosting compute-intensive applications, process large data before sending to the cloud, provide the cloud-computing capabilities within the radio access network (RAN) in close proximity to mobile users, and offer context-aware services with the help of RAN information. Therefore, MEC enables a wide variety of applications, where the real-time response is strictly required, e.g., driverless vehicles, augmented reality, robotics, and immerse media. Indeed, the paradigm shift from 4G to 5G could become a reality with the advent of {\color{black}new technological concepts}. The successful realization of MEC in the 5G network is still in its infancy and demands for constant efforts from both academic and industry communities. In this survey, we first provide a holistic overview of MEC technology and its potential use cases and applications. {\color{black}Then, we outline up-to-date researches on the integration of MEC with the new technologies that will be deployed in 5G and beyond. We also summarize testbeds and experimental evaluations, and open source activities, for edge computing.} We further summarize lessons learned from state-of-the-art research works as well as discuss challenges and potential future directions for MEC research.
\end{abstract}

}

\maketitle
\IEEEdisplaynotcompsoctitleabstractindextext
\IEEEpeerreviewmaketitle

\section{Introduction}	
\label{Sec:Introduction}
\begin{figure*}[!ht]
	\centering
	\includegraphics[width=1.00\linewidth]{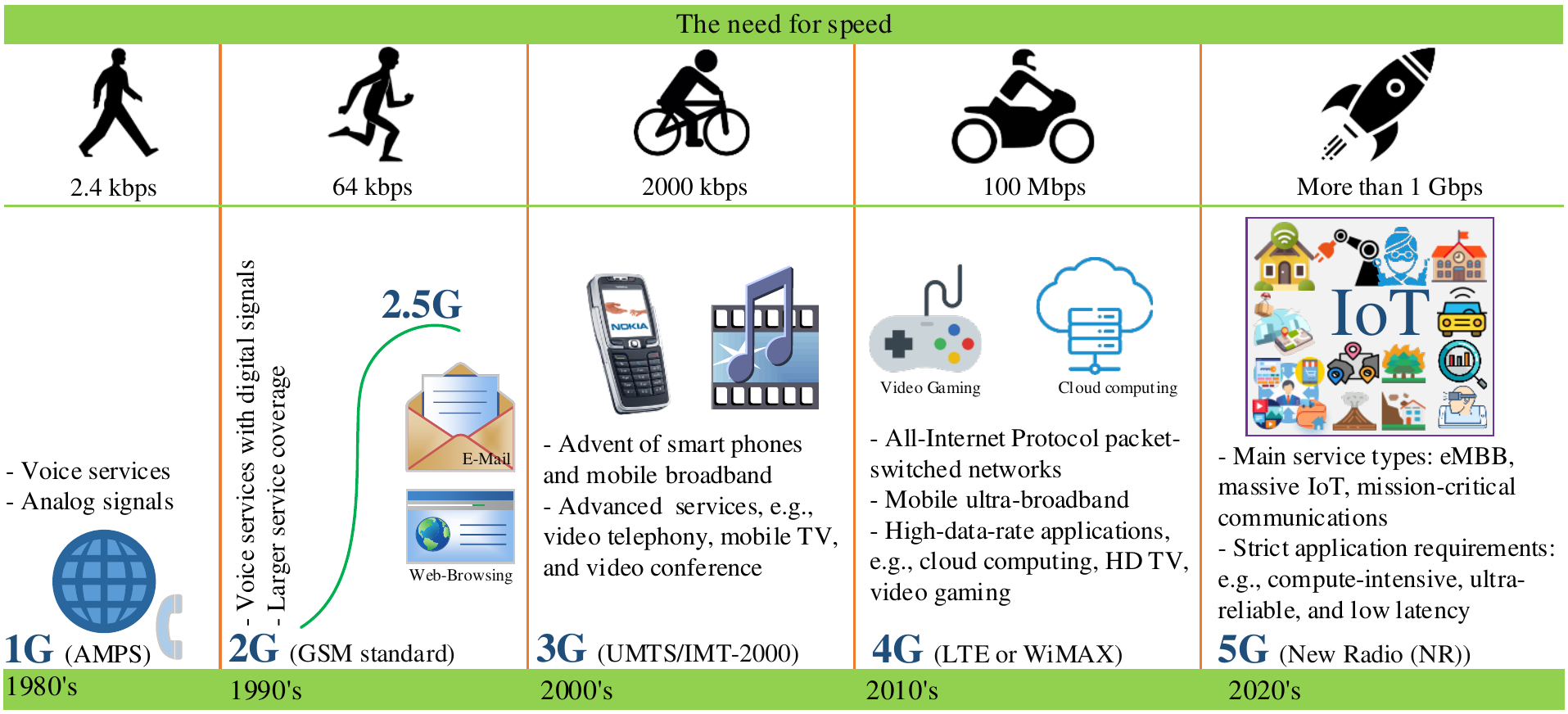}
	\caption{Evolution of wireless communication.}
	\label{Fig:Evolution_of_Wireless_Communication}
\end{figure*}

\IEEEPARstart{D}{uring} the last four decades, the evolution of wireless communication networks has changed every aspect of our lives, society, culture, politics, and economics. Since the commercialization of the first generation (1G) of cellular networks in early 1980's, generations have been launched with enormous differences in terms of the network architectures, key technologies, coverage, mobility, security and privacy, data, spectral efficiency, cost optimality, and so on. The brief summary of wireless communication evolution is shown in Fig.~\ref{Fig:Evolution_of_Wireless_Communication}. Now, both academic and industry communities are making tremendous efforts to finalize the 5G standardization and commercialization in 2019. 5G communications can be categorized into three categories: enhanced mobile broadband (eMBB), ultra-reliable low-latency communication (URLLC), and massive Internet of Things (IoT). Compared with previous generations, 5G will support not only communication, but also computation, control, and content delivery (4C) functions \cite{Mao2017_aSurveyMEC}. {\color{black}Moreover, many new applications and use cases are expected 
with the advent of 5G, for example, virtual/augmented reality (VR/AR), autonomous vehicle, Tactile Internet, and IoT scenarios. These applications are poised to induce a significant surge in demand for not only communication resources but also computation resources. To meet such ever-growing demands, various technological concepts have been developed for 5G in terms of radio access, network resource management, applications, network architectures and scenarios, power supply, and performance improvement \cite{Andrews2014What}. For example, non-orthogonal multiple access (NOMA), dense heterogeneous networks (HetNets), cloud radio access network (C-RAN), unmanned aerial vehicle (UAV), IoT, wireless power transfer (WPT) and energy harvesting (EH), and machine learning (ML), have been considered as key enabling technologies.}

{\color{black}The Cisco white paper \cite{Cisco2017VNI} showed that global data traffic will grow at a compound annual growth rate (CAGR) of 26 percent between 2017 and 2022 (i.e., increase more than threefold) and reach 122 exabytes (EB) per month by 2022. Mobile and wireless networks carried 11.51 EB per month in 2017, 28.56 EB per month in 2019, and 77.49 EB per month at the end of 2022. Moreover, traffic generated by new applications and services will increase at a much higher CAGR, for example, 12-fold for AR and VR, ninefold for Internet gaming, and sevenfold for Internet video surveillance. It is also anticipated that the number of connected things (e.g., sensors and wearable devices) will reach 28.5 billion by 2022, up from 21.5 billion in 2019.} However, most connected devices have limited communication and storage resources and finite processing capabilities, which show the mismatch between the stringent requirements for emerging applications and the actual device capabilities. Despite recent advancements in the hardware capability, mobile computing still cannot cope with the demand of many applications that need to generate, process, and store a massive amount of data and require large computing resources. One potential solution to these challenges is to transfer computations to centralized clouds, which can be, however, burdened by many issues, such as network congestion and privacy policies. This has driven the development of mobile edge computing (MEC). 

\subsection{From Centralized Clouds to Mobile Edges}
The very first computing paradigm that combines cloud computing, mobile computing, and wireless communication networks is {\color{black}mobile cloud computing (MCC). MCC enables developers and service providers to build more complex applications by moving the computing capabilities and data storage away from mobile devices and into the cloud \cite{Dinh2013aSurvey}.} However, MCC suffers from considerable disadvantages. First, MCC is not suitable for applications that desire the wide-spread distribution of users. Second, the large physical distance between users and the central cloud implies high latency, which fails to meet the stringent requirement of latency-sensitive applications. Third, privacy and security are crucial factors in the success of MCC since the high concentration of data information leaves the cloud highly susceptible to violent attacks and data offloaded to the cloud through wireless environments can be overheard by eavesdroppers. {\color{black}Finally, exchanging the big data between a number of users and the central cloud is challenging due to the extreme burden on bandwidth constraints and the network congestion \cite{Taleb2013FollowMC}.} 

One of the very first edge computing concepts, so-called cloudlet, was proposed by Satyanarayanan \textit{et al.} in 2009 \cite{Satyanarayanan2009TheCase}. Cloudlet at its simplest, refers to a trusted and resource-rich computer or a cluster of computers that are located in a strategic location at the edge and well connected to the Internet. The main purpose of cloudlet is to extend the cloud computing to the network edge and support resource-poor mobile users in running resource-intensive and interactive applications. Virtual machine (VM) provisioning makes cloudlets operable in the standalone mode without the intervention of the cloud  \cite{Mukherjee2018Survey_FogComputing}. In fact, the idea of cloudlet to distribute the cloud computing in close proximity to mobile users is similar to the WiFi concept in providing Internet access \cite{Mach2017Mobile}. The WiFi connection between users and cloudlets can be a serious drawback. In this way, users are unable to access cloudlets in the long distance and use both WiFi and cellular connection simultaneously \cite{Barbarossa2014Communicating}, i.e., users have to switch between the mobile network and WiFi when they use cloudlet services. {\color{black}Another drawback is that WiFi operates over the limited unlicensed frequency band.}

Fog computing, a term put forward by Cisco in 2012, refers to the extension of the cloud computing from the core to the network edge, thus it reduces the amount of data needed to transfer to the central cloud \cite{Chiang2016FogIoT}. 
Therefore, most intensive computations from and data collected by end users can be processed and analyzed by fog nodes (i.e., nodes in fog computing) at the network edge, thus reducing the execution latency and network congestion \cite{Atlam2018FogComputing, Gu2018Joint}. Since fog nodes are generally deployed in distributed locations, they are wide-spread and geographically available in large numbers \cite{Yousefpour2019AllOneNeeds}. Due to its characteristics, fog computing plays an important role in many use cases and applications \cite{Bonomi2012FogComputing}, e.g., smart cities, connected vehicle, smart grid, wireless sensor and actuator networks, smart buildings, and decentralized smart building control. However, a fog node cannot act as a self-managed cloud data center (DC) and needs the support of the cloud. 

{\color{black}The cloudlet and fog computing are similar in that cloudlets and fog nodes are not integrated into the mobile network architecture, thus fog nodes and cloudlets are commonly implemented and owned by private enterprises and  it is not easy to provide mobile users with quality of service (QoS) and quality of experience (QoE) guarantees \cite{Mach2017Mobile, Mukherjee2018Survey_FogComputing}. One of the concepts that capitalize on cloud computing is cloud radio access networks (C-RANs). In C-RANs, traditional base stations (BSs) are replaced by distributed remote radio heads (RRHs) and a centralized baseband unit (BBU), and the baseband signal processing functionalities in the traditional BS are moved to the central BBU \cite{Pham2017NetworkUtility, Peng2016RecentAdvances}.} By that decoupling, RRHs are responsible only for basic radio frequency functions and can support high capacity in hot spots while the central BBU provides the large-scale signal processing, e.g., centralized encoding and decoding, and joint beamforming and resource allocation \cite{Peng2016RecentAdvances}. Although C-RAN is promising in reducing the cost and power consumption, improving spectral efficiency and energy efficiency, and increasing the utilization of hardware \cite{Wu2015CloudRAN}, the centralization of C-RANs leads to massive demands for information exchange between the distributed RRHs and the centralized BBU, thus imposing stringent requirements on the fronthaul links. Moreover, the virtualized BBU is mainly utilized for centralized radio signal processing and resource allocation \cite{Peng2016RecentAdvances} and rarely used for computation execution \cite{Mach2017Mobile}.

\begin{table*}[t]
	\caption{Summary of existing surveys on multi-access edge computing.}
	\label{Table:Summary_Previous_Surveys}
	\centering
	\begin{tabular}{|p{2.2cm}|p{1.7cm}|p{10.0cm}|}
		\hline 
		\textbf{Theme} & \textbf{Reference} & \textbf{Major Contribution}\\ 
		\hline
		\hline
		\multirow{2}{2.2cm}{Architecture and Computation offloading} & \cite{Mach2017Mobile, Jiang2019TowardsCO}  & - Review of potential MEC architectures and computation offloading. \\ 
		\cline{2-3}
		& \cite{Taleb2017onMultiAccess} & - Introduction of MEC and its key enablers: NFV, SDN, and VM. \newline - Analysis of MEC reference architecture and orchestration deployment scenarios. \\ 
		\hline
		\multirow{2}{2.2cm}{Resource Allocation} &\cite{Mao2017_aSurveyMEC}  & - Survey of the basic MEC models from the communication perspective. \newline - Review of joint communication and computation resource allocation in MEC systems.  \\ 
		\cline{2-3} 
		& \cite{Wang2017aSurvey, Wang2018IntegrationOfCCC} & - Review of convergence and integration of communication, computation, and caching. \\
		\hline
		\multirow{1}{2.2cm}{Mathematical Frameworks} & \cite{Wu2018MultiObjective} & - Survey of computation offloading decisions using multi-objective optimization.   \\ 
		\cline{2-3}
		& \cite{Moura2019GameTheory} & - Fundamentals of game theory models and MEC. \newline - Review of game theoretical contributions to wireless networks and MEC systems. \\ 
		\hline		
		\multirow{4}{2.2cm}{Research Directions} & \cite{Abbas2018Mobile} & - Fundamentals of MEC, use cases, infrastructure, and security \& privacy issues.  \\ 
		\cline{2-3}
		& \cite{Khan2019219EdgeComputing, Liu2018Mobile} & - MEC concepts, applications, architectures, and open research challenges. \\
		\cline{2-3}
		& \cite{Porambage2018Survey, Premsankar2018Edge, Yu2018aSurvey} & - Review of how to exploit MEC and other edge computing paradigms for IoT applications. \\
		\cline{2-3}
		& \cite{Ronan2018Mobile, Makitalo2018Safe, Shirazi2017theExtended} & - Review and analyses of security and resilience of edge computing technologies. \\
		\hline
		\multirow{1}{2.2cm}{{\color{black}MEC with 5G Technologies}} & {\color{black}Our Survey} & - {\color{black}Survey on integration of MEC with 5G technologies, including NOMA, WPT and EH, UAV communications, IoT, and H-CRAN. \newline - Applications of ML to MEC: 4C optimization, security and privacy, big data analytics, and mobile crowdsensing.}  \\
		\hline		
	\end{tabular}
\end{table*}

In late 2014, the European Telecommunications Standards Institute (ETSI) Mobile Edge Computing Industry Specification Group (MEC ISG) initiated the MEC concept. As a complement of the C-RAN architecture, \emph{MEC aims to unite the telecommunication and IT cloud services to provide the cloud-computing capabilities within radio access networks in the close vicinity of mobile users} \cite{Patel2014MobileEdge}. Therefore, MEC enables a wide variety of applications, e.g., driverless vehicles, VR/AR, robotics, and immerse media. In order to reap additional benefits of MEC with heterogeneous access technologies, e.g., 4G, 5G, WiFi, and fixed connection, ETSI ISG officially changed the name of mobile edge computing to mean multi-access edge computing in 2017 \cite{Kekki2018MEC5G}. {\color{black}After this scope expansion, MEC servers can be deployed by the network operators at various locations within RAN and/or collocated with different elements of the network edge, such as BSs (aka eNB in 4G and gNB in 5G), optical network units, radio network controller sites, and WiFi access points. This transformation pushes intelligence towards the edge so that not only communication functionalities but also computation, caching, and control services can be better facilitated.} From this point, the correct name for MEC is multi-access edge computing and this paper uses that name. 

\subsection{Limitations of Prior Works and Our Contributions}
\label{Subsec:Limitations_Prior_Works}
Over the last few years, there have been a large number of studies focusing on either technical aspects of MEC architectures or reviews of attributes and application use cases of MEC. Many also consider the importance of MEC in 5G enabling technologies and applications and cover certain research aspects discussed in our article, for example, \cite{Abbas2018Mobile, Khan2019219EdgeComputing, Liu2018Mobile, Mach2017Mobile, Jiang2019TowardsCO, Mao2017_aSurveyMEC, Taleb2017onMultiAccess, Porambage2018Survey, Premsankar2018Edge, Yu2018aSurvey, Wang2017aSurvey, Wang2018IntegrationOfCCC, Wu2018MultiObjective, Moura2019GameTheory}. The previous surveys are summarized as follows. The surveys in \cite{Abbas2018Mobile, Khan2019219EdgeComputing, Liu2018Mobile} presented a general overview of MEC on definitions, architectures, advantages, deployment scenarios and testbeds, and security and privacy issues. The survey in \cite{Mach2017Mobile, Jiang2019TowardsCO} reviewed several edge computing concepts and focused on computation offloading. The authors in \cite{Mao2017_aSurveyMEC} reviewed joint communication and computation resource management in MEC systems. In \cite{Taleb2017onMultiAccess}, the authors described four fundamental enabling technologies for MEC including virtual machines and containers, network functions virtualization (NFV), software-defined networking (SDN), and network slicing. Moreover, the authors provided analyses of the MEC service orchestration, MEC service mobility, and joint optimization of virtual network functions and MEC services. Several works in \cite{Porambage2018Survey, Premsankar2018Edge, Yu2018aSurvey} revealed the role of MEC for IoT applications and realization. Recent studies in \cite{Wang2017aSurvey, Wang2018IntegrationOfCCC} focused on reviewing the integration of communication, caching, and computation. Some potential mathematical frameworks for optimization of MEC systems were reported in \cite{Wu2018MultiObjective, Moura2019GameTheory}. In particular, the authors in \cite{Wu2018MultiObjective} conducted a survey on the computation offloading decisions when multiple challenges, e.g., heterogeneous resources, large amounts of computation and communication, intermittent connectivity and network capacity, are considered (i.e., multi-objective optimization). The authors in \cite{Moura2019GameTheory} reviewed research works that applied theoretical games in addressing problems and challenges of MEC systems. In Table~\ref{Table:Summary_Previous_Surveys}, we provide a summary of the recently published surveys and reviews on MEC.

{\color{black}Previous surveys addressed important problems in MEC systems, while they have several limitations. These surveys are limited to specific aspects and potential use cases of MEC, for instance, MEC overview \cite{Abbas2018Mobile, Liu2018Mobile}, architecture and computation offloading \cite{Mach2017Mobile}, resource allocation \cite{Mao2017_aSurveyMEC}, and mathematical frameworks \cite{Wu2018MultiObjective, Moura2019GameTheory}. Indeed, these articles provide only high-level discussions of the problems and challenges of MEC in 5G. To the best of our knowledge, there is no existing survey to provide a discussion of MEC in the context of other 5G technologies. Furthermore, it is necessary to have an updated survey since MEC has gained popularity in years with  a fast-growing research trend and ETSI has released a set of phase 2 specifications, but almost all the articles mentioned in the related work were prepared and/or submitted quite long ago. Therefore, this paper sets to provide a comprehensive survey of the state-of-the-art research works which are focused on the integration of MEC and the forthcoming technologies that will be deployed in 5G and beyond network.} In a nutshell, contributions offered by our survey can be summarized as follows: 
\begin{itemize}
	\item We conduct an overview of MEC including fundamentals of MEC (e.g., characteristics, challenges, and market drivers), and MEC integration in the 5G network with potential use cases and applications. 
	\item We discuss the role of MEC in the 5G network architecture and undertake a holistic review of related literature published in the last few years for the integration of MEC with the forthcoming 5G and beyond technologies and scenarios including NOMA, WPT and EH, UAV, IoT, and heterogeneous C-RAN, and ML.
	\item We provide a concise summary of lessons learned from the state-of-the-art research works and describe potential future directions.
\end{itemize}

\subsection{Paper Organization}
\begin{figure}[t]
	\centering
	\includegraphics[width=1.0\linewidth]{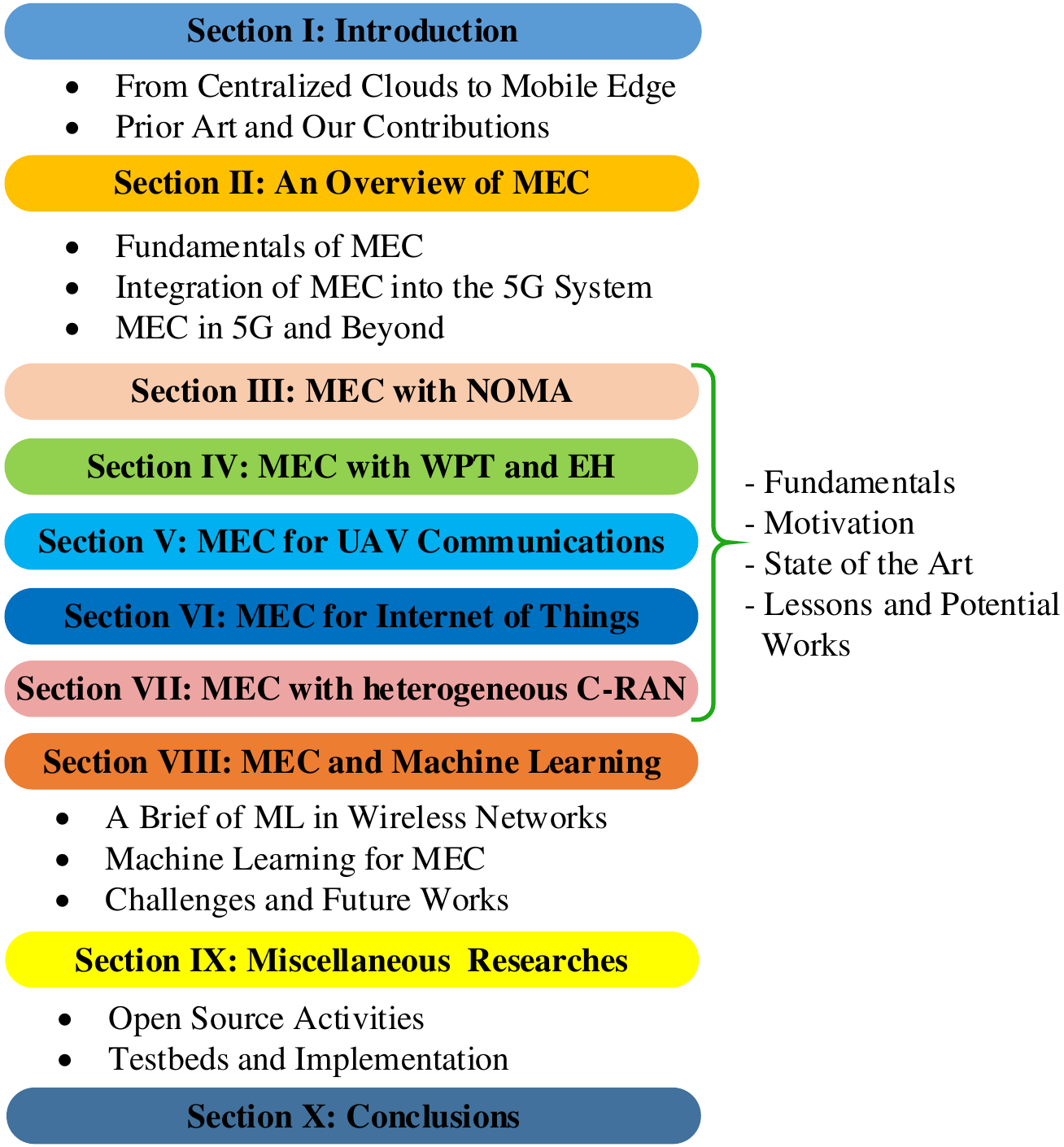}
	\caption{Diagrammatic view of the organization of this survey.}
	\label{Fig:PaperOrganization}
\end{figure}

The remaining of this paper is organized as follows. {\color{black}Sections~\ref{Sec:FundamentalBackground} provides the fundamentals of MEC, including its benefits, integrated architecture in the 5G scenario, and key use cases. The major part of this work is a review of MEC in the context of NOMA (Section~\ref{Sec:NOMA_MEC}), WPT and EH (Section~\ref{Sec:WPT_MEC}), UAV communications (Section~\ref{Sec:UAV_MEC}), IoT (Section~\ref{Sec:IoT_MEC}), heterogeneous C-RAN (Section~\ref{Sec:HetNet_MEC}), and machine learning (Section~\ref{Sec:MEC_MachineLearning}).} For each section, we first outline background, then provide motivations for the integration, and finally outline learned lessons and potential directions. The paper is concluded in Section~\ref{Sec:Conclusion}. For the sake of clarity, Fig.~\ref{Fig:PaperOrganization} shows the organization of this paper and a reading map for the readers.  

\section{Overview of MEC researches}
\label{Sec:FundamentalBackground}
{\color{black}We present fundamentals of MEC by listing the main features and discussing design challenges of MEC and the benefits offered by MEC. We also show the interactions between MEC in the forthcoming 5G technologies and further illustrate MEC use cases with representative examples.}

\subsection{Fundamentals of MEC}
\label{Subsec:Fundamentals_MEC}


\begin{figure*}[t]
	\centering
	\includegraphics[width=1.00\linewidth]{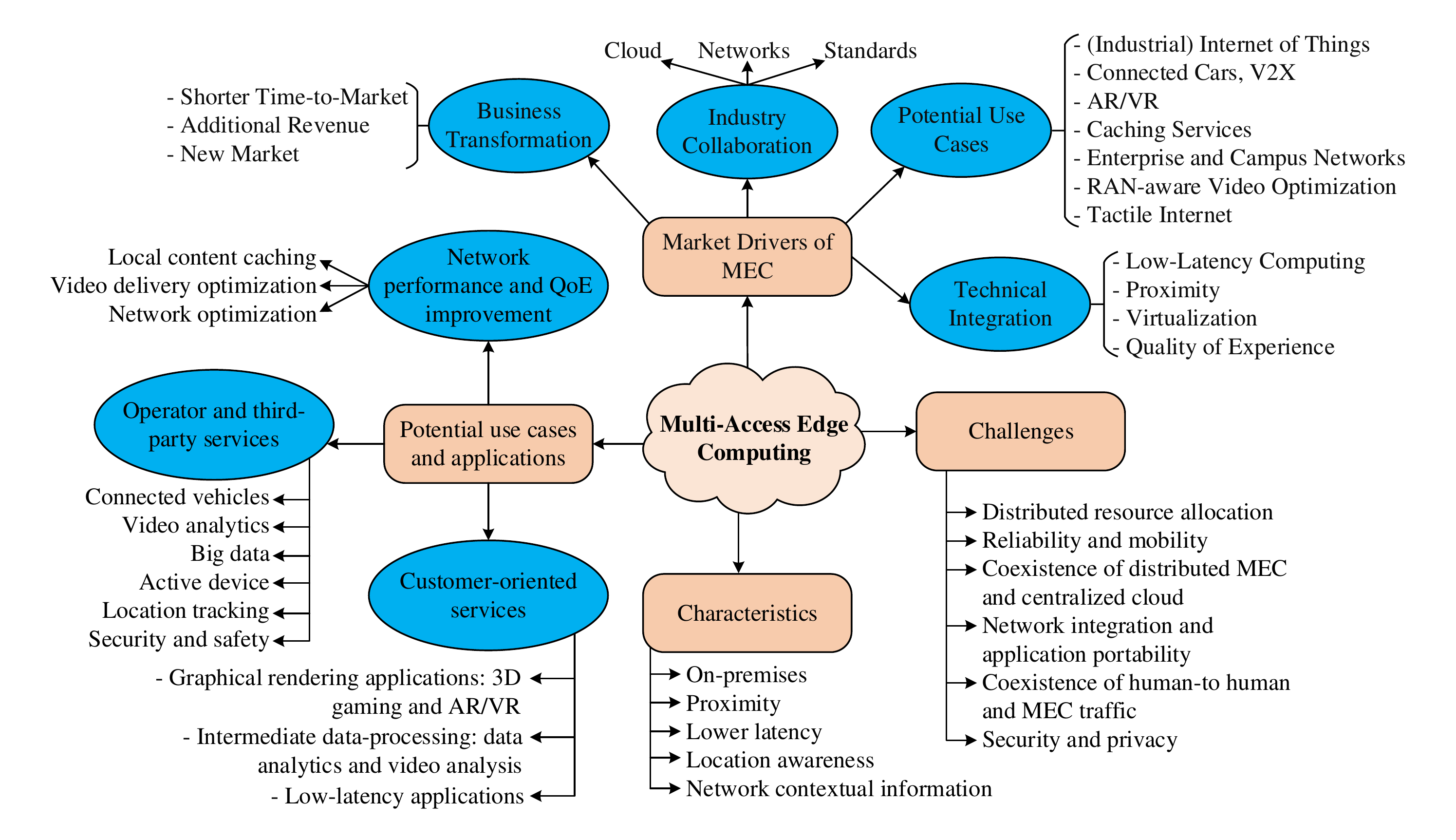}
	\caption{An overview of MEC: challenges, characteristics, potential use cases and applications, and market drivers.}
	\label{Fig:Overview_Of_MEC}
\end{figure*}
The key idea of MEC is ``providing an IT service environment and cloud-computing capabilities at the edge of the mobile network, within the RAN and in close proximity to mobile subscribers" \cite{ETSI2018MEC_UseCases}.
The demand for MEC has been driven by many factors, such as the increasing pervasiveness of smart and IoT devices, rapid increase in the data volume and velocity, the increasing need for the rapid development of new high-bandwidth and low-latency applications, introduction of new wireless technologies, and increasing requirement of QoE and QoS. Among those factors, low-latency computing is considered as the primary driven factor for the development of MEC. The demand for low-latency computing is increasing rapidly as low latency is a fundamental metric for network performance and is required by many emerging applications (e.g., VR, interactive gaming, and mission-critical controls). The development of MEC is further fortified by great opportunities for business transformation. On the one hand, mobile network operators (MNOs) need to shorten the time-to-market of new applications and services to maximize the overall revenue. On the other hand, the success and widespread deployment of MEC are guaranteed only when there is the participation of multiple stakeholders (e.g., mobile operators, service providers, vendors, and users) as well as their collaboration. As suggested in \cite{Hu2015Mobile}, the key growth drivers in the MEC market can be classified into four major categories: technical integration, potential use cases, business transformation, and industry collaboration (see Fig.~\ref{Fig:Overview_Of_MEC}). In the foreseeable future, MEC will open up new markets for different industries and sectors by enabling a wide variety of use cases, e.g., IoT, Industry 4.0, Vehicle-to-everything (V2X) communication, smart city, and Tactile Internet. {\color{black}A complete picture of MEC, including challenges, characteristics, use cases and applications, and market drivers, is pictorially illustrated in Fig.~\ref{Fig:Overview_Of_MEC}.}

According to the ETSI white paper \cite{Patel2014Mobile}, MEC can be characterized by some features, namely on-premises, proximity, lower latency, location awareness, and network context information. These features can be shortly explained as follows:
\begin{itemize}[leftmargin=*]
	\item \emph{On-premises}: MEC can operate in standalone environments (i.e., MEC can run isolated from the rest of the network) and has access to local resources.
	
	{\color{black}
	\item \emph{Proximity}: MEC servers are usually positioned in the close vicinity of mobile users, thus MEC can capture information from mobile users for further purposes such as data analytics and big data processing.
	
	\item \emph{Lower latency}: although an MEC server has a finite computation power, it is usually sufficient to process emerging compute-intensive applications in real time. MEC has the potential of shortening the communication and propagation latency, which makes MEC a promising enabler for latency-critical 5G applications. MEC also opens up the opportunities to alleviate the burdens on the fronthaul and backhaul links and to accelerate the content and service responsiveness by appropriately caching popular and locally-relevant contents at the network edge.
	
	\item \emph{Location awareness}: Due to the close proximity, MEC can utilize signaling information received from end users to estimate their precise locations. This becomes particularly important for MEC location-based services.
	
	\item \emph{Network contextual information}: characterized by proximity, MEC can utilize the knowledge of real-time radio network conditions and local contextual information to optimize the network and QoS. For example, real-time and contextual information can be used to improve user experience via personalized services \cite{Hu2015Mobile}.} 
\end{itemize}

{\color{black}In spite of several opportunities and potentials, many challenges need to be studied in order to create an edge ecosystem where all network players (i.e., IoT users, service/infrastructure providers, and mobile operators) can benefit from edge services. The discussion can be summarized as follows.}
\begin{enumerate}[leftmargin=*]
	\item \textit{Distributed resource management}: Resource allocation is a key challenge for the success of MEC due to finite resources, growing number of applications, and explosive increase in the mobile traffic \cite{Pham2016ResourceAllocation}. The optimization of resource allocation may be multi-objective that varies in different situations due to diverse nature of applications, heterogeneous MEC servers, various user demands/characteristics, and channel connection qualities. With massive users, the wireless channel would be bottlenecked and the competition among users for scarce computing resources becomes highly intense \cite{Lyu2017Multiuser}. Although the centralized approach can achieve competitive performance, it has the weakness of high computational complexity and huge reporting overhead. Therefore, the centralized approach is not suitable for distributed MEC systems \cite{Pham2018Decentralized, Pham2017Fairness}. Additionally, there may not exist a dedicated backhaul for information exchange and computation offloading and even if there is, the wireless backhaul could be congested due to the high burden of huge data sharing \cite{Ge2015_5GBackhaul}. All of these points call for efficient and distributed MEC resource allocation schemes. 
	
	\item \textit{Reliability and mobility}: Densification is a key block for the 5G network and is expected to gain enormous benefits. However, managing mobility and ensuring reliability are quite challenging in such environments. First, under the coverage of multiple small-scale servers, user mobility can cause frequent handovers, which introduce the service disruption problem and affect the overall network performance \cite{Sung2013Predictive}. Next, users (e.g. vehicles) may move to new locations during the computation offloading period. In such case, users may not be able to receive the computational result since they already move out of the service coverage of their serving servers. Therefore, efficient computation offloading models are necessary for the application accomplishment. Further, the dynamic change in the number offloading users results in random uplink interference and time-varying computing resources \cite{Dong2016Mobility}. Last, providing reliable MEC services in mobile environments is really challenging owing to time-varying dynamics of wireless connections and user mobility. For instance, AR-based applications usually require the real-time response and ultra-reliable connection between the server and users. However, these requirements would not be well fulfilled on account of dynamic channel qualities and intermittent connections. 
	
	\item \textit{Network integration and application portability}: Depending on the underlying technologies, technical and business requirements, MEC servers can be deployed at different places within the RAN. Thus, another critical challenge is the seamless integration of MEC into the underlying network architecture and existing interfaces \cite{Patel2014Mobile}. The existence of MEC and enabled applications should not affect the standard specifications of the core network and end devices. According to \cite{Kekki2018MEC5G}, the key component of the MEC integration is the ability of MEC to interact with 5G networks in routing the traffic and receiving relevant control information. Furthermore, the application migration 
	necessitates a so-called application portability requirement. This removes the need for app developers to design multiple versions for different MEC platforms. 
	 
	\item \textit{Coexistence of distributed MEC and centralized cloud}: Cloud DCs, with abundant computing resources, can process big-data applications in near zero time and support a large number of users. However, distributed MEC is highly desired since the computation at the network edge can not only meet the user requirement but also reduce the end-to-end delay caused by the traffic congestion and transmission delay. By analogy to the HetNet architecture, it is highly beneficial to implement MEC in a hierarchical manner, i.e., user, edge-computing, and cloud-computing layers. In this way, the MEC vendor also injects computing resources to the small-eNBs so that the advantages of HetNets can be exploited for diversifying radio transmissions and spreading computing demands \cite{Huynh2020Efficient}. We note that distributed MEC may not have enough computing resources to process all computation requests and complete reliance on the cloud poses challenges of providing latency-critical services. Therefore, it is intuitive to distribute big-data/latency-critical computations to distributed MEC servers while transferring compute-intensive and delay-tolerant tasks to the cloud DC \cite{Zhang2018OptimalTask}. The coexistence of distributed MEC and centralized cloud is an important issue and more research is needed for their interactions. 
	
	\item \textit{Coexistence of human-to-human and MEC traffic}: Incorporating both conventional Human-to-Human (H2H) traffic (e.g., voice, data, and video) and MEC traffic in 5G is a challenging task due to massive IoT connections coupled with the diverse QoS requirements and unique characteristics of MEC traffic \cite{Levesque2014CoexistenceAnalysis}. For instance, the IoT system comprises of human-type devices (HTDs) and machine-type devices (MTDs) that may run different kinds of applications, e.g., MTD with sensors and smart homes, and HTD with video game. While MTDs have a mixed set of QoS requirements, such as latency, reliability, and energy efficiency, HTDs typically require a high-speed rate with the limited energy budget \cite{Abuzainab2017CognitiveHierarchy}. Similarly, the MEC system should be designed in a way that the QoS requirements of H2H traffic are satisfied while unique characteristics of M2M traffic (e.g., real-time response and context awareness) are maintained. 
	 
	\item \textit{Security and privacy}: Although MEC has the capability to improve security and privacy compared with MCC, MEC has its own security and privacy challenges. First, MEC can be collocated with different heterogeneous network elements, thus making the conventional privacy and security mechanisms, which have been already operated in MCC, inapplicable to MEC systems. Second, the task offloading over wireless channels may not be secure since computation tasks can be overheard by malicious eavesdroppers. The transfer of compute-intensive applications can be secured by encryption at the user side and decryption at the destination server side. This, however, can increase the propagation delay as well as execution delay, thus reducing the application performance \cite{Ahmed2017Mobile}. Physical layer security and blockchain have emerged as effective solutions to secure MEC systems \cite{Xu2019Exploiting, Nguyen2019Integration}. Finally, the sharing of the same storage and computation resources among multiple mobile users raises issues of private data leakage and loss.
\end{enumerate}

{\color{black}
\begin{figure*}[ht]
	\centering
	\includegraphics[width=0.85\linewidth]{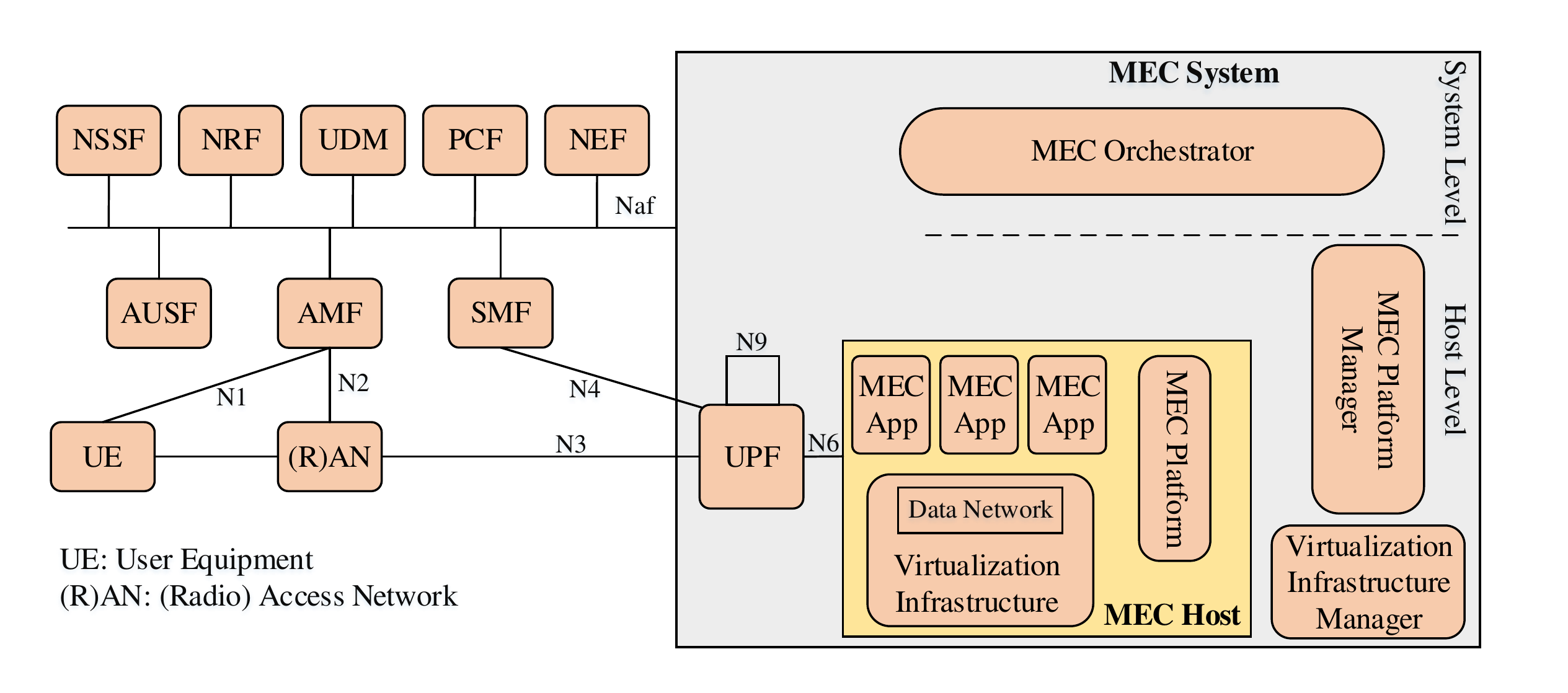}
	\caption{MEC integrated architecture in 5G \cite{Kekki2018MEC5G}.}
	\label{Fig:IntegratedArchitecture}
\end{figure*}

\subsection{Integration of MEC into the 5G System}
\label{Subsec:MEC_in_5G}
After initializing the MEC concept, the ETSI ISG and many members in the value chain have spent a great deal of efforts for the development of MEC specifications based on industry consensus. At the time of writing this paper, there are 64 members and 32 participants in the ETSI consortium\footnote{The complete list of MEC members and participants is available at \url{https://portal.etsi.org/TB-SiteMap/MEC/List-of-Members}.}, which are not only mobile operators but also manufacturers, service providers, and universities, e.g., Vodafone, IBM, Intel, NTT Corporation, University CarlosIII de Madrid, etc. Their involvement plays a major role in ensuring an open and interoperable MEC environment, and MEC is beneficial to various stakeholders including MNOs, application developers, over-the-top players, independent software vendors, telecom equipment vendors, IT platform vendors, system integrators, and technology providers. The ETSI ISG has published a set of standards and specifications focusing on, for example, framework and reference architecture \cite{ETSI2019MEC_Framework}, MEC in the NFV environment \cite{ETSI2018MEC_DeploymentNFV}, and collocating C-RAN and MEC \cite{Reznik2018MEC_CRAN}.

The 3GPP started including MEC in the 5G network standardization in the technical specification 3GPP TS 23.501 \cite{3GPP2018SystemArchitecture}. Recently, in \cite{Kekki2018MEC5G} and based on functional enablers defined in \cite[clause 5.13]{3GPP2018SystemArchitecture}, the 3GPP clarified how to deploy MEC in and seamlessly integrate MEC into 5G, which can be illustrated in Fig.~\ref{Fig:IntegratedArchitecture}. Actually, the architecture comprises two parts: the 5G service-based architecture (SBA) on the left and an MEC reference architecture on the right.

The network functions defined in the 5G architecture and their roles can be briefly summarized as follows.
\begin{itemize}
	\item \emph{Access and Mobility Management Function} (AMF): establishes mobility and access procedures, e.g., connection management, reachability management, mobility event notification, termination of the RAN control plane, and access authentication/authorization.
	\item \emph{Session Management Function} (SMF): performs functionalities related to session management, e.g., session establishment, termination of interfaces towards policy control functions, and downlink data notification.
	\item \emph{Network Slice Selection Function} (NSSF): executes the allocation of slicing resources and AMF set to serve users. 
	\item \emph{Network Repository Function} (NRF): supports the discovery of network functions and their supported services.
	\item \emph{Unified Data Management} (UDM): handles user subscription and identification services.
	\item \emph{Policy Control Function} (PCF): unifies the network policies and provides policy rules to control plane functions.
	\item \emph{Network Exposure Function} (NEF): acts as a service-aware border gateway for providing secure communication with the services supported by the network functions.
	\item \emph{Authentication Server Function} (AUSF): performs authentication procedures. 
	\item \emph{User Plane Function} (UPF): provides functionalities to facilitate user plane operations, e.g., packet routing and forwarding, data buffering, and allocation of IP address. 
\end{itemize}
More details of the SBA and the 5G network functions can be found in 3GPP TS 23.501 \cite{3GPP2018SystemArchitecture}.

The MEC reference architecture is composed of the MEC system level and host level \cite{ETSI2019MEC_Framework}. The MEC orchestrator (MECO) is the core component of the MEC system level, which maintains information on deployed MEC hosts (i.e., servers), available resources, MEC services, and topology of the entire MEC system. The MECO is also responsible for selecting of MEC hosts for application instantiation, on-boarding of application packages, triggering application relocation, and triggering application instantiation and termination. The host level management consists of the MEC platform manager and the virtualization infrastructure manager (VIM). The MEC platform manager carries out the duties on managing the life cycle of applications, providing element management functions, and controlling the application rules and requirements. The MEC platform manager also processes fault reports and performance measurements received from the VIM. Meanwhile, the VIM is in charge of allocating virtualized resources, preparing the virtualization infrastructure to run software images, provisioning MEC applications, and monitoring application faults and performance. Finally, the MEC host comprises an MEC platform and a virtualization infrastructure. The former includes the set of functionalities needed to run MEC applications on a particular virtualization infrastructure and the latter includes the data plane functionalities of executing the traffic rules received by the MEC platform and steering the traffic among applications and networks. 

New functional enablers were defined in \cite{3GPP2018SystemArchitecture} to integrate MEC into the 5G SBA, which can be explained as follows.
\begin{itemize}
	\item \emph{User Plane Reselection and Selection}: The 5G core network supports the UPF (re)selection for selective traffic routing to the data network. Parameters used for the UPF selection mechanism is dependent on the UPF deployment scenario and MEC service operator configuration.
		
	\item \emph{Local Routing and Traffic Steering}: The UPF enables various traffic routing schemes for MEC applications in the 5G network. Moreover, application functions (AFs) may affect the UPF (re)selection and make specific traffic routing rules for a particular user.
	
	\item \emph{Local Area Data Network} (LADN): The support for LADN is enabled by the flexibility in the UPF location. Then, MEC hosts can be deployed on the N6 interface that is between the UPF and a data network. The user using MEC services may discover LADN availability during the registration procedure based on LADN information received from the AMF. 
	
	\item \emph{Session and Service Continuity} (SSC): The support for SSC is essential to enable user and application mobility. The 5G architecture allows MEC applications to select one among three SSC modes \cite{3GPP2018SystemArchitecture}. Particularly, SSC mode 1 provides the stable network connectivity to the user, SSC mode 2 may release the current connectivity to the user before making a new one, and SSC mode 3 ensures service continuity for the user by changing the new user plane before disconnecting the existing one.
	
	\item \emph{Network Capability Exposure}: The 5G architecture allows both direct access to network functions for the authorized MEC and indirect access via the NEF. Examples of exposed capabilities are exposure of user events, exposure of user behavior provisioning to external functions, and exposure of analytics to external parties.
	
	\item \emph{QoS and Charging}: The PCF in the 5G SBA defines QoS and charging rules for the user traffic routed to the LADN.

\end{itemize}
}

\begin{figure*}[ht]
	\centering
	\includegraphics[width=0.90\linewidth]{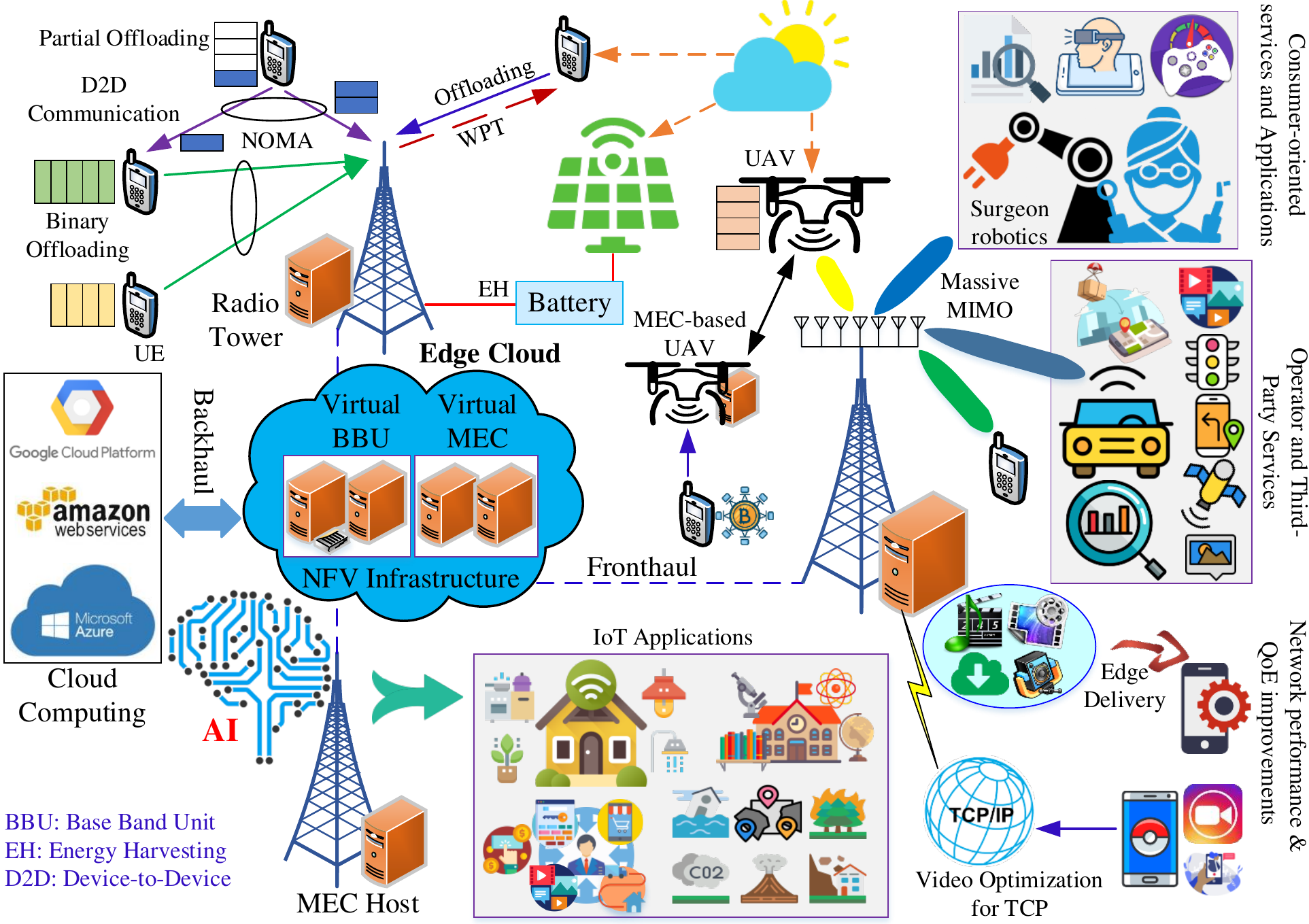}
	\caption{Integration of MEC with the forthcoming 5G technologies.}
	\label{Fig:MEC_5G_Scenario}
\end{figure*}

\subsection{MEC in 5G and Beyond}
\label{Subsec:MEC_Integration}

{\color{black}
Many services and applications will be supported in 5G and beyond, which can derive substantial benefits from MEC by being executed at the distributed edge servers. No matter what the service is, MEC use cases can be classified under three main categories, namely \textit{consumer-oriented services}, \textit{operator and third-party services}, and \textit{network performance and QoE improvements} (see Fig.~\ref{Fig:Overview_Of_MEC}) \cite{ETSI2018MEC_UseCases}. The fact is that the MEC ecosystem should support all these categories to create a myriad of new services and applications at the edge of mobile networks. Generally, the classification is dependent on who could reap the advantages and benefits. First, the use case ``consumer-oriented services" aims to bring direct benefits to users through the capability of running computation-heavy and latency-sensitive applications at the network edge. By means of computation offloading, users can exploit substantial computing resources on the edge server \cite{Ryu2019MEC}. Applications and services under the first category can include graphical rendering applications (e.g., 3D gaming, AR/VR, assisted reality, and cognitive assistance), intermediate data-processing (e.g., data analytics and video analysis), and low-latency applications (e.g., remote surgery on tactile Internet, AR/VR, video games, and interactive applications), and location-based service recommendation. Under the second category, operators and third parties take advantages of MEC computing and storage facilities to place their own applications and services on the network edge. This is enabled by the ``open and interoperable environment" nature of MEC, and is to encourage innovation and development in MEC from multiple parties and overcome obstacles (e.g., deployment difficulties and operational costs) in providing MEC services at the hard-to-reach areas \cite{Kekki2018MEC5G}. Applications and services offered by operators and third-party vendors can include V2X applications (e.g., safety, convenience, and driving assistance), big data, active device location tracking, security, safety, data analytics, and indoor precise positioning and content pushing. Finally, the services under ``network performance and QoE improvements" group intend to optimize operations of the network, thus improving the network performance and QoE. Examples of the third category are local content caching at the edge, mobile backhaul optimization, traffic deduplication, video delivery optimization for TCP, multi radio access technology (RAT) computation offloading, and network congestion in dense-network environments. Further details about the description of these use cases can be found in \cite{ETSI2018MEC_UseCases} and an earlier version of this survey at https://arxiv.org/abs/1906.08452.

To support the aforementioned applications and services, new architectures and technologies will be introduced in the 5G network. As shown in Fig.~\ref{Fig:MEC_5G_Scenario}, the integration of MEC with the forthcoming 5G technologies is necessary to achieve added values in MEC systems. A brief description of MEC in the 5G scenario is given as follows. 
\begin{enumerate}
	\item NOMA, millimeter Wave (mmWave), and massive multiple-input multiple-output (MIMO): As a multiple access technology to meet the demand for massive connectivity, the integration of NOMA into MEC systems is an important research issue, which needs more attention in the years to come. Moreover, the coexistence of MEC with mmWave massive MIMO is necessary to enable massive wireless connectivity with high data rates, low-latency, and large computing capabilities. These schemes are provided in Section~\ref{Sec:NOMA_MEC}. 
	 
	\item EH and WPT: Thanks to EH and WPT, the design of self-sufficient and self-sustaining wireless communications (aka green communication) becomes a reality. The combination of EH and/or WPT with MEC in a single system offers great potential to solve fundamental limitations of traditional systems, e.g., limited battery lifetime, unstable grid power supply, and low computing capability. To understand these issues more clearly, research works on EH and WPT MEC systems are surveyed in Section~\ref{Sec:WPT_MEC}.
	
	\item UAV communications: UAVs can be exploited to enable many potential applications due to their features of flexibility, mobility, maneuverability, and low cost. On the one hand, UAVs can be aerial edge servers to perform heavy computations offloaded from ground users. On the other hand, UAVs can act as new aerial users and associate with ground BSs to offload their tasks. The integration of UAV into MEC systems is a new and promising research topic, which will be summarized in Section~\ref{Sec:UAV_MEC}.
	
	\item IoT: IoT devices are quite resource-constrained to run compute-intensive tasks due to their limited computing capability and battery capacity. MEC is a powerful solution to solve these limitations. Inversely, IoT expands MEC services into more scenarios and objects like sensors, actuators, and mechanized agriculture. We will provide a survey on recent MEC-enabled IoT applications in Section~\ref{Sec:IoT_MEC}.
	
	\item H-CRAN: With the realization of NFV, the collocation of MEC and heterogeneous C-RAN (H-CRAN) is expected to bring potential benefits. In such scenario, the edge host (i.e., MEC server) in MEC and the BBU pool in H-CRAN can be collocated with each other to share the same virtualization infrastructure. We will study the integration of H-CRAN and MEC further in Section~\ref{Sec:HetNet_MEC}.
	
	\item Machine Learning:  The massive amount of mobile data, together with recent breakthroughs in ML and the non-convexity nature of resource allocation in a complex network, inspires many creative solutions for wireless communications and networking problems. ML plays a central role in the design of MEC mechanisms as we will elaborate in Section~\ref{Sec:MEC_MachineLearning}. 
	
	\item VM, SDN, NFV, and Network Slicing: MEC systems primarily rely on four enablers: VM, SDN, NFV, and network slicing. VM virtualization enables transient customization of MEC infrastructure, while SDN, NFV, and network slicing provide greater flexibility and agility of multi-tenant MEC ecosystems. For more information on these issues, we refer the interested readers to \cite{ETSI2019MEC_Framework, Taleb2017onMultiAccess, Baktir2017HowCan} and references therein.
\end{enumerate}

In summary, we focus on the following aspects in MEC systems: radio access (NOMA, mmWave, and massive MIMO), network architectures and scenarios (H-CRAN and UAV), applications (IoT, V2X, and UAV), power supply (EH and WPT), and performance improvement (ML). In the following sections, these researches are discussed in more details. 

}

\section{MEC with Non-Orthogonal Multiple Access}
\label{Sec:NOMA_MEC}
\subsection{Fundamentals of NOMA}
\label{SubSec:Fundamentals_NOMA}

Non-orthogonal multiple access (NOMA) has been considered as an essential principle for the design of radio access techniques in the emerging 5G network \cite{Dai2015NOMA, Pham2018AlphaFairness}. 
The key idea of NOMA is the use of the superposition coding technique at the BS side and interference cancellation techniques (e.g. multiple user detection and successive interference cancellation) at the user side. \color{black}{Compared to the conventional orthogonal multiple access (OMA), NOMA can enable multiple users to share the same time-frequency resource to achieve higher spectral efficiency. There are two main NOMA categories: power-domain NOMA and code-domain NOMA. Power-domain NOMA exploits the channel gain differences between users and multiplexes users in the power-domain while code-domain NOMA uses user-specific sequences for sharing the entire available radio resource \cite{Vaezi2019Multiple}. 
Typical examples of code-domain based access strategies are low-density spreading CDMA, low-density spreading-based OFDMA, sparse code multiple access (SCMA), and multi-user shared access (MUSA).} NOMA has the potential to accommodate more users than the number of available subcarriers, which leads to various potentials, including massive connectivity, lower latency, higher spectral efficiency, and relaxed channel feedback \cite{Shin2017NOMA}. The comparison between OMA and NOMA is summarized in Table~\ref{Tab:OMAvsNOMA} \cite{Vaezi2019Multiple}.

\begin{table}
	\caption{Comparison between OMA and NOMA.}
	\label{Tab:OMAvsNOMA}
	\centering
	\begin{tabular}{ | p{0.75cm} | p{3.2cm} | p{3.45cm} | }
		\hline
		& Advantages & Disadvantages  \\ \hline
		\hline
		\multirow{3}{*}{OMA} 
		& - Simpler receiver detection  & - Lower spectral efficiency  \\
		&    & - Limited number of users  \\
		&    & - Unfairness for users  \\
		\hline
		\multirow{5}{*}{NOMA} 
		& - Higher spectral efficiency & - Increased complexity of receivers. \\
		& - Higher connection density & - Higher sensitivity to channel uncertainty. \\
		& - Enhanced user fairness  &   \\ 
		& - Lower latency  &   \\ 
		& - Supporting diverse QoS  &   \\ 
		\hline
	\end{tabular}
\end{table}

Although NOMA is able to support a large number of users simultaneously and surpasses OMA in several aspects, various challenging problems associated with NOMA must be addressed before this technology can be employed in real networks. Islam \textit{et al.} in \cite{Islam2017PowerDomain} and Dai \textit{et al.} in \cite{Dai2018aSurveyOfNOMA} provided some research directives for NOMA in their survey: dynamic user pairing, impact of transmission distortion, channel and interference estimation, etc. \color{black}{NOMA can be flexibly combined with many existing wireless technologies and emerging ones including MIMO, massive MIMO, mmWave communications, cognitive and cooperative communications, visible light communications, physical layer security, energy harvesting, wireless caching, and so on \cite{MVNOMA2019}.} To gain a deeper understanding of the benefits and opportunities that NOMA offers as well as its challenges and application scenarios, interested readers are recommended to refer to NOMA research works, such as, \cite{Islam2017PowerDomain, Dai2018aSurveyOfNOMA, Ding2017ApplicationOfNOMA, Dai2015NOMA, Ding2017aSurveyOnNOMA, Vaezi2019Multiple,FangMagazine2018,FangTVT2019,FangWCL2019}.

\subsection{Motivation to combine NOMA and MEC}
\label{SubSec:Motivation_NOMA}
Both NOMA and MEC are considered as the key enabling technologies in 5G due to their enormous potentials and wide-range applications.  \color{black}{There are many benefits of MEC and NOMA, including increasing the number of connecting users, reducing the transmission latency and the energy consumption of end users and providing high performance for more complex network scenarios, i.e. mmWave massive MIMO.}
\begin{itemize}
	\item \color{black}{The combination of MEC and NOMA can significantly improve the user satisfaction and network performance through the provision of golden opportunities.} While NOMA offers several advantages at improving the spectral efficiency and cell-edge throughput, relaxing the channel feedback requirement, and reducing the transmission latency, MEC brings considerable benefits to not only users, but also operators and third-parties, and enables to improve overall network performance as well. It is expected that 5G will support a massive increase in device connections, high-speed transmissions of 1--10 Gbps, and greatly reduced latency and high reliability. 
	\item \color{black}{The combination of MEC and NOMA can reinforce the services and applications that are supported by the 5G network.} On the one hand, NOMA is expected to vastly increase the number of users in various scenarios where rank deficiency can occur \cite{Dai2018aSurveyOfNOMA}. On the other hand, edge computing in MEC indicates that computing resources are provided for end users in close proximity and at the edge of RANs. Therefore, MEC is capable of widely distributing computing resources from centralized cloud to the network edge and immediately serving a large number of users, hence MEC has the potential to support massive connectivity and distributed computation. 
	\item \color{black}{The combination of MEC and NOMA can provide low-latency transmission.} Because the 5G network will not completely rely on a single technology, we must optimize the network from multiple perspectives, e.g., air interface, network architecture, and enabling technologies. To cope with demands for lower latency, MEC and NOMA are two promising solutions. MEC moves the cloud services and functions to the network edge, where data is mostly generated and handled. Hence, MEC empowers the services running at the edge to better meet the lower latency requirements of end users compared to the cloud computing. In a similar sense, flexible scheduling and grant-free access in NOMA enables lower transmission latency for users in the 5G network. 
	\item \color{black}{NOMA and MEC can be flexibly combined with many existing wireless technologies, e.g., MIMO, massive MIMO, mmWave communications, etc., to further increase connectivity, spectral efficiency, energy efficiency and computing capability. For example, massive MIMO can drastically increase the spectral efficiency of wireless networks via excessive spatial multiplexing, thus Massive MIMO-NOMA can support massive connectivity and high spectral efficiency. To support gigabits-per-second data rates, mmWave bands can be used for wireless communications. The large path-losses caused by mmWave can be compensated by high gains, which can be obtained by massive MIMO. As a result, NOMA MEC can be deployed jointly with mmWave massive MIMO to enable multiple mobile devices to offload tasks simultaneously with high uploading/downloading data rates.}
\end{itemize}
Promoted by a variety of opportunities and advantages offered by MEC and NOMA, both academic and industrial communities have conducted extensive researches to design the 5G network with MEC and NOMA \cite{Ding2017aSurveyOnNOMA, Ding2019Impact, Tran2017Collaborative}. 
However, the state-of-the-art MEC researches still have not explored the full potential benefits of NOMA in the context of MEC. 
NOMA and MEC are both conceived as the bids to fill the gap between IoT devices and IoT applications and services. 
On the one hand, MEC empowers resource-constrained IoT devices with significant additional computational capabilities through computation offloading, thus bringing new applications and services to IoT devices. Similarly, with IoT, the scope of MEC services and applications is applicable to not only mobile phones, but also a wide range of smart objects ranging from sensors and actuators to smart vehicles. 
On the other hand,  NOMA is capable of substantially improving on system capability since it enables multiple users to transmit using a dedicated orthogonal channel resource. 
Furthermore, motivated by the benefits of NOMA over OMA, it appears utterly reasonable that one can exploit NOMA to further improve the use of MEC in IoT networks, as compared to the performance of conventional OMA-based MEC approaches. 

Apparently, NOMA can be exploited to increase the efficiency and performance of multi-user MEC systems. In the following, we present an overview of research works that have explored the combination of NOMA and MEC and then discuss fundamental challenges and open directions. 
\begin{table*}[ht!]
	\caption{Summary of existing works on NOMA MEC}
	\label{Table:NOMA MEC}
	\centering
	\begin{tabular}{|p{2cm}|p{2.9cm}|p{1.8cm}|p{9.5cm}|}
		\hline 
		\textbf{Topic} & \textbf{Designed frameworks} & \textbf{References}&\textbf{Contributions}\\ 
		\hline		\multirow{2}{2.2cm}{Architecture of
			\newline NOMA MEC} & Uplink NOMA MEC and downlink NOMA MEC&\cite{ZDing2018TCOM,MVNOMA2019}  & NOMA MEC outperforms OMA MEC on lower latency and lower energy consumption than the traditional OMA scheme. \\ 
		\hline
		\multirow{2}{2.2cm}{Energy consumption minimization} &\emph{Partial offloading}&\cite{FWang2017Globecom,YPanCL2018}  & Resource allocation schemes are proposed to minimize the energy consumption for a uplink and downlink NOMA enabled MEC system. \\ 
		\cline{2-4} 
		& \emph{Binary offloading}
		& \cite{ZDing2018WCL,AKiani2018JIOT,XCao2019,LPSIC2018IoT} & A hybrid NOMA MEC system is proposed to minimize energy consumption. \\
		\hline
		\multirow{4}{2.2cm}{Task delay minimization} &\emph{Partial offloading}&\cite{FangGlobecome2019,YWuJSTSP2019,XCao2019} & Optimal task and power allocation is proposed to minimize the task delay in NOMA MEC system.   \\ 
		\cline{2-4}
		& \emph{Binary offloading}& \cite{ZDing2018WCLDely}& Dinkelbach and Newton’s methods are compared to minimize task delay for the hybrid NOMA MEC system. \\ 
		\hline		
	\end{tabular}
\end{table*}

\subsection{State of the Art}
\label{SubSec:State_of_the_Art_NOMA}
While the use cases of NOMA or MEC have been widely studied in the literature, there are only a few studies on MEC-NOMA scenarios. The advantages of NOMA and MEC have motivated several studies supporting the application of NOMA to MEC \cite{ZDing2018TCOM,FWang2017Globecom,ZDing2018WCL,ZDing2018WCLDely,AKiani2018JIOT,QPham2019TVT}.
When NOMA uplink transmission is applied to the MEC system, multiple users can offload their tasks to the MEC server simultaneously via the same frequency band. By applying the successive interference cancellation (SIC) technology at the MEC server, the MEC server can remove the interference from the user whose data has been decoded before on the same frequency band. When NOMA downlink transmission is applied to the MEC system, one user can utilize NOMA to offload multiple tasks to multiple MEC servers simultaneously via the same frequency band. The performance comparison of NOMA-MEC and OMA-MEC systems was conducted in \cite{ZDing2018TCOM}, which reveals that the NOMA-MEC system can achieve superior performance in reducing latency and energy consumption.

Most existing research works focus on resource allocation i.e., computation resource and communication resource. Specifically, in \cite{FWang2017Globecom}, partial offloading assignment (i.e., each user can partition the computation task into two parts for local computing and offloading) and power allocation were investigated to minimize the weighted sum of the energy consumption for a multi-user NOMA-MEC system. 
In this work, an efficient algorithm for user's task partitioning, local computing CPU frequency and transmit power allocation was proposed to achieve the minimum energy consumption for multi-user NOMA-MEC networks. Unlike OMA-MEC and pure NOMA-MEC systems (i.e., both the users offload all of their tasks at the same time) proposed in \cite{FWang2017Globecom,ZDing2018TCOM}, a hybrid NOMA  strategy (i.e.,  a user  can  first  offload  parts  of  its  task within time slot allocated to other user and then offload the remaining of its task during a time slot solely occupied by itself) was proposed in \cite{ZDing2018WCL}, in which power allocation and time allocation were optimized to minimize the energy consumption for an MEC-enabled NOMA system. Subsequently, the delay minimization was investigated for the hybrid NOMA-MEC system \cite{ZDing2018WCLDely}. Different from partial offloading tasks, the authors in \cite{AKiani2018JIOT} considered that the offloading tasks are independent and non-separate. Then the communication resource (i.e., frequency bands and transmit powers) and the computing resource (i.e., computing resource blocks) were jointly optimized to minimize the energy consumption for the NOMA-MEC system \cite{AKiani2018JIOT}, in which an efficient heuristic algorithm of user clustering and frequency and resource block allocation was proposed to address the energy consumption minimization problem per NOMA cluster.
\color{black}{In \cite{QPham2019TVT}, the computing offloading scheme was investigated in the NOMA MEC system where a distributed algorithm based on game theory was proposed to improve the system performance. Moreover, the delay minimization problem was investigated in \cite{FangGlobecome2019,YWuJSTSP2019}. In \cite{YWuJSTSP2019} an efficient algorithm of the offloading workload, offloading and downloading duration optimization was proposed to minimize the overall delay of the computation tasks. The energy efficient power allocation, time allocation and task assignment were proposed to minimize the energy consumption for MEC networks \cite{XCao2019,YPanCL2018}. Besides the computational resource, SIC decoding order was optimized to reduce the task delay for NOMA enabled narrowband Internet of Things (NB-IoT) systems \cite{LPSIC2018IoT}. The summary of the existing works on NOMA MEC is provided in Table~\ref{Table:NOMA MEC}.}

\subsection{Learned Lessons and Potential Works}
\label{SubSec:Lesson_and_Potential_Works_NOMA}
Because of limited researches advocated to coexisting MEC-NOMA scenarios there are many key open problems that must be investigated. \color{black}{The potential works of NOMA and MEC can be viewed from the following four aspects: 1) Joint resource optimization; 
2) Secure communications; 3) Cooperative NOMA MEC; 4) Coexistence of NOMA MEC and mmWave massive MIMO.}
\begin{itemize}
	\item \emph{Joint resource optimization}: 
	\color{black}{Resource allocation plays an important role to improve the performance of the wireless networks. Thus in MEC-NOMA networks, the communication and computing resource can be jointly optimized to enhance the system performance, i.e., sum rate and energy efficiency. In other words, the scheduler may need to decide the computation load that the user can offload to the MEC server, and the remaining can be computed locally to minimize the latency. Moreover, computation capacity (i.e. processing speed of MEC servers or mobile devices) and communication resource (i.e. transmit power) are also important factors to reduce the computation latency. Joint optimization of these factors presents an open and challenging research problem. When NOMA uplink transmission is applied to the MEC system, multiple users can offload their tasks to the MEC server at the same time. Therefore, the total latency experienced by the multiple users can be investigated. By controlling the offloaded computation load and transmit power of each user, the optimal and suboptimal strategies can be developed to minimize the total latency of the system by considering the total energy consumption. The proposed solution can be extended to the NOMA downlink MEC system. Moreover, user grouping or user association can be another trend in resource optimization of MEC-NOMA systems, where matching theory \cite{FangTCOM2016} or game theory can be exploited to group users into different groups which use different subchannels to offload their tasks. Besides, the performance of the SIC technology is sensitive to the availability of channel state information (CSI). Thus  another possible direction to address this issue is to rely on partial CSI. The application of partial CSI in downlink NOMA system was investigated in \cite{ZYangTCOM2016,JCuiSPL2016,FangJSAC2017,PXuTWC2016}, which can be investigated in resource allocation for MEC-NOMA systems.	}
	\item \emph{Secure communication}: Security and privacy-preserving communication attract lots of research attention, specially when NOMA is applied to the MEC system. For example, two users are offloading tasks to an MEC server at the same time by using the NOMA principle. When SIC is performed, one user can decode the other user's message. During this period, an eavesdropper or an attacker may attempt to decode the mobile user's message. To address the scenario with external eavesdropper, the physical layer security can be utilized to cope with this challenge for the NOMA-MEC system \cite{YZhangCL2016,YLiuTWC2017}. The combination of PLS and NOMA-MEC is a promising research topic.
	\item \emph{Cooperative NOMA-MEC system}: To improve the connectivity of the NOMA-MEC network, the cooperative MEC can be adopted to enable computation offloading to the main MEC server. In this scenario, the mobile device transmits the superimposed signals to the primary MEC server and the helper MEC server, which acts as a relay helping MEC server \cite{DingWCL2016,GLiuTWC2017}. Considering the local computing capacity of the mobile users and energy consumption constraint, the task assignment and transmit power allocation can be optimized to improve the performance of NOMA MEC system.
	\item \color{black}{\emph{Coexistence of NOMA MEC and mmWave massive MIMO}: Massive MIMO-NOMA is another scenario to support massive connectivity and high spectral efficiency \cite{MVNOMA2019}. To further improve the transmission data rate, the mmWave bands (30 GHz to 300 GHz) have been proposed to provide gigabit-per-second data rates. Therefore, the integration of NOMA MEC into mmWave MIMO based wireless networks can improve the computing capability, spectrum efficieny and reduce the task delay, where multiple mobile devices can transmit tasks simultaneously via the mmWave bands. Inspired by the challenges of the traditional MIMO transmission scheme, an efficient approach of joint beamforming design and communication and computing resource allocation will be a major challenge to tackle. Moreover, the user grouping needs to be well investigated to further enhance the system performance. }
\end{itemize}

\section{MEC with Energy Harvesting and Wireless Power Transfer}
\label{Sec:WPT_MEC}
\subsection{Fundamental of Energy Harvesting and Wireless Power Transfer}

\color{black}{The current industrial landscape is becoming increasingly aware of the need to optimize energy use and management for all domains, including telecommunications. Among others, EH, also known as energy scavenging or power harvesting, is a promising technique for 5G systems since EH is an alternative solution to traditional energy supply sources \cite{Liu_nwrk_15}. The basic concept of EH is to capture various available energy from different sources to power the energy-constrained devices for prolonging their lifetime \cite{S_Sudevalayam_CST_2011,G_Piro_2013,Kamalinejad_ComMag_2015}. 
Together with the traditional energy grid, EH can help to fulfill the energy requirements of the different tiers of 5G networks including the sensors in IoTs, the mobile devices, the eNBs in HetNets, assisting relays in D2D systems, and the computing servers \cite{M_Imran_Arxiv2019}.
 Additionally, the recent development in advanced materials and hardware designs helps realize the EH circuits for small portable consumer
 electronic devices which accelerate the adoption of EH for the IoTs \cite{Dong_Ma_Arxiv2019,F_Unlu_EDNA18}.}

\begin{figure*}[t]
	\centering
	\includegraphics[width=0.90\linewidth]{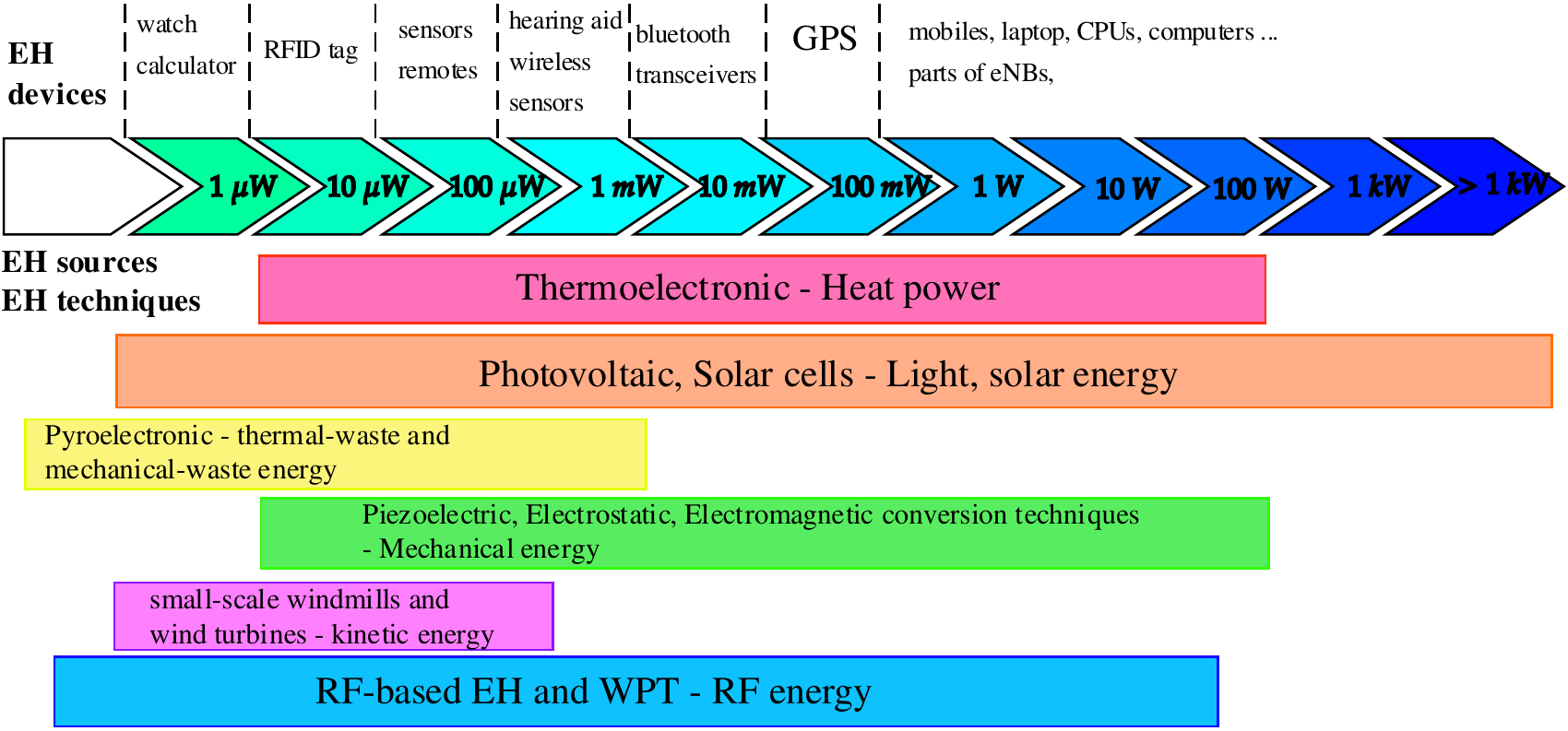}
	\caption{EH technologies with generated intermittent power and power consumption for various devices, adapted from \cite{M_Imran_Arxiv2019,Dong_Ma_Arxiv2019,F_Unlu_EDNA18}.}
	\label{Fig:Energy_devices}
\end{figure*}
 
\color{black}{EH is simple in concept, but more complex in implementation which strongly depends on the type of EH power sources. 
The harvestable energy can be scavenged from natural or human-made sources which are controllable or uncontrollable \cite{Lumpkins_CEMag_2014}.
As illustrated in Fig.~\ref{Fig:Energy_devices}, various EH techniques can be employed to leverage the corresponding energies \cite{M_Imran_Arxiv2019,F_Unlu_EDNA18}.}
Compared with the traditional natural energy sources, RF signals are less affected by weather or other external environmental conditions. As a result, these signals can be efficiently controlled and designed, so RF-based EH has great potential to provide stable energy to low-power energy-constrained networks including wireless sensor networks (WSNs), IoTs, and extremely remote area communication (eRAC) use cases in 5G networks \cite{CR_Valenta_mW_Mag_2014}. Specifically, RF-EH can be employed in indoor, hostile, and harsh environments, e.g., sensors inside a building or human body, toxic environment, and so on \cite{CR_Valenta_mW_Mag_2014}.  RF-EH can scavenge wireless energy from 1) ambient sources (e.g., WiFi, AM, and FM) which can be predictable or unpredictable or 2) dedicated sources which are deployed to provide an energy supply to meet the requirements of the mobile/edge nodes. RF-EH exploiting the ambient sources normally requires an intelligent process to monitor the communication frequency bands and time periods for harvesting opportunities. RF-EH with proper management of dedicated energy sources between the emitters and the harvesters can be considered as WPT.

WPT was first proposed by Nikola Tesla in 1899 \cite{Lumpkins_CEMag_2014} and continuously studied by both industry and academic communities.
Existing WPT technologies can be categorized into three classes: inductive
coupling, magnetic resonant coupling, and RF-based WPT. 
The first two technologies rely on near-field electromagnetic (EM) waves, which cannot support mobility for the energy-limited wireless communication devices due to the limited wireless charging distances (a few meters) and the required alignment of the EM field with the EH circuits \cite{Duong_Trung_book_2019}. 
In contrast, RF-based WPT exploits the far-field properties of EM waves over long distances (hundreds of meters).
In the RF-based WPT system, embedding the modulated information (e.g., phase-embedded information) into the RF-based WPT signals forms the concept of simultaneous wireless communication and power transfer (SWIPT) which was proposed and studied in \cite{X_Zhou_TCom_2013} from an information theoretical perspective. 

 \color{black}{Recently, the works \cite{Liu_nwrk_15,M_Imran_Arxiv2019} have demonstrated that integrating EH/WPT into typical 5G systems including IoTs, device-to-device (D2D) networks, HetNets, and cognitive radio networks (CRNs), can bring benefits in improving energy and spectral efficiency.} However, integration of EH/WPT in 5G architecture also raises some technical challenges as follows.
\begin{itemize}
\item  \color{black}{How to cover the unstable and intermittent characteristics of the ambient resources, i.e, power, spectrum, periods, is a challenging problem which should be considered in designing an EH systems.}
\item  \color{black}{How to allocate the network resources to well balance between harvested energy and consumed power is another issue. Towards this end, one must well understand the generating environment and the characteristics of energy source, the power consumption properties of different elements in the system, the 
coverage area, the communication distance, the data rate, and underlying application specifics.}
\item  \color{black}{In WPT systems, since the energy harvesting process may affect the modulated information, joint resource allocation for the EH and data transmission should be investigated to improve the network performance.}
\end{itemize}
An in-depth study of these challenges is required to design an efficient wireless network, which must consider different factors, like features of power generators, transducers, power storage, power management methods and application requirements.
For deeper understanding of the benefits and opportunities offered by EH and WPT techniques as well as their challenges and potential application scenarios interested readers are recommended to refer to EH and WPT surveys, such as \cite{G_Piro_2013,S_Ulukus_JSAC_2015,X_Lu_CST_2015,AHMED2015234,B_Clerckx_JSAC_2019}.

\begin{figure*}[t]
	\centering
	\includegraphics[width=1.00\textwidth]{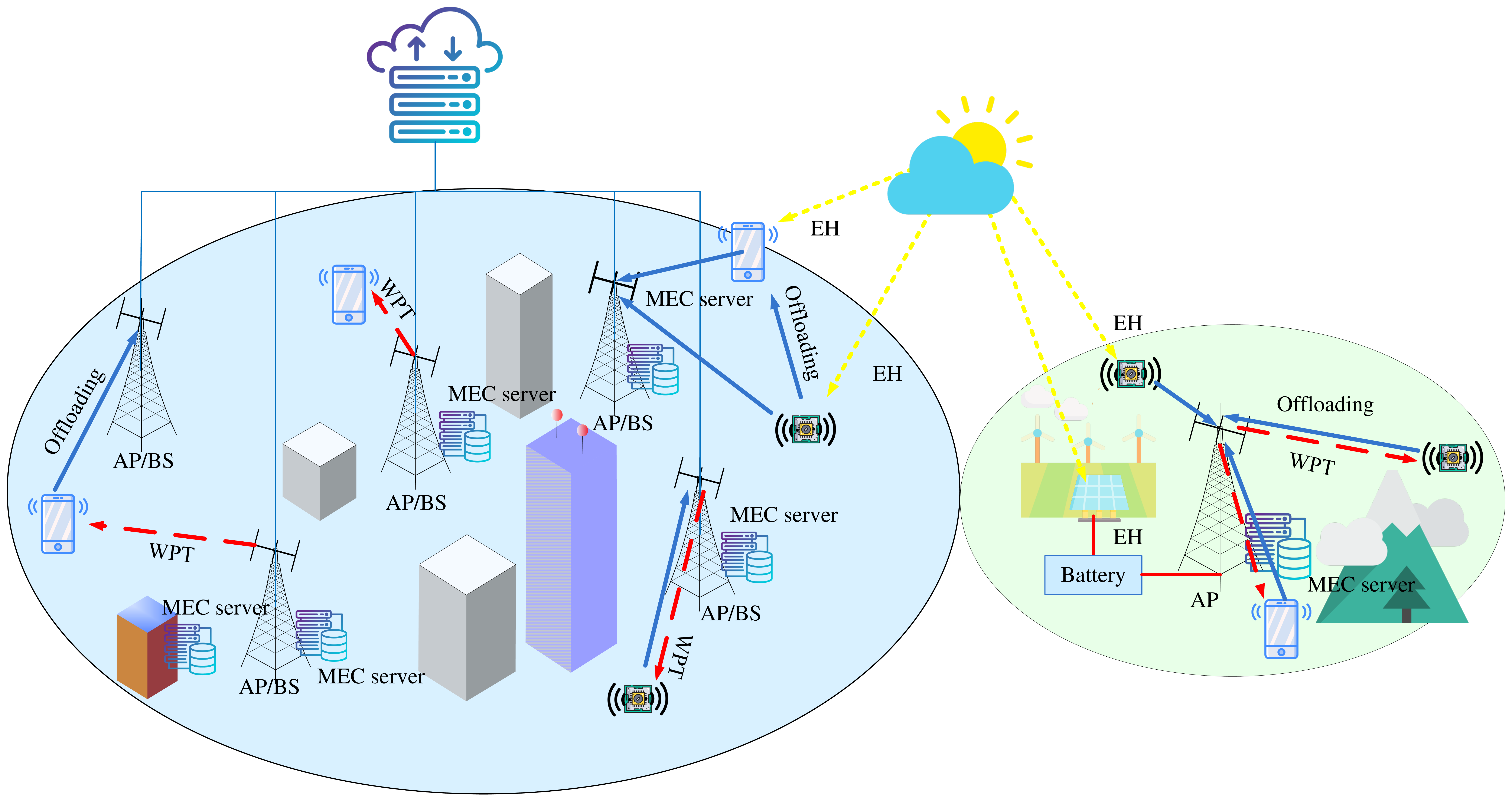}
	\caption{EH/WPT enabled MEC Access Networks.}
	\label{Fig:EH_MEC}
\end{figure*}

\subsection{Motivation}
\label{SubSec:Motivation_WPT_EH}

\color{black}{Both EH/WPT and MEC have been considered as promising technologies for the 5G networks, which can improve the energy efficiency of mobile/edge devices and prolong their battery lifetime of communication nodes at remote areas.
While MEC enables to detach the end devices from heavy computation workloads for saving their energy consumption, EH/WPT techniques allow them exploit the energy in their surrounding environment for re-charging their batteries.
Hence, integrating these two technologies in the future wireless communication systems can significantly improve network performance by leveraging the strengths of both underlying technologies.}
EH, WPT and MEC technologies will lead to the following benefits:
\begin{itemize}
\item  \color{black}{EH/WPT techniques can power the edge devices in the MEC systems to enlarge the set of options for computation offloading which will result in improving the network performance \cite{M_Min_TVT_2019}.
Specially in the IoT context, important use cases of MEC-enabled 5G networks, scavenging the ambient environment and utilizing WPT provide promising solutions to perpetually support the massive number of electronic sensors \cite{Dong_Ma_Arxiv2019}.
Moreover, EH/WPT modules by leveraging green energy (e.g. solar and/or wind) can be employed to power MEC servers.
Especially, EH/WPT provides a great solution for the eRAC use cases of MEC-based 5G where the MEC servers and mobile edge devices can be located outside the coverage of the electric grid for reasons such as deployment constraints, reliability requirements, carbon footprint, weather, disasters, and maintenance expenses.}
\item  \color{black}{The distributed computing power of MEC systems can be leveraged to learn the time-varying properties of the energy source for optimizing the network performance \cite{J_Xu_TCCN_2017}.}
\item  \color{black}{A MEC server can be deployed to support a cluster of mobile/sensor nodes in EH-enabled wireless networks. At the node level, MEC can help each EH device reduce processing time and reserve more time for EH by offloading its heavy workloads to fog servers \cite{Balasubramanian_WFIoT_2018}. At the network-level, MEC can allow to deploy a centralized EH strategy to tune the functionality of all devices for better EH and performance \cite{S_Sudevalayam_CST_2011}.}
\end{itemize}

A simple EH-enabled and wireless powered MEC system is illustrated in Fig.~\ref{Fig:EH_MEC} in which the EH and WPT are employed at MEC servers and mobile devices.
While EH and WPT bring many benefits as discussed above, the MEC system also faces many challenges including communication resource allocation, computing resource allocation, latency minimization, and security problem.
In the following, we describe some major challenges which must be addressed for efficient integration of EH/WPT into the MEC system.
\begin{itemize}
\item In general, mobile devices have limited battery size and computation capability. Integrating EH/WPT into MEC-based wireless networks facilitates mobile devices with an external power source for processing heavy workload but this requires additional processing workload on controlling the EH function. Thus, such integrated system must cope with a more complicated resource allocation problem.  Major research issues along this line include resource allocation and offloading design to well balance between harvested energy and consumed energy consumption. Specifically, how to perform the energy-efficient computation offloading in EH/WPT-MEC system considering practical constraints in the harvesting process remains a challenge issue. 
\item In the scenarios where MEC servers are primarily powered by renewable energy, the availability of energy source in the space and time domains would follow a unstable and non-uniform distribution. Moreover, these harvested energy level may vary over space which leads to load imbalance among servers. Hence, load balancing among all edge servers is also an interesting research problem which should be addressed in engineering EH/WPT-based MEC systems.
\end{itemize}

\subsection{State of the Art}
\label{SubSec:State_of_the_Art_WPT_EH}

\begin{table*}[!t]
	\caption{Summary of EH/WPT-MEC works.}
	\label{Table:EH_WPT_works}
	\centering
	\begin{tabular}{|l|l|l|l|l|}
		\hline
		& EH-enabled devices                                                    & EH-enabled MEC server                                           & WPT                                                                 & SWIPT                                       \\ \hline
		Resource Allocation & \cite{W_Chen_TSC_2018,G_Zhang_TII_2018}                               & \cite{S_Ulukus_JSAC_2015,Jie_Xu_Arxiv_2017,F_Gou_INFOCOM_2018} & \cite{S_Bi_TWC_2018,Y_Dong_Access_2019,L_Ji_JIoT_2019,D_Wu_GC_2018} & \cite{H_Zheng_JIoT_2019,Janatian_WCNC_2018} \\ \hline
		Offloading Designs  & \cite{W_Chen_TSC_2018,G_Zhang_TII_2018,L_Liu_IoT_2018,L_Liu_IoT_2018} &                                                                & \cite{S_Bi_TWC_2018,Y_Dong_Access_2019,L_Ji_JIoT_2019,D_Wu_GC_2018} & \cite{H_Zheng_JIoT_2019,Janatian_WCNC_2018} \\ \hline
		Load Balancing      &                                                                       & \cite{S_Ulukus_JSAC_2015,Jie_Xu_Arxiv_2017,F_Gou_INFOCOM_2018} &                                                                     &                                             \\ \hline
	\end{tabular}
\end{table*}

\color{black}{This section provides a survey on some recent works for efficient integration of EH/WPT into MEC systems which are summarized in Table~\ref{Table:EH_WPT_works}. Existing works on combining EH/WPT and MEC mainly consider three schemes.  In the first two schemes, the EH and WPT techniques are implemented at mobile devices in MEC-enabled wireless communication networks. These schemes can be applicable to WSNs, IoTs, eRAC, and D2D systems in the 5G network which support a massive number of small battery-operated devices connecting wirelessly to MEC servers for offloading and data processing. Because these devices typically have very limited batteries to supply power for their communication, EH and WPT technologies can be employed to provide valuable additional powers for their long-term operations such as sensing, reporting data, or offloading the heavy computation load. To do so, the edge devices need to estimate the power and time consumed by their operation. The resource allocation and offloading decision design for these devices become more complicated due to the additional energy harvesting stages in EH/WPT enabled MEC systems which are promising research issues. The third scheme focuses on the scenarios in which connecting the MEC servers of MEC-enabled systems to the electric grid is costly and even imposible in certain situations such as natural disasters, remote locations. Then, on-site renewable energy is mandated as a major or even sole power supply source for these MEC servers \cite{Jie_Xu_Arxiv_2017}. In these cases, efficient load balancing design among all MEC servers under the unpredictable and unstable harvested energy has attracted a lot of attention from both industry and academic societies.
}

\subsubsection{Offloading and Resource Allocation for MEC-enabled systems using EH technique}
Recently, \cite{W_Chen_TSC_2018} considered the multi-user multi-task computation offloading problem which aims to
maximize the overall revenue of the wireless EH-enabled devices. 
The Lyaponuv optimization approach was adopted in this work to devise the energy harvesting policy and the task offloading schedule.
The tradeoff between energy consumption and execution delay for the MEC system with EH capability was studied in \cite{G_Zhang_TII_2018}
in which the authors proposed an online dynamic task scheduling to minimize the average weighted energy consumption and execution delay subject to constraints on the stability of buffer queues and battery level.
Employing the game theoretic approach, authors in \cite{L_Liu_IoT_2018} studied the impact of the EH technique at mobile
devices in the computation offloading design. The work aimed to minimize the social group execution cost. Different queue models are applied to model the energy cost and delay performance, based on which a dynamic computation offloading scheme was designed.
Using the deep learning (DL) approach, \cite{M_Min_TVT_2019} proposed a reinforcement learning based offloading scheme, where each EH-based IoT device selects its MEC server and the offloading rate without knowledge of the MEC model based on the current battery level, the previous radio transmission rate to each server, and the predicted harvestable energy. 

\subsubsection{Offloading and Resource Allocation for MEC-enabled systems using WPT technique}
Considering the WPT-enabled MEC systems, \cite{S_Bi_TWC_2018} aimed to maximize the (weighted) sum computation rate of all wireless devices
in the network by jointly optimizing the individual offloading decision and the time allocation for transmission.
Similarly, \cite{Y_Dong_Access_2019} considered the time division strategy for the two-way data exchange between the fog node and the mobile user in WPT-based MEC systems. 
The closed-form average age of information for both directions as well as the achievable data rate of the mobile user were described in this paper, based on which the trade-off between the downlink and uplink performance was investigated.
The cooperation among edge users was studied in \cite{L_Ji_JIoT_2019,D_Wu_GC_2018}. Specifically, the work \cite{L_Ji_JIoT_2019} aimed to maximize energy efficiency (EE) to ensure the fairness of users by encouraging the near user (NU) forwarding the far user’s (FU) tasks to the edge cloud.  
While \cite{D_Wu_GC_2018} enabled the surrounding idle devices as the helpers to use their opportunistically scavenged wireless energy to help remotely execute active users’ computation tasks. 
The work tried to maximize the computation rate by jointly optimizing the transmit energy beamforming at the ET, as well as the communication and computation resource allocations at both the user and its helpers.
\cite{F_Zhou_JSAC_2018} considered an WPT-based UAV-assisted MEC system in which an UAV acts as an MEC-enabled BS offering WPT and offloading services to a number of EH-enabled ground mobile devices. The work aimed to maximize the system computation rate under both partial and binary computation offloading modes, subject to the energy-harvesting causal constraint and the UAV’s speed constraint. 
On another approach, \cite{H_Zheng_JIoT_2019} investigated the power splitting problem for information transmission and power transfer in the SWIPT-based MEC system. 
Specifically, the authors proposed a new algorithm to minimize the required energy under the constraints on required information transmissions and processing rates. 
\cite{Janatian_WCNC_2018} considered the time division protocol for the SWIPT-based fog computing system that consists of a multi-antenna access point (AP), an ultra-low power (ULP) single antenna device and a fog server. In this work, the time slots devoted to EH, ID and local computation as well as power required for the offloading are optimized to minimize the energy cost of the ULP device.

\subsubsection{Load balancing design for multiple EH-based MEC servers}
In the EH-based MEC systems where the computation servers are mainly powered by the uncontrollable and unpredictable energy sources (e.g, solar, wind), individual MEC servers may be overloaded at any moment due to the limited harvested power and computing capacity \cite{S_Ulukus_JSAC_2015}.
Hence, energy prediction and load balancing among all EH-based servers are important research issues which must be tackled to achieve effective MEC operations.
In particular, \cite{Jie_Xu_Arxiv_2017} considered a joint geographical load balancing and admission control for EH-based MEC networks which aims to minimize the long-term system cost due to violating the computation delay constraint and dropping data traffic.
To deal with this geographically load balancing (GLB) optimization problem, Xu et al. developed an algorithm, called GLOBE, by leveraging the Lyapunov stochastic optimization technique.
In particular, the algorithm enables MEC-enabled BSs to make GLB decisions without requiring future system information. 
Integrating the EH into MEC-enabled HetNets, authors in \cite{F_Gou_INFOCOM_2018} investigated the joint load management and resource allocation problem that maximizes the number of offloading users utilizing the limited energy and
computation resources, via managing the load and distributing the resources to the users. 
To solve the underlying complicated problem, a distributed three-stage iterative algorithm was proposed. 
In the first stage, the load balancing among all BSs is conducted for a given channel allocation and computation resource allocation scheme.
In the remaining two stages, the channel allocation and computation resource allocation were optimized in turn based on the load traffic achieved in the first stage. Finally, an iterative algorithm is developed to obtain the joint load balancing and resource allocation solution.

\subsection{Learned Lessons and Potential Works}
\label{SubSec:Lesson_and_Potential_Works_WPT_EH}
Due to the great benefits offered by MEC and EH/MEC as well as their complementary properties, it is convinced that
the combination of MEC and EH/WPT is beneficial in the future. Although various problems and issues in EH/WPT-MEC systems have been intensively studied, there are still several challenges. In the following, we discuss some challenges in EH/WPT-MEC systems and outline the open research directions.
\begin{itemize}
\item{\textit{Energy Prediction:}} Most of the renewable energy sources are unpredictable. For example, clouds can appear or disappear which can affect the solar harvesting process. Other kinds of harvestable energy sources, e.g., wind, heat, and vibration, vary over the time. In the WPT systems, channel characteristics practically vary depending on the environment in which the level of interference and the number of paths cannot be known in advance.  Thus, understanding the surrounding ambient environment is critical for efficient implementation of the EH and WPT techniques. Recently, advanced machine-learning and deep-learning methods have been utilized to predict the arrival energy based on the historical and geographic data. Notwithstanding considerable benefits, ML/DL mechanisms and big data analytics raise some several challenging issues for implementation, such as, collecting data, large computation resources required to process the high-dimensional big data, which can be overcome by employing the MEC concept. Exploiting learning at MEC servers to extract useful information collected by all EH-enabled devices can reduce the time caused by sending the data to a remote cloud server; hence, the predicted information can be achieved on-time for high efficient EH, which can extend the capability of EH-enabled devices. 

\item{\textit{EH/WPT-based MEC for IoT/dense networks:}}
An IoT network aims at supporting a massive number of connections from machine-type devices which are small, fabricated and deployed at very low cost, and are expected to operate in a self-sufficient manner for a long time. The large number of connecting devices and their low power operation require an advanced wireless access networks, such as, dense access points or multi-hop data transmissions. MEC systems can play a relevant role in this scenario to manage functionality of individual nodes in terms of synchronization, reliability, efficiency of utilizing channel resource and energy, to exploit the available harvestable energy source, to cooperate with others for WPT, data transmission and offloading. The other challenge in successful large-scale deployment of devices in an IoT infrastructure is to minimize their impact on human-body and the environment \cite{Kamalinejad_ComMag_2015}. The presence of multiple devices implementing various EH technologies corresponding to different kinds of energy sources, WPT and SWIPT over different frequency bands in the dense-users networks also require efficient and scalable offloading and resource allocation designs. 
\end{itemize}

\section{MEC for UAV Communications}
\label{Sec:UAV_MEC}
\subsection{Fundamentals of UAV}
\label{SubSec:Fundamentals_UAV}
\begin{figure*} [ht]
	\centering
	\includegraphics[scale=0.60]{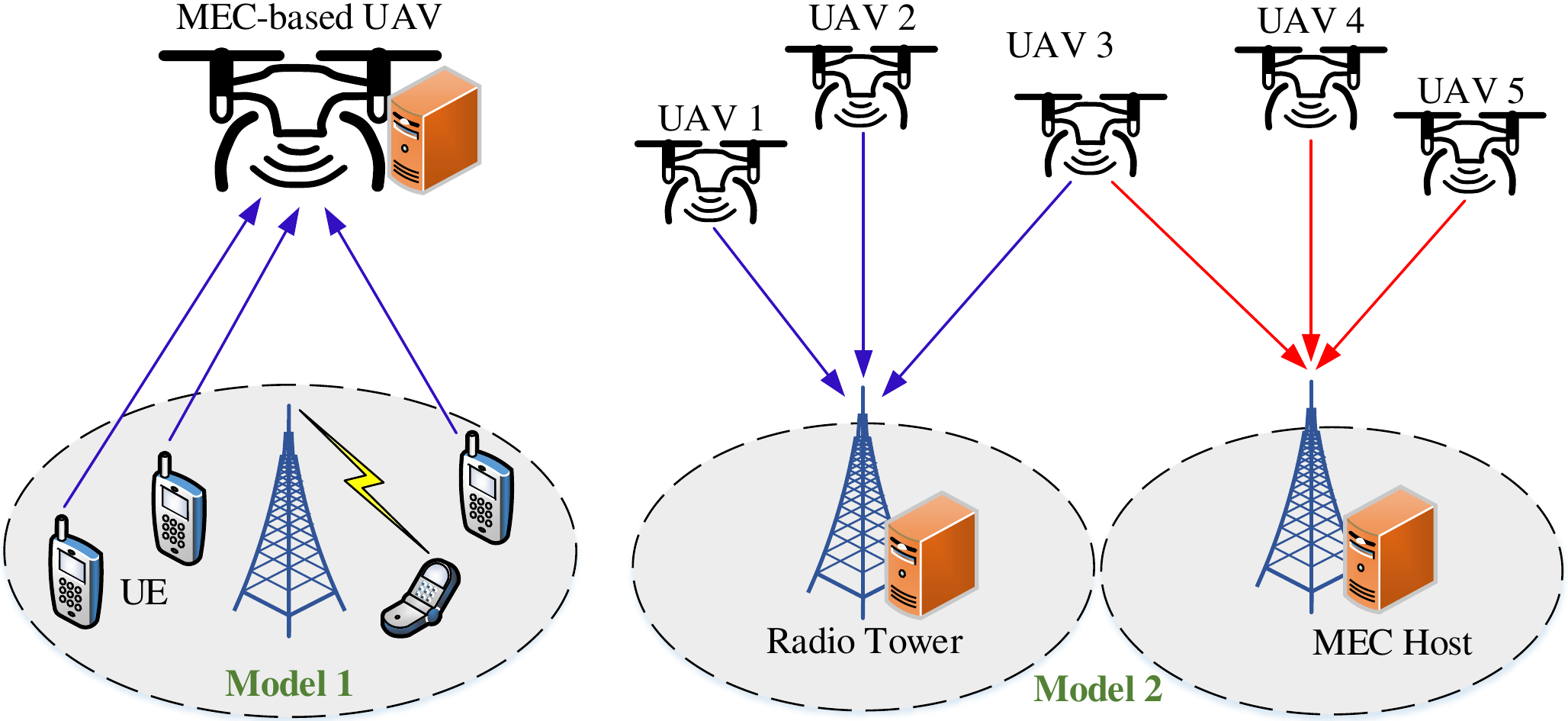}
	\caption{MEC-enabled UAV Networks Architecture.}
	\label{Fig:UAVMEC}
\end{figure*}
Historically, UAVs have been considered as enablers of various applications including military, surveillance and monitoring, telecommunications, delivery of medical supplies, and rescue operations, owing to their autonomy, flexibility and broad range of coverage \cite{RWSUA2012,RYICC2017}. However, in those applications, UAVs mainly focused on navigation, control, and autonomy. As a result, the communication challenges of UAVs have typically been either neglected or considered as part of the control and autonomy components \cite{Mozaffari2018ATutorial}. UAVs are commonly known as drones or remotely piloted aircrafts, and have several key potential applications in wireless communication systems due to its high mobility, flexibility, adaptive altitude and low cost \cite{YZengMag2016}. Specifically, small UAVs are more easily accessible to the public recently due to its continuous cost reduction and device miniaturization, thus small UAVs can be used in weather monitoring, forest fire detection, traffic control, emergence search and rescue, cargo transport etc. In recent years, UAV-based wireless communication systems attract lots of attention thanks to their cost-effective wireless connectivity in scenarios without infrastructure coverage, which is caused by severe shadowing by urban or mountainous terrain, or damage to the communication infrastructure caused by natural disasters \cite{Merwaday2015}. Among the UAV applications in wireless communication systems, UAV mainly serves as two important roles: 1) aerial base station, 2) flying mobile terminals. In the first scenario, when UAV serves as an aerial base station, it can provide communications in emergency and public safety situations to enhance coverage, capacity, reliability and energy efficiency of the wireless networks. In the second scenario, UAV can serve as a flying mobile terminal within the cellular networks to deliver real time video stream.

\color{black}{For UAV classifications, several factors such as outlook and application goals, need to be taken into account. The different types of UAVs depend on their functions, and capabilities. From their outlook characteristics, UAVs can be broadly classified into two categories: fixed-wing UAVs and rotary-wing UAVs. 
From the UAV application and goals, one alternative classification of UAVs can be done to meet various QoS requirements, the nature of the operation environment and federal regulations. To properly classify the applications and use of UAVSs, UAVs' flying altitude and capabilities can be taken into account. Among these factors, flying altitude can be utilized for UAVs classification: high altitude platforms (HAPs) and low altitude platforms (LAPs) \cite{YZengMag2016}. HAPs, e.g., balloons, usually operate in the stratosphere that is 17 km above the Earth’s surface. 
On the contrary, LAPs, flying at altitudes not exceeding several kilometers, have several important advantages: fast movement and more flexibility compared to LAPs. One application of LAPs is to collect data from ground sensors. Compared to HAPs, which are designed to have longer flying endurance (e.g., up to few months), LAPs have low cost and can be readily recharged or replaced if needed during the flying time. }
The benefits of UAVs application in wireless communications can be summarized as follow:
\begin{itemize}
	\item \emph{Cost-effective, fast, flexible and efficient deployment:} \color{black}{UAVs can provide cost effective wireless communications and can be more flexibly deployed for unexpected or limited-duration missions. One of the main applications is that UAVs can serve as aerial base station. It is well known that building a conventional terrestrial base station, including expensive towers and infrastructure deployment, is very expensive. In this case, UAV aided base station can provide on-the-fly communications at low cost since UAVs do not require highly constrained and expensive infrastructures.}  
	\item \emph{Line-of-sight (LoS) link:} \color{black}{Compared with conventional terrestrial base stations, a UAV-aided flying base station is able to offer on-the-fly communications and to establish LoS communication links to ground users. Especially in low-altitude UAVs, the established LoS communication links can improve the performance of the network significantly. LoS communication can facilitate high frequency (e.g., mmWave). Combined with other 5G technologies, e.g., mmWave communications, MIMO, UAV aided base stations can establish LoS communication links to achieve high data rates.} 
	\item \emph{Coverage and capacity enhancement:} \color{black}{In the downlink communications, UAV aided flying base stations can rapidly reconfigure UAV-to-ground user links to provide a large coverage network due to its maneuverability. 
	Specifically, in the uplink communications, the UAV-aided flying base station can also collect delay-tolerant information from the distributed wireless devices within the coverage. Since UAVs experience good channels, e.g., LoS link, they can provide higher transmit data rates. Moreover, the speed of UAVs can be manually adjusted to support wireless connectivity to the ground terminals. The benefits of large coverage and capacity improvement make UAV-aided wireless communication a promising integral component of the 5G wireless systems and beyond.} 
	\item \emph{Complementary network for emergency situations and disaster relief, search and rescue:} Compared to the traditional network scenarios (e.g., 4G long term evolution (LTE) and WiFi), UAVs aided wireless communication network can provide a complementary network to the existing networks in emergency situations. For example, UAVs can act as hotpots for an ultra dense network, where the ground base station is overloaded. When the ground base station is damaged or even completely destroyed by natural disasters (e.g., earth quake, floods, severe hurricanes and snow storms), UAV aided wireless networks enable to provide effective communications and help rescue lives. 
\end{itemize}

\subsection{Motivation to combine MEC and UAV}
\label{SubSec:Motivation_UAV}
\color{black}{Due to the features of UAV, such as mobility, maneuverability, and flexible development, UAVs can be integrated into wireless communication systems to provide seamless, reliable, low delay and cost-effective communication \cite{RZhang2018,RZhang20182}. To further improve the computation capacity, the combination of UAV and MEC has been proposed in existing works. There are two typical scenarios as shown in Fig.~\ref{Fig:UAVMEC}. In Mode 1 of Fig.~\ref{Fig:UAVMEC}, UAVs serve as aerial base stations \cite{SeongahIEEETVT2018}. In this scenario,  UAV can be equipped with an MEC server. Thus, MEC-enabled UAV servers provide opportunities for ground mobile users to offload heavy computation tasks. After computation, the mobile users can download the computation results from UAV based MEC servers via reliable, cost-effective wireless communication links.
In Mode 2 of Fig.~\ref{Fig:UAVMEC}, UAVs serve as new aerial mobile users of the \emph{cellular-connected} network, where the MEC server based BS is able to provide the seamless and reliable wireless communications for UAVs 
to improve the computation performance. 
MEC has strong computing capability which can be complementary to the UAVs enabled wireless communications systems. The combination of UAV and MEC technology will lead to the following benefits:}
\begin{itemize}
	\item \emph{UAV based MEC server}: In this scenario, UAVs can be used as mobile cloud computing  systems, in which the UAV based MEC server can provide offloading opportunities to ground mobile users. Due to its flexibility and mobility, UAVs can receive the offloaded tasks especially when the territorial MEC servers are not available. For example, when the emergency relief or disaster happened, the mobile device with limited processing capability can benefit from the moving UAV aided MEC server to execute tasks, e.g., analyzing assessment of the status of victims, enemies and hazardous terrain \cite{SeongahIEEETVT2018}. Thanks to LoS links between UAVs and ground mobile users, the offloading and downloading capacity can be largely enhanced. Moreover, the coverage can be improved by the UAVs based MEC communication system. 
	\item \emph{UAV-UE MEC system}: Different from the traditional scenario where the mobile user is associated with a fixed GBS over the complex fading channel, the UAV-UE MEC system enables the high-mobility UAV-UEs to offload their computation tasks to the number of
	optimized GBSs simultaneously leveraging more reliable LoS links. There are two advantages of this scenario. On the one hand, the trajectory of the UAV can be jointly designed with the resource allocation (offloading task scheduling) as it has controllable mobility in 3D airspace. On the other hand, UAVs are associated with a group of GBSs simultaneously over LoS links to exploit their distributed computing resources to improve the computation capability.
\end{itemize}

Despite the promising benefits from the combination of UAV and MEC, there are several technical challenges existing in the MEC-enabled UAV systems. \color{black}{On the one hand, the main challenges in the UAV-BS scenario include the optimal 3D deployment of UAVs, the flight time optimization and the trajectory optimization. 
On the other hand, the challenges faced in MEC including communication resource allocation, computing resource allocation and security problem, need to be addressed. }Therefore, combining UAV with MEC system may raise to the following challenges:

\begin{itemize}
	\item \emph{Mobility control and trajectory optimization}: Since UAV has limited flight time, the optimal path planning for UAVs MEC system is an important research issue. For the UAV-based MEC server, the location and flying path must be optimized to provide better offloading opportunities for the mobile devices. Similar with the UAV-UE scenario, the location and flying path must be optimized to better offload computation tasks to a group of GBSs to provide seamless communication with other UAVs. In both scenarios, the mobility control has a significant impact on the quality of the network. It is challenging to optimize the trajectory of UAV as it typically requires to solve non-convex continuous optimization problems. The channel variation and energy consumption and maximum flying speed are required in this design. In addition, coupled with other optimization factors, such as QoS metric, the trajectory optimization is challenging to tackle.
	\item \emph{Communication and computation resource optimization}: In the UAV based MEC server communication system, UAVs act as flying base stations equipped with MEC servers. The communication resource (i.e., offloading power) and computation resource (i.e., task offloading ratio) need to be jointly optimized considering potentially different objectives, e.g., relay minimization and energy consumption minimization. In the UAV-UE MEC system, UAVs act as high-mobility relay users to offload their computation-intensive tasks to the MEC server deployed at GBSs for remote execution. In this case, trajectory of UAVs can be jointly optimized with the communication and computation resource allocation, which would be more challenging compared with the fixed user and base station cases.
\end{itemize}

\begin{table*}
	\caption{Summary of existing works on UAV MEC.}
	\label{Tab:UAVMEC}
	\centering
	\begin{tabular}{ | p{2.80cm} | p{2.3cm} | p{8.75cm} | }
		\hline
		\textbf{Topic} & \textbf{References}&\textbf{Scenarios or design objectives}\\
		\hline \multirow{1}{2.80cm}{UAV-based MEC server}&\cite{FZhou2018ICC,YDouTVT2019,LFang2018,QHuIoT2019}& UAV acts as flying base station with MEC server.\\
		\hline
		\multirow{1}{2.80cm}{BS-based MEC server}&{\cite{XCaoIEEE2018,TBaiTVT2019}}& {UAV is served by multiple ground base stations with MEC servers}\\
		\hline \multirow{1}{2.80cm}{Energy efficient design}&\cite{SeongahIEEETVT2018,FZhou2018ICC,TBaiTVT2019}& To minimize the energy consumption of UAV MEC systems.\\
		\hline \multirow{1}{2.80cm}{Delay minimization}&\cite{YDouTVT2019,LFang2018,QHuIoT2019,XCaoIEEE2018}& To minimize the flying time of UAV.\\
		\hline \multirow{1}{2.80cm}{UAV flight design}&\cite{WCNC2019}& To optimize UAV trajectory, e.g., minimize the total flying distance of UAV.\\
		\hline
	\end{tabular}
\end{table*}

\subsection{State of the Art}
\label{SubSec:State_of_the_Art_UAV}
There are two scenarios for which UAVs can be combined with MEC in communication systems. In the first scenario, UAVs act as flying base stations equipped with MEC servers offering offloading opportunities for the users on the ground \cite{SeongahIEEETVT2018}. This scenario is quite common in practice. For example, the moving MEC enabled UAV plays an important role in disaster response and emergence scenario, in which the ground base station (GBS) cannot provide any service due to the damages caused by a sudden disaster, e.g. earthquake. Mobile devices with limited processing capabilities can benefit from the UAV based MEC server. In the scenario of UAV-based MEC server \cite{SeongahIEEETVT2018}, the UAV can act as a moving MEC server in the sky to help execute the computation tasks offloaded by multiple ground users. This work aimed to minimize the total energy consumption considering the QoS requirement. By means of successive convex approximation (SCA) methods, the bit allocation was studied to minimize the mobile energy for OMA uplink and NOMA downlink in the UAV based MEC system.  An energy consumption minimization problem was investigated for the UAV-enabled MEC system in \cite{FZhou2018ICC}. To address the limited computing capacity and finite battery life time of the mobile device, the UAV based MEC server was proposed to provide offloading opportunities to the mobile device. An alternative algorithm was proposed to minimize the UAV's energy consumption by optimizing the offloaded computation bits and the CPU frequency of the users and the trajectory of the UAV with the maximum speed limitation. Simulation results in this work showed that the proposed scheme outperforms the benchmark schemes. \color{black}{In \cite{YDouTVT2019}, the computing resource allocation and UAV hovering time were optimized to minimize the total energy consumption of the UAV. Moreover, the CPU's computational speed was considered in the optimization of UAV's trajectory and task assignment to minimize the energy consumption in \cite{LFang2018}. Delay minimization is also an important issue for UAV-MEC communication. In \cite{QHuIoT2019}, the delay minimization among all users was studied by jointly optimizing the UAV trajectory, the ratio of offloading tasks and the user scheduling.}

In the second scenario, a cellular-connected UAV is served by multiple ground base stations that are equipped with MEC servers \cite{XCaoIEEE2018}. In this scenario, UAV needs to complete certain computation tasks during the flying time over some given locations. Thus the tasks can be offloaded to some selected ground base station. The work \cite{XCaoIEEE2018} aimed to minimize the UAV's mission completion time by jointly optimizing its trajectory  and computation task scheduling considering the maximum speed constraint of the UAV and the computation capacity of the GBSs. It turns out that the formulated problem is nonconvex, thus it is difficult to find the global optimal solution in polynomial time.  Therefore, the alternating optimization and SCA were exploited to obtain a high-quality suboptimal solution. \color{black}{In \cite{WCNC2019}, the total travel distance of UAV was minimized and two different solutions were proposed, i.e., MEC-ware UAV's path planing (MAUP) based integer linear programming and accelerated MAUP.  }
\color{black}{Physical-layer security was investigated in \cite{TBaiTVT2019}, where the optimal solutions based on the condition of three offloading options and the computational overload event from a physical-layer perspective were provided. The summary of exiting works on UAV MEC is provided in Table~\ref{Tab:UAVMEC}.}

\subsection{Learned Lessons and Potential Works}
\label{SubSec:Lesson_and_Potential_Works_UAV}
Thanks  to the  great  benefits  from the combination of MEC and UAVs as well as  their  limited resource, it can be concluded that MEC-UAV is an inevitable trend in the future wireless communication systems.  Although some existing works have been done to engineer MEC-UAV systems, there are still several challenges to address. In the following, we discuss key open problems in MEC-UAV systems:
\begin{itemize}
	\item \emph{Performance analysis of UAV-MEC systems}: A fundamental performance analysis is required for the UAV-MEC system. In particular, the coverage probability, throughput, delay or reliability can be investigated to evaluate the impact of each design parameter on the overall system performance. Due to the 3D development and short flight duration of UAVs and the delay awareness of MEC, the performance analysis for the UAV-MEC system is challenging. 
	\item \emph{Energy-aware resource allocation}: The flying time and the resource of UAVs are limited because UAVs typically have small sizes, weight and limited power. Therefore, the trajectory and resource allocation (i.e., communication resource and computation resource) need to be optimally designed to reduce the energy consumption. However, most existing works only considered designing trajectory and optimizing resource allocation separately, which cannot achieve the highest network performance. Hence, jointly optimizing the path planning and resource allocation for MEC-UAV system is an open challenging problem. It becomes more challenging when other optimization factors, such as, QoS requirement, offloading power allocation and task assignment together with the channel variation, delay constraint and maximum flying speed, are considered in such design.
	\item \emph{User grouping and UAV association}: In the UAV based MEC server communication system, each UAV acts as a flying MEC-enabled base station. The ground users need to offload their tasks to one UAV or multiple UAVs simultaneously. Thus the user group  problem must be solved by using suitable approaches, e.g., matching theory, game theory and convex optimization methods. On the contrary, in the UAV MEC systems, UAVs need to offload tasks to GBSs for remote computation. The subchannel allocation and UAVs association can be investigated.
	
\end{itemize}

\section{MEC for Internet of Things}
\label{Sec:IoT_MEC}
\subsection{Fundamentals of IoT}
\label{SubSec:Fundamentals_IoT}
\begin{figure*}[!t]
	\centering
	\includegraphics[width=1.00\linewidth]{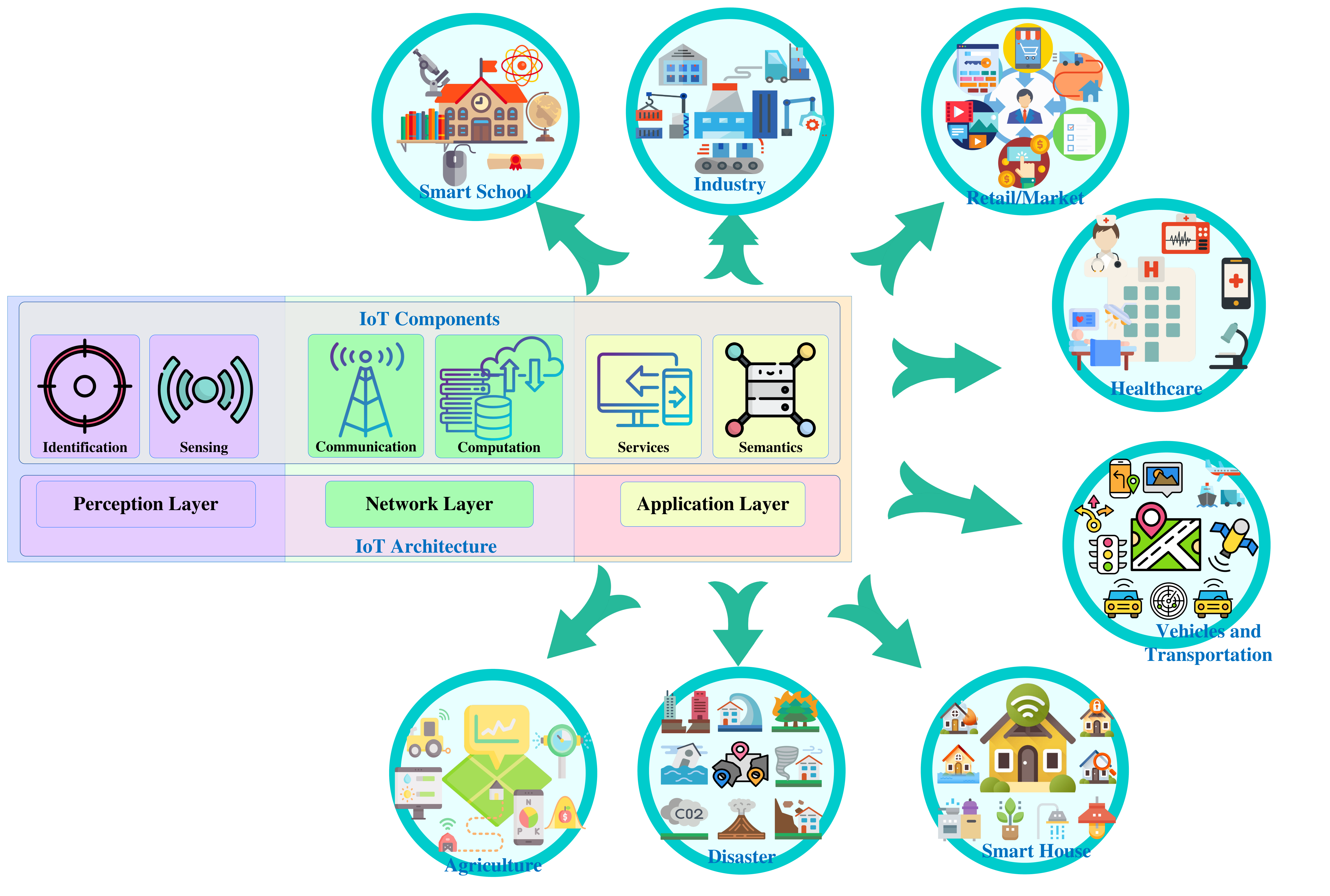}
	\caption{The overall picture of IoT applications and architecture.}
	\label{Fig:IoT}
\end{figure*}
Thanks to significant advancement in computation and storage technologies, and communication networks, billions of devices with their every domain-specific applications are able to connect to the Internet to generate/collect data, to exchange important messages amongst themselves, and to coordinate decisions via complex communication networks \cite{CISCO_IOT_11}. 
This phenomenon has opened new era of Internet, the so-called the IoT \cite{CISCO_IOT_11}. 
The basic concept of IoT is that anything can be interconnected with the global information and communication infrastructure at any time and any place \cite{ITU-TY.2060}. 
Things can be physical things existing in the physical world 
or virtual things existing in the information world. 
IoT has been playing a significant role in solving various challenges of modern society effectively and improving the quality of human life, such as, safer, healthier, more productive, and more comfortable \cite{Gu2018Joint}.
The fundamental characteristics of IoT can be condensed as following:
\textit{1) inter-connectivity,}
\textit{2) things-related services,}
\textit{3) heterogeneity,} 
\textit{4) dynamic changes}, see \cite{ITU-TY.2060} for details.

A basic architecture of IoT as well as its specific every-domain applications can be summarized in Fig.~\ref{Fig:IoT}. In particular, the IoT basic architecture consists of three layers: Perception, Network, and Application \cite{Al-Fugaha_CST15,Lin_JIoT17}. In the first layer, the physical sensors collect useful information/data from things or the environment 
which are then transformed into digital form
and it marks all objects with unique address identification.
The principal responsibility of the second layer is to help and secure data transmission between the perception and the application layers \cite{Silva_IETE18}.
The third layer is to provide the personalized based services according to users' relevant needs and to link the major gap between users and applications. 
It combines the industry to attain the high-level intelligent solutions for IoT specific every-domain applications such as
the disaster monitoring, healthcare, smart house, transposition, production controling, health care, retail, education.
In other aspects, the third layer can be further divided into three sub-layers: \textit{1) The
service management layer,}
\textit{2) The application layer,}
\textit{3) The Business layer}.
Due to the high-level requirement of some applications and services, one more layer has been potentially added  between the application and network layers which consists of MEC and fog computing servers to perform some specific distributed computation duty or pre-data processing.

\color{black}{IoT is also one of the motivations for developing the promising 5G technologies to allow the massive connections from a large numbers of ``things'' to the Internet via wireless networks. Inversely, 5G is considered a basic platform to facilitate emerging IoT applications \cite{Akpakwu_Access18}. As expected, manifold data traffic (typically of Gbps order), low latency transmission can be provided by 5G communication networks which can support a tremendous increase in dense connected “things” in wireless networks, including high-mobility IoT/UEs, embedded sensors in the human body (or clothing), wearable devices, equipment for monitoring biometrics, or even autonomous cars (also called V2X communications). Furthermore, by exploiting spectrum resources in very high-frequency bands and providing the coexistence of multiple numerologies, 5G networks can realize Tactile Internet requiring ultra-low latency with extremely high availability, reliability, and security \cite{M_Maier_Access18}.} Although, IoT can potentially benefit modern society, many technical issues as well as requirements remain to be addressed as follows \cite{ITU-TY.2060}:
\textit{1) identification-based connectivity, 
2) interoperability,
3) autonomic networking,
4) autonomic services provisioning,
5) location-based capabilities,
6) security and privacy protection,
7) high quality and highly secure human body related services,
8) plug and play,
9) manageability.
}
For more information on the techniques and future trends of IoT, we invite the reader to further refer to the following references \cite{Al-Fugaha_CST15,Lin_JIoT17,Silva_IETE18}.

\subsection{Motivation to use MEC for IoT and challenges}
ETSI, in its report \cite{ETSI_15}, has distinguished IoT as one of the most important MEC application instances. 
There are many benefits of employing MEC into IoT systems, including but
not limited to, lowering the amount of traffic passing through the infrastructure and reducing the latency for applications and services \cite{Taleb_CST17}.
 \color{black}{Among these, the most significant is the low latency introduced by MEC which is suitable for 5G Tactile Internet applications requiring round-trip latency in the millisecond range \cite{Y_Xiao_JSAC18}.
MEC technologies are envisioned to work as gateways placed at the middle layer of IoT architecture which can aggregate and process the small data packets generated by IoT services and provide some additional special edge functions before they reach the core network; hence, the end-to-end delay can be reduced.
Additionally, these techniques are also able to lower the energy consumption of small-size IoT devices and prolong their battery-life by supporting significant additional computational capabilities through intelligent computation offloading strategies.
Furthermore, MEC platforms will be offered
and deployed by the network operator at any tiers of 5G networks, e.g., eNBs, multi-RAT aggregation points, neighbor mobile devices, which can be made open to authorised developers and content providers to deploy versatile and uninterrupted services on IoT applications \cite{Mach2017Mobile}.
In addition, based on the context and platforms of MEC, artificial intelligence (AI) on the edge can gain the huge benefit to realize distributed IoT applications and intelligent system management, which is now considered as a part of beyond 5G standardization \cite{T_Leppanen_EoT2019}.}
Inversely, IoT also energizes MEC with mutual advantages. 
In particular, IoT expands MEC services to all types of smart objects ranging from sensors and actuators to smart vehicles.
Integrating MEC capabilities to the IoT systems come with an assurance of better performance in terms of quality of service and ease of implementation. 

To realize the benefits of MEC for IoT applications, several technical aspects such as scalability, communication, computation offloading and resource allocation, mobility management, security,
privacy, and trust management, should be considered. 
To gain the deeper understanding of the technical requirements in MEC-enabled IoT, interested readers are recommended to refer to the existing survey papers on MEC-IoT topic which are listed in 
Table~\ref{Table:Summary_MECIoT_Surveys}.

\begin{table*}[ht!]
	\caption{Summary of existing survey papers on MEC-IoT topic.}
	\label{Table:Summary_MECIoT_Surveys}
	\centering
	\begin{tabular}{ |l|p{110mm}|l|l|l|l|l| }
		\hline
		\multirow{2}{*}{\textbf{Ref.}} & \multirow{2}{*}{\textbf{Contributions}} & \multicolumn{5}{c|}{\textbf{Aspects}}  \\ \cline{3-7}
		& & {\rotatebox[origin=c]{90}{Research Directions}} & {\rotatebox[origin=c]{90}{Taxonomy}} & {\rotatebox[origin=c]{90}{Architecture}} & {\rotatebox[origin=c]{90}{Applications}} & {\rotatebox[origin=c]{90}{Tech. Aspects}} \\ \hline
		\cite{Porambage2018Survey} & {A holistic overview on the exploitation of MEC technology for the realization of IoT applications and their synergies. Technical aspects of enabling MEC in IoT and provide some insight into various other integration technologies therein.} & $\checkmark$ & $\checkmark$ & $\checkmark$ & $\checkmark$ & $\checkmark$ \\ \hline
		\cite{Li_JIII18} & {The current research state-of-the-art of 5G IoT, key enabling technologies, and main research trends and challenges in 5G IoT.} & $\checkmark$ & $\checkmark$ & & & \\ \hline
		\cite{Ansari_IEICE18} & {Proposing a Mobile Edge Internet of Things architecture by leveraging the fiber-wireless access technology, the cloudlet concept, and the software defined networking framework.} & $\checkmark$ & $\checkmark$ &  $\checkmark$ & $\checkmark$ & \\ \hline
		\cite{Verma_CST17} & {Explaining the shortcomings of the network
			methodologies to support them. Discussing the relevant network methodologies which may
			support the real-time IoT analytics. Presenting research problems and future research directions.} & $\checkmark$ & $\checkmark$ & $\checkmark$ & $\checkmark$ & \\ \hline 
		\cite{Sun_ComMag16} & {Proposing a hierarchical fog computing architecture in each fog node to provide flexible IoT services while maintaining user privacy: each user's IoT devices are associated with a proxy VM (located in a fog node), which collects, classifies, and analyzes the devices' raw data streams, converts them into metadata, and transmits the metadata to the corresponding application VMs (which are owned by IoT service providers).} & $\checkmark$ & & $\checkmark$ & $\checkmark$ &$\checkmark$ \\ \hline
		\cite{Abbas_ITJ18} & {A comprehensive survey on the recent research and technological development in the area of MEC and its application domains, research challenges, and open issues in IoT.} &  $\checkmark$ & $\checkmark$ & $\checkmark$ & $\checkmark$ & $\checkmark$ \\ \hline
		\cite{Sabella_CEMag16} &  {Proving a brief tutorial on MEC technology, an overview of the MEC framework, architecture, and its role in IoT.} & $\checkmark$  & & $\checkmark$  & $\checkmark$  & $\checkmark$ \\ \hline
		\cite{Premsankar_JIoT18} & {Advocating edge computing for emerging IoT applications that
			leverage sensor streams to augment interactive applications.} &  &  & $\checkmark$ & $\checkmark$ & \\ \hline
		\cite{Manogaran_FGCS18} & {A new architecture is proposed to store and process scalable sensor data.} & & $\checkmark$ & $\checkmark$ & $\checkmark$ &  \\ \hline
		\cite{Ai_DCN17} & {A concise tutorial of three edge computing technologies including MEC, cloudlets, and fog computing.} & $\checkmark$ & $\checkmark$ & $\checkmark$ & & $\checkmark$ \\ \hline
		\cite{Maurizio_FI_2019} & {The hardware architectures of typical IoT devices and sums up many of the low power techniques which make them appealing for a large scale of applications.} &  &  & $\checkmark$ & $\checkmark$ & $\checkmark$ \\ \hline
		\cite{Puliafito_ACM_2019} & {A comprehensive survey on the employment of fog computing to support IoT devices and services.} & $\checkmark$ & $\checkmark$ & $\checkmark$ & & $\checkmark$ \\ \hline
		\cite{J_Ni_Network_2019} & {An overview on edge-assisted data processing for IoT from security and efficiency perspectives.} & $\checkmark$ & $\checkmark$ & $\checkmark$ & $\checkmark$ & $\checkmark$ \\ \hline
	\end{tabular}
\end{table*}

\subsection{State of the Art - MEC-enabled IoT Application Scenarios}
 \color{black}{ As illustrated in Table~\ref{Table:Summary_MECIoT_Surveys}, the surveys on MEC and IoT have been introduced in some existing papers, i.e., \cite{Porambage2018Survey,Abbas_ITJ18}. Hence, this section focuses on providing a survey on recent MEC-enabled IoT works in application scenarios related to 5G uses cases. The technical aspects and application scenarios of these works are summarized in Table~\ref{Table:MECIoT_tech_acspect_apps}.}

\begin{table*}[ht!]
	\caption{Summary of MEC-enabled IoT papers on different application scenarios and technical aspect.}
	\label{Table:MECIoT_tech_acspect_apps}
	\centering
	\begin{tabular}{|l|l|l|l|l|l|l|}
	\hline
		& Smart city                                                                       & Healthcare                                                   & V2X                                  & \begin{tabular}[c]{@{}l@{}}Industrial\\ Internet\end{tabular}                                   & \begin{tabular}[c]{@{}l@{}}Wearable IoT/\\ AR and VR\end{tabular}               & \begin{tabular}[c]{@{}l@{}}Mechanized \\ Agriculture\end{tabular}    \\ \hline
		\begin{tabular}[c]{@{}l@{}}Offloading/\\ Resource\\ allocation\end{tabular} & \cite{Singh_Gill_JSS_2019,Yassine_FGCS_2019}                                     & \cite{Abdellatif_ESA_2019}                                   & \cite{Jingjing_Zhao_Arxiv_2018}      & \cite{W_Sun_TII_2018,Li_JIoT18,G_Li_TII_2018,C_Lai_TII_2019,J_Xu_TII_2019,Xiaomin_Li_FGCS_2018} & \cite{L_Tao_Access_2018,X_Yang_Access_2018,Y_Sun_TCom2019,Y_Li_SEC_2018,Y_liu_TMul2019} & \cite{Fan_IOP_2018,Zamora-Izquierdo_bioEng_2019}                     \\ \hline
\begin{tabular}[c]{@{}l@{}}Energy\\ Management\end{tabular}                                                         & \cite{Singh_Gill_JSS_2019,YLiu_network_2019}                                     & \cite{Abdellatif_ESA_2019}                                   & \cite{Jingjing_Zhao_Arxiv_2018}      & \cite{W_Sun_TII_2018,Li_JIoT18,X_Li_Wan_TII_2019}                                               & \cite{L_Tao_Access_2018,Y_Li_SEC_2018}                                          &                                                                      \\ \hline
		Safety                                                                    & \cite{Z_Samia_PAIS_2018,Pacheco_CONIITI_2018,Pratama_IES_ETA_2017,Mochamad_2019} &                                                              & \cite{Dam_2018}                      &                                                                                                 & \cite{Trilles_SubCom_2019}                                                      &                                                                      \\ \hline
		Security                                                                  & \cite{Z_Samia_PAIS_2018,Pacheco_CONIITI_2018,MARahman_Access_2019}               & \cite{X_Li_Access_2019,Soraia_Sensors_2018}                  & \cite{Dam_2018}                      &                                                                                                 &                                                                                 &                                                                      \\ \hline
		Privacy                                                                   & \cite{MARahman_Access_2019}                                                      &                                                              & \cite{Hochstetler_SEC_2018}          &                                                                                                 &                                                                                 &                                                                      \\ \hline
		Convenience                                                          & \cite{Pratama_IES_ETA_2017,L_Zhao_ComMag_2019,Mochamad_2019}                     & \cite{Soraia_Sensors_2018}                                   & \cite{Dam_2018} & \cite{Li_JIoT18,J_Xu_TII_2019}                                                                  &                                                                                 &                                                                      \\ \hline
		\begin{tabular}[c]{@{}l@{}}Monitoring/\\ Controlling\end{tabular}         & \cite{L_Zhao_ComMag_2019}                                                        &                                                              & \cite{Jingjing_Zhao_Arxiv_2018}      & \cite{G_Li_TII_2018,C_Lai_TII_2019,X_Li_Wan_TII_2019}                                           &                                                                                 & \cite{Fan_IOP_2018,Zamora-Izquierdo_bioEng_2019,Trilles_SubCom_2019} \\ \hline
		Reliability                                                               &                                                                                  & \cite{Pace_TII_2019,Abdellatif_ESA_2019,Soraia_Sensors_2018} & \cite{Dam_2018}                      & \cite{G_Li_TII_2018,Xiaomin_Li_FGCS_2018,EE_Ugwuanyi_Access18,M_Chowdhury_JIoT17,J_Xu_TGCN19,Y_Xiao_JSAC18}                                                     & \cite{L_Tao_Access_2018,Y_liu_TMul2019}                                                        & \cite{Zamora-Izquierdo_bioEng_2019}                                  \\ \hline
		Latency                                                                   &                                                                                  & \cite{Pace_TII_2019,Abdellatif_ESA_2019}                     & \cite{Jingjing_Zhao_Arxiv_2018}      & \cite{Li_JIoT18,G_Li_TII_2018,Sodhro_TII_2019,Y_Xiao_JSAC18,M_Chowdhury_JIoT17,J_Xu_TGCN19}                                                  & \cite{X_Yang_Access_2018,Y_Sun_TCom2019,Y_Li_SEC_2018,Y_liu_TMul2019}                   &                                                                      \\ \hline
		Scalibility                                                               &                                                                                  &                                                              &                                      & \cite{G_Li_TII_2018}                                                                            &                                                                                 &                                                                      \\ \hline 
	\end{tabular}
\end{table*} 

\subsubsection{Smart home and Smart city}
One of the most important use cases of IoT is smart city and its important subset smart home/building \cite{Taleb_CommMaga17,Khan2019EdgeCE}.
\color{black}{Recently, the MEC contexts and novel 5G technologies have been enabled to emerge the judicious edge big data analysis and wireless access for IoT systems to further improve the urban quality of life for citizen with many aspects including security, privacy, energy management, safety, convenient life, ect..}
For energy management, an fog-based IoT automation mechanism was validated in \cite{Singh_Gill_JSS_2019} to optimize the resource management for smart building systems. By leveraging the fog-enabled cloud computing environments, the novel implemented smart home systems can reduce 12\% utilized network bandwidth, 10\% response time, 14\% latency and 12.35\% in energy consumption.
For monitoring and controlling the smart home/buildings, innovative analytics on IoT captured data from smart homes was presented in \cite{Yassine_FGCS_2019} employing the fog computing nodes.
This fog-based IoT system can address the challenges of complexities and resource demands for online and offline data processing, storage, and classification analysis in home/building environment.
The MEC-enabled IoT frameworks in \cite{Z_Samia_PAIS_2018,Pacheco_CONIITI_2018} focus on behaviour features by monitoring the student's location and activities in school environment for safety aspect. 
In particular, \cite{Z_Samia_PAIS_2018} designed a platform to identify any student activities that occur at the classroom level in which the raw indoors environment data is processed at an edge computing server (Raspberry Pi) for detecting the presence of individuals in classroom while
\cite{Pacheco_CONIITI_2018} exploited the DL algorithms in an MEC-enabled IoT smart classroom for person recognition.

For the smart city use cases, the security and privacy aspects were considered in \cite{MARahman_Access_2019} where a blockchain-based smart contract services for the sustainable IoT-enabled economy is proposed for smart cities by employing AI solutions in processing and extracting significant event information at the fog nodes, and then utilizing blockchain algorithms to save and deliver results.
Recent work in \cite{YLiu_network_2019} studied the energy management aspect in smart city where the deep reinforcement learning methods were employed into MEC-enabled IoT system to manage the energy grid efficiently. 
\cite{Pratama_IES_ETA_2017,L_Zhao_ComMag_2019} both considered the safety and convenience aspects where Pratam et al. 
\cite{Pratama_IES_ETA_2017} implemented a Raspberry Pi-based MEC system on school shuttle buses for tracking the locations of students and vehicles while \cite{L_Zhao_ComMag_2019} developed a smart routing for crowd management based on deep reinforcement learning algorithms to satisfy the latency constraints of service requests from the people. 
A platform to detect potholes and road monitoring was studied in \cite{Mochamad_2019} to cope with flooding on the roads in rainy seasons for traffic safety.

\subsubsection{Healthcare}
Healthcare solutions with more intelligent and prediction capabilities have been developed and implemented based on the rapid developments of IoT and cyber physical systems \cite{RAY20191,MUTLAG201962}. MEC-enabled IoT has shown a huge potential in improving the performance of healthcare systems which includes but not limit to the mobile monitoring healthcare scheme. 
In this system, the MEC-enabled gateways can offer several higher-level services such as local storage, real-time local data processing, embedded data mining, etc. beside controlling the data transmission \cite{Rahmani_FGCS18}.
These enable to empower the system to deal with many challenges of managing the remote devices, i.e., security, reliability, latency, energy efficiency issues.
Freshly, Li et al. in \cite{X_Li_Access_2019} considered the security issue in mobile healthcare systems by proposing a secure and efficient data management system named EdgeCare in which healthcare data and facilitating data trading are processed at edge servers with security considerations. 
Focusing on improvement of latency and reliability performance, \cite{Pace_TII_2019} proposed BodyEdge, a novel body healthcare architecture consisting of a tiny mobile client module and an edge gateway for collecting and locally processing data coming from different scenarios. 
Sharing the same view, \cite{Abdellatif_ESA_2019} implemented an accurate and lightweight classification mechanism employing the edge computing to 
detect the seizure at network edge based on the information extracted from the vital signs with precise classification accuracy and low computational requirement. The implementation results show that the proposed system outperforms conventional non-MEC remote monitoring systems by: 1) achieving 98.3\% classification accuracy for seizures detection, 2) extending battery lifetime by 60\%, and 3) decreasing average transmission delay by 90\%.
For emergency department systems, Soraia et al. \cite{Soraia_Sensors_2018} proposed a resource preservation net framework integrated with cloud and edge computing where the key performance indicators such as patient length of stay, resource utilization rate and average patient waiting time are modeled and optimized considering high reliability, efficiency and security.

\subsubsection{Vehicle-to-Everything (V2X) IoT}
In \cite{3GPP_22.886}, 3GPP has identified 5G as the key technology supporting the V2X concepts in several use cases: Information (state map, environment, traffics) sharing, vehicle platooning, remote driving, grouping-based cooperative driving, communication between vehicles, cooperative collision avoidance, dynamic ride sharing. The QoS requirements in data rate and communication range may vary in different V2X applications \cite{Giust2018MultiAccess}. However, the crucial factors such as ultra low latency, high reliability, and security have to be improved due to the safety in most use cases, which can be fulfilled by employing MEC technologies \cite{Kabalci_Springer2019}. 
Recently, the security aspects in V2X were considered in \cite{Dam_2018} which enabled a cooperative intelligent transportation system by deploying MEC-equipped cell towers hosting local communication to increase the safety on roads and the traffic efficiency with smoother flow.
\cite{Jingjing_Zhao_Arxiv_2018} focused on the latency in MEC based dense mmWave V2X networks by optimizing the offloaded computing tasks and transmit power of vehicles and road side units to minimize the energy consumption under delay constraint resulting from vehicle mobility.
The work in \cite{Hochstetler_SEC_2018} enabled the object recognition enhancement with DL algorithms at the edge side with MEC deployment in V2X networks to improve the information sharing and communication performance. Specifically, an Intel Movidius Neural Compute Stick along with Raspberry Pi 3 Model B is used as an edge computing server to analyze the objects contained in real-time images and videos.

\subsubsection{Industrial Internet}
\color{black}{MEC yields a significant paradigm shift in industrial Internet of Things (IIoT), well-known as Industry 4.0 
- a use-case of 5G technologies, by bringing computing resources close to the lightweight IIoT devices in IIoT domain \cite{Brito_TETT18,W_Sun_TII_2018}. In IIoT, there are many application scenarios such as, factory automation, process automation, human-machine interfaces, production IT, logistics and warehousing, monitoring and maintenance. Intelligently managing the edge resources, MEC enables to power the IIoT system to address some significant technical issues, e.g. latency, resilience, connectivity, and security. 
}

\color{black}{To make MEC an enabler for latency-critical IIoT applications, time-sensitive networking (TSN)\footnote{TSN includes a set of protocols to provide timing guarantees for latency-critical applications. The IEEE 802.1 TSN's home page is available at https://1.ieee802.org/tsn/ and its overview paper can be found in \cite{Nasrallah2019ULL}.} is a vital solution. \cite{Chen2018Edge, Pop2018Enabling} proposed TSN-based configuration architectures of MEC that can support real-time IIoT applications. 
Considering system resources, \cite{W_Sun_TII_2018} reported that enabling MEC in IIoT systems can improve the system efficiency by jointly designing resource allocation and offloading based on an auction-based method where both claimed bids and asked prices were given by the MEC servers. Additionally, Li et al. in \cite{Li_JIoT18} employed MEC servers in SDN for IIoT systems to dynamically optimize the routing path considering the aggregation of time deadline, traffic load balances, and energy consumption to provide better solution for IIoT data transmission in terms of average time delay, throughput, energy efficiency, and download time. \cite{G_Li_TII_2018} proposed a service popularity-based smart resource partitioning scheme for fog computing-enabled IIoT. By demonstrating the notable performance improvements on delay time, successful response rate and fault tolerance, the authors confirm the significant benefit of enabling fog computing to cope with the large-scale IIoT services. While \cite{Sodhro_TII_2019} implemented DL at the edge servers to enhance the range and computational speed of IIoT devices remarkably in the MEC-based IIoT framework for increasing the energy efficiency and battery lifetime at acceptable reliability (around $95$ \%). \cite{EE_Ugwuanyi_Access18} focused on obtaining higher reliability of network interactions by proposing a deadlock avoidance resource provisioning algorithm for Industrial IoT devices using MEC platforms.}

\color{black}{Aiming at improving quality of industrial production, \cite{C_Lai_TII_2019} implemented parallel computing with MEC servers to improve the efficiency of equipment identification. In particular, adopting the long short-term memory to analyze big data features and build a non-intrusive load monitoring system with MEC can enlarge the average recognition rate to over 80\%. MEC can also be applied for smart IoT-based manufacturing to improve performance of edge-equipment network, information fusion, and cooperative mechanism, based on which the excellent real-time, satisfaction degree and energy consumption performance of the manufacturing system can be significantly improved \cite{X_Li_Wan_TII_2019}. On another view, to achieve higher goodput, \cite{Xiaomin_Li_FGCS_2018} enabled the MEC platform to improve the caching management for IIoT system. For the security purpose, \cite{J_Xu_TII_2019} employed a smart blockchain-based platform with many MEC servers in IIoT systems to effectively solve the network congestion caused by transferring raw data (e.g., pictures or video clips) between a publisher and workers.}

\subsubsection{Wearable IoT, AR and VR}
The newly emerging applications corresponding to mobile AR, VR, and wearable devices, e.g., smart glasses and watches,
are anticipated to be among the most demanding applications over wireless networks so far, but there is still lack of sufficient capacities to execute sophisticated data processing algorithms.
To overcome such challenges, the emergence of MEC and 5G techniques would pose the longer battery lifetime, powerful set of computing and storage
resources, and low end-to-end latency \cite{Melike_AdNet_2018,Y_Sun_TCom2019}.
Sharing this view, \cite{L_Tao_Access_2018} presented Outlet system to explore the available computing resources from user's ambience, e.g., from nearby smart phones, tablets, computers, Wi-Fi APs, to form an MEC platform for executing the offloading tasks from wearable devices.
Promising performance achieved by Outlet, e.g., mostly within 97.6\% to 99.5\% closeness of the optimal performance, has demonstrated the advantage of enabling edge computing technique into wearable IoT systems.
Applying MEC on VR devices, \cite{X_Yang_Access_2018} presented an effective solution to deliver VR videos over wireless networks minimizing the communication-resource consumption under the delay constraint. 
This work also demonstrated the interesting tradeoffs among communications, computing, and caching. 
In \cite{Y_Sun_TCom2019}, a novel delivery framework enabling field of views caching and post-processing procedures at the mobile VR device was proposed to save communication bandwidth while meeting low latency requirement.
Impressively, an implementation of MEC concepts over Android OS and Unity VR application engine in \cite{Y_Li_SEC_2018} enabled to reduce more than 90\% computation burden, and more than 95\% of the VR frame data being transmitted to mobile devices by letting MEC servers adaptively store the previous results of VR frame rendering of each user and considerably reuse them for others to reduce the computation load.
On a diffrent view, Liu \textit{et al.} in \cite{Y_liu_TMul2019} illustrated the advantage of implementing MEC in panoramic VR system to maintain the high quality of the video streaming by intelligent balancing the link adaptation, transcoding-based chunk quality adaptation, and viewport rendering offloading.

\subsubsection{Mechanized Agriculture with IoT}
IoT emerging the use of low-cost hardware (sensors/microcontrollers) and 5G communication technologies for eRAC has opened new era for cultivating soil, namely \textit{``smart agricultural''} \cite{Ferrandez-Pastor_Sensors16, A_Jukan_CompMag19}. 
Many advanced abilities, e.g.,  predictive analytic, weather forecasting for crops or smart logistics and warehousing, can be offered by enabling MEC technologies in this scenario \cite{Xiaojie_sensors_2019}.
Recently, there are some works on emergence of MEC and IoT in agriculture.
In particular, \cite{Fan_IOP_2018} proposed an intelligent agricultural water monitoring system with advanced MEC technology to effectively manage the data collected by the sensors.
As a part of EU DrainUse project, \cite{Zamora-Izquierdo_bioEng_2019} presented a local/edge/cloud three-tier platform for monitoring and managing soil-less agriculture in full re-circulation greenhouses using moderately saline water. 
In this platform, the edge plane is deployed to increase system reliability against network access failures while the data analytic modules are located in the cloud. 
To protect the plant on vineyard fields, \cite{Trilles_SubCom_2019} implemented a disease alerting platform using a low-cost sensors in the municipality of Vilafam\'{e}s (Castell\'{o}, Spain).
In this platform, the edge computing is deployed to improve the capablity of monitoring meteorological phenomena collect (temperature, humidity, ...) based on that an alert disease model on the cultivation of the vine was developed for improving the product quality.

\subsubsection{Tactile Internet}
\color{black}{Tactile Internet is defined by the International Telecommunication Union (ITU) as the next evolution of IoT that combines ultra low latency with extremely high availability, reliability and security \cite{ITU_TI_14}. Encompassing human-to-machine and machine-to-machine interaction, Tactile Internet will combine multiple technologies including 5G and MEC, i.e., 5G may be employed for the data transmission with low delay and high reliability while MEC efficiently exploit computing resources close to the end users for better QoE. The applications related to Tactile Internet can be automation, robotics, tele-presence, tele-operation, AR, VR \cite{M_Maier_Access18,ITU_TI_14}. The works employing MEC in these scenarios considering low latency and high reliability can be found in Table~\ref{Table:MECIoT_tech_acspect_apps} and introduced in the previous parts. The following summarizes the recent works focusing on the technical aspects involving to the MEC implementation in Tactile Internet. \cite{Y_Xiao_JSAC18} considered an energy-efficient design of fog computing networks that support low service response time of end-users in Tactile Internet applications and efficiently utilize the power of fog nodes. The trade-off between the latency and required power was presented and then extended to fog computing networks leveraging cooperation between fog nodes. \cite{M_Chowdhury_JIoT17} exploited the MEC systems including cloud, decentralized cloudlets, and neighboring robots equipped with computing resource collaborative nodes for computation offloading in support of a host robot's task execution. Then, a proper task allocation strategy by combining suitable host selection and computation task offloading was proposed to meet the required task execution time. The work also showed that the MEC-based collaborative task execution scheme outperforms the non-collaborative scheme in terms of task response time and energy consumption efficiency. Recently, Xu et. al in \cite{J_Xu_TGCN19} designed a hybrid edge caching scheme for Tactile Internet which can reduce latency and achieve better performance in overall energy efficiency than existing ones.}

\subsection{Learned Lessons and Potential Works}
\label{SubSec:Lesson_and_Potential_Works_IoT}

Several research works and implementations in the literature have demonstrated that MEC is an ideal solution for IoT systems. In many applications and use cases, exploiting MEC resources for managing the data collection or pre-processing the massive data at the edge networks is able to lead to significant advantages.
These advantages include but not limit to 
reducing the radio resource consumption (i.e., 12\% in \cite{Singh_Gill_JSS_2019}), 
shortening the reaction time (i.e., 10\% in \cite{Singh_Gill_JSS_2019}), 
lessening the system latency (i.e., 14\% in \cite{Singh_Gill_JSS_2019}, 90\% in \cite{Abdellatif_ESA_2019}), 
and diminishing the overall energy consumption (i.e., 12.35\% in \cite{Singh_Gill_JSS_2019}). 
In addition, MEC also helps offloading the computational burden at IoT devices, which results in prolonging their battery life (i.e., 60\% in \cite{Abdellatif_ESA_2019}), increasing the accuracy rate of task processing (i.e., improving the seizures detection rate over 98\% in \cite{Abdellatif_ESA_2019}), mitigating the amount of transmission data (i.e., 95\% in \cite{Y_Li_SEC_2018}), and lowering the computation load (i.e., 90\% in \cite{Y_Li_SEC_2018}).
However, to maximize benefits of MEC in IoT applications, 
one requires the more efficient management of the MEC resources and access networks, and capacities as well as abilities of the IoT components or elements.
These demands open many potential research directions to effectively governance MEC in IoT systems.
The future works considering technical aspects of IoT and MEC, i.e., scalability, communication, computation offloading and resource allocation, mobility management, security, privacy, and trust management, have been well indicated and manifested in some recent MEC-IoT surveys, such as, \cite{Porambage2018Survey,Abbas_ITJ18,Premsankar_JIoT18,Maurizio_FI_2019,Puliafito_ACM_2019,J_Ni_Network_2019}to which the interested readers are recommended to refer.
In the following, we discuss key open problems in MEC-enabled IoT systems which are different to the  mentioned challenging technical aspects. 
\begin{itemize}
\item \textit{Effective cooperation in dense MEC-based IoT networks:} 
Currently, each MEC server is deployed by the infrastructure providers to supply the computing and radio access services to a specific set of distributed edge IoT nodes at the IoT network edge. 
In addition, a provisioning set of computation or networking functions including data analyzing, compressing, caching, routing, etc., are installed at a distributed MEC server to serve its set of devices from the aspect of their applications.
In dense IoT-based smart cities, massive heterogeneous IoT devices running diverse advanced services corresponding to various domains of city life \cite{L_Zhao_Network_2019}. This leads to a huge number of devices with diverse service requirements from different infrastructure providers locating in a same geographical area. Although a new service (i.e., out-set-of-function service)
can be supported by MEC by offloading raw data to cloud for processing, this may lead to huge cost of energy and time.
In addition, non-cooperative edge servers deployed by different infrastructure provider may result in severe under-utilization of resources.
Hence, enabling cooperative edge computing environment can open the resource of many types of edge computing servers for serving the diverse requirements in the dense IoT networks. 
However, to realize the cooperation among the edge nodes to maximize their benefits, several particular challenges should be solved: The trade-off between the cloud and the edge; The optimization of the service placement on distributed and limited edge resources; The contradiction between the computation-intensive edge services and the limited edge resources \cite{Xiaofeng_Cao_Arxiv_2019}.
\item \textit{Employing AI techniques in MEC-based IoT systems:}
Recently, AI techniques with ML/DL have been considered as important tools for processing big data in the IoT-based environment.
The integration of ML/DL and AI algorithms at the network edge can provide efficient data analysis, make accurate decisions, predict tasks at the network edge, optimize the mobile edge caching, computation offloading, and preserve network security and data privacy. 
In addition, adopting AI techniques for MEC-enabled IoT system 
can extract the behaviors of physical/networking resources and users in different time and scenarios, 
dynamically monitor and adjust the configuration of network resources, and realize real-time data collection of loT, efficient processing of computation, based on which the intelligent services for heterogeneous IoT devices can be optimized \cite{Guo_Kun_Sensors_2018}.
However, to apply the AI technology regularly requiring big data processing at the edge nodes which are commonly equipped limited computation, storage resource, one needs novel ML/DL-based algorithm with distributed computing and data access which is an challenging issue for the future works. For more detail on ML/DL for MEC applications, we invite the readers to refer to Section~\ref{Sec:MEC_MachineLearning}.
\end{itemize}

\section{MEC with Heterogeneous Cloud Radio Access Network}
\label{Sec:HetNet_MEC}

\begin{figure*} [ht]
	\centering
	\includegraphics[width=0.75\linewidth]{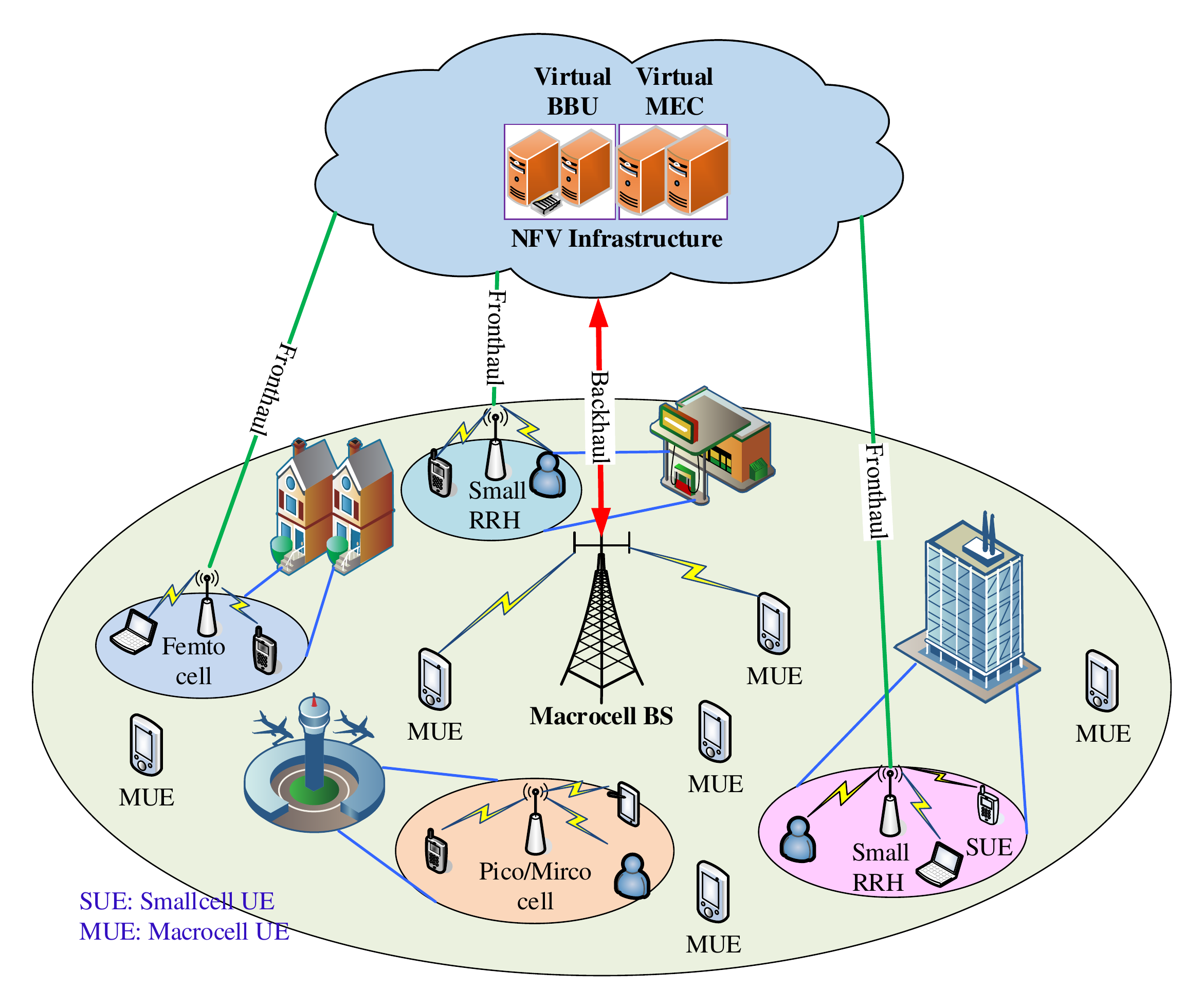}
	\caption{H-CRAN MEC Architecture.}
	\label{Fig:HeterogeneousNetworks}
\end{figure*}

\subsection{Fundamentals of Heterogeneous C-RANs}
\label{SubSec:Fundamentals_HetNet}
\color{black}{
To meet the unprecedented increase in the network traffic volume and the massive number of connected devices, network densification has become the cornerstone of the 5G networks, where more base stations and access points are added and spatial spectrum reuse is exploited. HetNet is defined as an integration of higher-tier macrocells and lower-tier small cells, for example, picocells, femtocells, and relay nodes  \cite{Pham2018Energy, Ngo2014Architectures}. HetNets have been developed because of its following benefits: 1) better coverage and capacity, 2) improved macrocell reliability, cost benefits, and 3) reduced cost and subscriber turnover \cite{Chandrasekhar2008Femtocell}. However, the deployment of dense HetNets has several challenges: 1) severe interference, 2) unsatisfactory energy efficiency, and 3) inflexibility and unscalability.
To overcome these challenges, another new promising network infrastructure, C-RAN, is proposed to provide a high transmission data rate and high energy efficiency performance, which attracts a lot of attention from academic and industrial communities \cite{CRANMagazine14}. In \cite{PR2014Magazine}, the challenges and
requirements of C-RAN were studied to enable network densification and centralized operation of the radio access network over heterogeneous backhaul networks. In C-RANs, shown in Fig. \ref{Fig:HeterogeneousNetworks}, a large number of low-cost low-power RRHs connecting to the BBU pool through the fronthaul links, are randomly deployed to enhance the wireless capacity in hotspots. RRHs operate as soft relay by compressing and forwarding the received signals from users to the BBU pool via wire/wireless fronthaul links. As a result, the combination of HetNets and C-RANs, known as heterogeneous C-RANs (H-CRANs), is proposed as a potential solution to provide high spectral and energy efficiency \cite{HCRAN2017}. In order to support more 5G applications and reduce the investment cost of MEC deployment, MEC was proposed to be combined with CRAN in \cite{Reznik2018MEC_CRAN}, where MEC services enable to exploit C-RAN by using the planned BBU pool. Even though CRANs and MEC can be perfectly paired to provide low latency for the IoT applications in HeNets, the co-location of MEC and C-RAN results in some challenges (e.g., network management), especially in HetNets.}

\subsection{Motivations and Challenges}
\label{SubSec:Motivation_HetNet}
\color{black}{H-CRANs can provide large coverage and high energy efficiency, while MEC can provide the considerable computing capability for the low-latency applications. Collocating these two key technologies can help support more applications in 5G. Considering the computational and storage resources in the BBU pool and the distribution of the RRHs, H-CRAN can be combined with MEC to facilitate the implementation of the MEC system. Therefore, the combination of MEC with H-CRANs can bring following benefits:
\begin{itemize}
\item The investment of MEC deployment can be significant reduced by collocating MEC and H-CRAN. As we all know, it is significant investment to deploy a sufficiently extensive MEC network. One way to mitigate the investment cost is to bootstrap MEC deployment to the C-RAN deployment. In this case, the cost of providing additional task calculation across the existing BBU pool or RRHs will be reduced.
\item The combination of MEC and H-CRAN can provide operational flexibility and network re-configurability, which can be offered by virtualization of H-CRAN. The H-CRAN can facilitate a faster radio deployment by reducing the time needed in the conventional deployments, e.g., standard General-Purpose Processors. Since CRAN virtualizes much of the RAN functions, thus MEC can also benefits large coverage, the energy savings, network simplicity and high security from H-CRAN.
\item H-CRAN MEC can be flexibly deployed across different locations. For example, C-RAN can process the task signals any locations, e.g., cell-tower co-located hut. Since H-CRAN deployment requires a substantial amount of processing power, it can automatically becomes an MEC server to calculate the tasks from the mobile users. 
\end{itemize}}

\color{black}{In addition to the above benefits, there exist several challenges in H-CRAN MEC systems that can be induced by co-location of MEC and H-CRAN, e.g., deployment scenarios design.} In the following, the major challenges of H-CRAN MEC systems are discussed \cite{Oo2017Offloading, Teng2018Resource, Reznik2018MEC_CRAN}.
\begin{itemize}
\item \color{black}{In the H-CRAN MEC system, the balance of the deployment and the network performance should be well investigated. Since H-CRAN supports a dynamic capacity of the H-CRAN, how far the C-RAN/MEC site is located to cell-sits  will affect the performance of MEC systems, e.g., how well it can support the applications. For example, locating CRAN/MEC site in a central office can reduce the cost significantly but it causes high latency \cite{Reznik2018MEC_CRAN}. In this case, use-cases should be carefully studied to run which applications at which sits.}
\item \color{black}{Most resource management methods for MEC consider the computation resource at MEC servers \cite{KWangMECCRAN2018,XWang2018} and thus can be applied in H-CRAN MEC directly. However, it is still challenging to jointly optimize computing resource and scheduling network resource in H-CRAN \cite{Tran2017Collaborative}. Especially in HetNets, the cross-layer and inter-cell interference needs to be considered. Moreover, based on NFV of C-RAN, the dynamic resource management scheme may need to be redesigned to elastically schedule virtual computation resources under different network sizes and task arrival rates.}
\item \color{black}{Security is another issue to be addressed in H-CRAN MEC systems. Since MEC service supports various kinds of applications, such as third party applications, which are not controlled by mobile network operator directly. There may be risks that these applications will exhaust resources or offer hackers to affect the functions of the network. Therefore, the service of performing integrity assurance checks on applications should be considered at installation or upgradetion.}
	\item \color{black}{Due to the existence of inter-carrier interference, the resource allocation problem in H-CRAN MEC networks is much more challenging than that in traditional MEC systems \cite{Tran2017Collaborative}.} To mitigate this effect, the spectrum resource within each cell can be divided into orthogonal subchannels, which should be efficiently allocated to mobile users (i.e., which subchannel a user should use to offload its computation task to the MEC server). \color{black}{In H-CRAN MEC networks, various types of resources need to be considered to reduce the ICI, including not only conventional wireless resources (e.g., subchannel, transmit power, time, and space) but also contra costs (e.g., backhaul spectrum, harvested energy, computing capabilities, and caching storage). The major challenges of dense H-CRAN MEC systems are user association, computation offloading, interference management, and resource allocation. More importantly, these problems are tightly coupled and must be solved jointly.}  
	\item \color{black}{On the one hand, it is foreseeable that a massive number of MEC servers will be widely deployed in the near future, which can be distinctly different in sizes (computing units) and configurations (computational speeds). On the other hand, the association between users and MEC servers (BBUs) greatly depends on the deployment locations of the MEC servers (BBUs). User mobility can be ignored whenever the UE moves inside the geographical area covered by the centralized BBUs. The type of BBU centralization determines the system efficiency and the user experience.} 
   
\end{itemize}

\subsection{State of the Art}
\label{SubSec:State_of_the_Art_HetNet}
\color{black}{The majority of the existing studies have focused on Heterogeneous MEC (Het-MEC) and C-RAN MEC. For Het-MEC network,  there are several papers working on interference management in dense Het-MEC systems \cite{Sardellitti2015Joint, AlShuwaili2017JointUplinkDownlink, Zhang2016EnergyEfficient, Zhang2018JointComputation, Zhao2018UplinkRA, Sun2017EMM}. In \cite{Sardellitti2015Joint}, the authors investigated a joint problem of radio and computational resources to minimize the total energy consumption of all mobile users under transmit power budget, latency, and maximum computing capability constraints. Similarly, Al-Shuwaili \textit{et al.} in \cite{AlShuwaili2017JointUplinkDownlink} considered several issues in single-server multi-cell Het-MEC systems: (1) the management of uplink and downlink interference, (2) the allocation of backhaul capacity for task offloading, and (3) the allocation of computing capabilities at the cloud for offloading users. Moreover, the joint optimization of offloading decisions and resource allocation has been extensively investigated to improve the network performance \cite{Zhang2016EnergyEfficient,Zhang2018JointComputation}. In order to realize the potential benefits of dense Het-MEC networks, a new technical challenge is mobility management. According to \cite{Wang2017LocalizedMobility, Semiari2018Caching}, there are several key issues for mobility management in Het-MEC systems. First, users may experience frequent handover when they move across different small-size and small-coverage smallcells/ MEC servers, thus increasing the overhead and interrupting the MEC services \cite{LopezPerez2012Mobility}. Second, continuously performing handover measurements and processing, which is needed to discover new target MEC servers in dense Het-MEC systems, is power- and radio resource-consuming, especially for battery-limited users. Third, in traditional dense HetNets, handover decision is mainly based on the quality of radio signals between users and potential eNBs. In addition, due to the lack of future information, e.g., channel conditions, available computing resources, task arrivals, the offloading and handover decisions should be known without prior information and be optimized in a long-term manner \cite{Sun2017EMM}. Due to its critical importance, an extensive body of work has appeared in the literature to address the challenges of mobility management in conventional dense HetNets \cite{Wang2017LocalizedMobility, Semiari2018Caching, LopezPerez2012Mobility, Prasad2013EnergyEfficient, Xenakis2014Mobility, Giust2015Distributed, Zhang2017NetworkSlicing, Qiao2017JointDeployment}. For example, two localized mobility management schemes for dense HetNets were proposed in \cite{Wang2017LocalizedMobility}, a cache-enabled mobility management framework in mmWave-microwave HetNets was studied in \cite{Semiari2018Caching}, various energy-efficient cell discovery techniques were discussed in \cite{Prasad2013EnergyEfficient}, a comprehensive review of mobility management was provided in \cite{Xenakis2014Mobility}, and the adoption of distributed mobility management was presented in \cite{Giust2015Distributed}. Although interesting, the body of work in \cite{Wang2017LocalizedMobility, Semiari2018Caching, LopezPerez2012Mobility, Prasad2013EnergyEfficient, Xenakis2014Mobility, Giust2015Distributed, Zhang2017NetworkSlicing, Qiao2017JointDeployment} solely focused on mobility management in HetNets. Taking challenges of mobility management in dense Het-MEC systems into consideration, the study in \cite{Sun2017EMM} optimized the association (which MEC server is selected for remote execution) and handover (i.e., when task migration is needed) decisions to minimize the average delay with the long-term energy budget constraint. Simulation in \cite{Sun2017EMM} indicated that without complete future information, the proposed algorithm for energy-efficient mobility management can still achieve close-to-optimal performance while guaranteeing the long-term energy budget constraint.}

\color{black}{There are several research works on the combination of C-RAN MEC systems \cite{KWangTVT2019MECCRAN,KWangMECCRAN2018,XWang2018}.
In \cite{KWangTVT2019MECCRAN}, the authors focused on C-RAN MEC systems to minimize energy by the proposed two algorithms, i.e., decentralized local decision algorithm and centralized decision and resource allocation algorithm. 
To deal with the resource-limited mobile user with computation intensive tasks, C-RAN with MCC was combined to provide high energy efficiency performance \cite{KWangMECCRAN2018}, in which a joint computational resource and transmit power allocation allocation scheme was proposed to minimize the energy consumption under the constraints of task latency, and fronthaul capacity. To further enhance the capabilities of mobile devices, C-RAN with MEC was proposed to be combined with each other to efficiently address the increasing mobile traffic issue \cite{XWang2018}. Different with previous work, in \cite{XWang2018}, a jointly network resource framework was proposed for power-performance tradeoff of mobile service provider. In this work, Lyapunov technique was exploited to dynamically make online decisions in consecutive time slots for task request. The proposed algorithm can achieve close to optimal performance. In \cite{ZjianAccess2019}, the profit function based on revenue and cost analysis was maximized by jointly optimization of offloading strategy, communication and computation resource. 
MEC was applied to ultra dense networks (UDNs) \cite{MChen2018}, where the authors investigated the task offloading policy in MEC-enabled UDN and introduced the software defined networking technology to manage the computation resource in edge cloud with centralized controller. Furthermore, there are other resource allocation scheme were proposed for other C-RAN MEC scenarios, i.e., Vehicular Fog-RANs \cite{KXiongTVT2019}, Near-Far Computing Enhanced C-RAN and \cite{YCai2019}.}

\subsection{Learned Lessons and Potential Works}
\label{SubSec:Lesson_and_Potential_Works_HetNet}
\color{black}{Due to the great benefits offered by MEC and H-CRAN, it is envisioned that the combination of MEC and H-CRAN is unavoidable in the future. Although various problems and issues in H-CRAN MEC systems have been intensively studied, there are still several challenges. In the following, we discuss some challenges in dense H-CRAN MEC systems and outline the open research directions.}
\begin{itemize}
	\item \textit{Computational complexity and signaling overhead}: \color{black}{It is obvious that the centralized optimization is usually easy to implement compared to distributed approaches and can provide the optimal/near-optimal solution with the desired performance guarantee. However, in H-CRAN MEC systems such centralized approaches are not scalable due to the explosive increase in the numbers of mobile users, eNBs, and MEC servers.} As a result, there is a need for lightweight and effective algorithms. In these schemes, distributed approaches can offer many benefits as they do not need any central entity and the algorithms are based on only local information or small amounts of signaling overhead. However, it is hard to guarantee the solution optimality with distributed approaches due to the lack of complete information. Therefore, one needs to tradeoff between the computational complexity and solution optimality. An effective way is to decompose the entire network into several regions and assign the responsibility for executing the algorithm to distributed MEC servers, that is the underlying problem is decomposed into subproblems, which are executed distributively at different MEC servers. This would significantly reduce the amount of information which need to be exchanged between the central entity and all users; hence, the network overhead can be also degraded. 
	\item \textit{Mobility management}: Ensuring the benefits of mobile users through computation offloading while taking into account user mobility is a challenging issue. \color{black}{Most existing studies in (Het-/CRAN-) MEC systems ignore the effect of user mobility due to its difficulty and intractability. In the proposed H-CRAN MEC systems, users may change their positions while using MEC services, e.g., they can move out of the coverage area of their source MEC servers and are in the serving coverage of other ones. This will result in user association in H-CRAN MEC since the scheduler may need to re-associate the user to a different RRH and then the offloaded task can be calculated by BBU pool with MEC server. In this case, scheduler (BBU pool) needs to be aware of user mobility in order to maintain service continuity. Thus the dynamic user association and resource allocation can be well studied in the future work. For example, some ML algorithms can be exploited to address the user mobility issue in the resource allocation for H-CRAN MEC.} 
	 Another potential solution to deal with user mobility is enabling MEC servers to continuously update the user context and then designing context-aware algorithms. Instead of using one-shot optimization, long-term optimization can be used to tackle the challenges of user mobility. To illustrate this point, we consider the following example with a mobile user, which is located far from the MEC server. The short-term optimization for computation offloading decision is not offloading, that is local execution. However, fixing this short-term decision is not always optimal since the user can move to a new position with better channel quality. Moreover, the short-term offloading decision affects not only the instant performance but also the long-term energy budget. In summary, there is a big room for researches into mobility management in dense Het-MEC systems. 
	\item \textit{Interference management and joint resource allocation}: \color{black}{Inherited from dense HetNets, the spectrum reuse among cells incurs severe mutual interference, which may significantly reduce the
expected system spectrum and energy efficiency. Therefore, the challenges for interference management in H-CRAN MEC systems remain to be solved for many reasons. Heterogeneity of mobile users and BBU pool with the MEC server makes the interference problem more challenging due to various transmit power budget of users in the uplink. Moreover, the network scheduling resource, communication resource and computing resource at BBU pool are coupled with each other, which makes the resource allocation more challenging. The various computation task characteristics require different priorities for users in accessing radio and MEC resources. Finally, interference management is highly coupled with other domains, such as resource allocation and network planning. Hence, more sophisticated interference management schemes incorporating features of H-CRAN MEC systems would be highly required for improving the users QoS with MEC services.} 
	\item \textit{Wireless backhaul limitation}: \color{black}{In H-CRAN MEC scenarios, the capacity of backhaul and fronthaul is of an important issue. For example, in case that backhaul is limited, the transmission time via backhaul links should be taken into consideration, thus affecting the offloading decisions of users (and other optimization variables as well). Most research works assume that small cells are connected with the central location (where vBBU and MEC servers are located) through high-speed wired links, e.g., fiber links. As a result, the scenario with wired backhaul/fronthaul may be simple and limited to implement for H-CRAN MEC networks, and then discuss their proposed approach in such network settings. The wireless bachhaul and fronhaul can be further investigated to enhance the networks performance. For example,  the authors in \cite{Pham2018Mobile}  focused on MEC with wireless backhaul; however, the network setting in this literature is simple, comprising a small-eNB and an MEC server collocated at the macro-eNB. This work is served as a fundamental study for more complex frameworks, e.g., the extension to dense Het-MEC systems and the consideration of mixed wireless and wired backhaul links.}
	\item \textit{Physical security}: \color{black}{In H-CRAN MEC networks, security will be a significant issue since MEC applications will run on the same physical platforms as some network functions. Therefore, to reduce the risk of that the external eavesdroppers/hackers who may affect the network functions, the physical layer security can be studied for H-CRAN MEC systems, which will be promising research topic.}
\end{itemize}

\section{MEC and Machine Learning}
\label{Sec:MEC_MachineLearning}
This section reviews the fundamentals and applications of ML in addressing various MEC problems: edge caching, computation offloading, joint optimization, security and privacy, big data analytics, and mobile crowdsensing. We also identify challenges and potential directions to energize further studies on applications of ML in MEC. 

\subsection{A Brief Review of Machine Learning in Wireless Networks}
\label{Subsec:Overview_ML}

\begin{figure*} [t]
	\centering
	\includegraphics[width=0.60\linewidth]{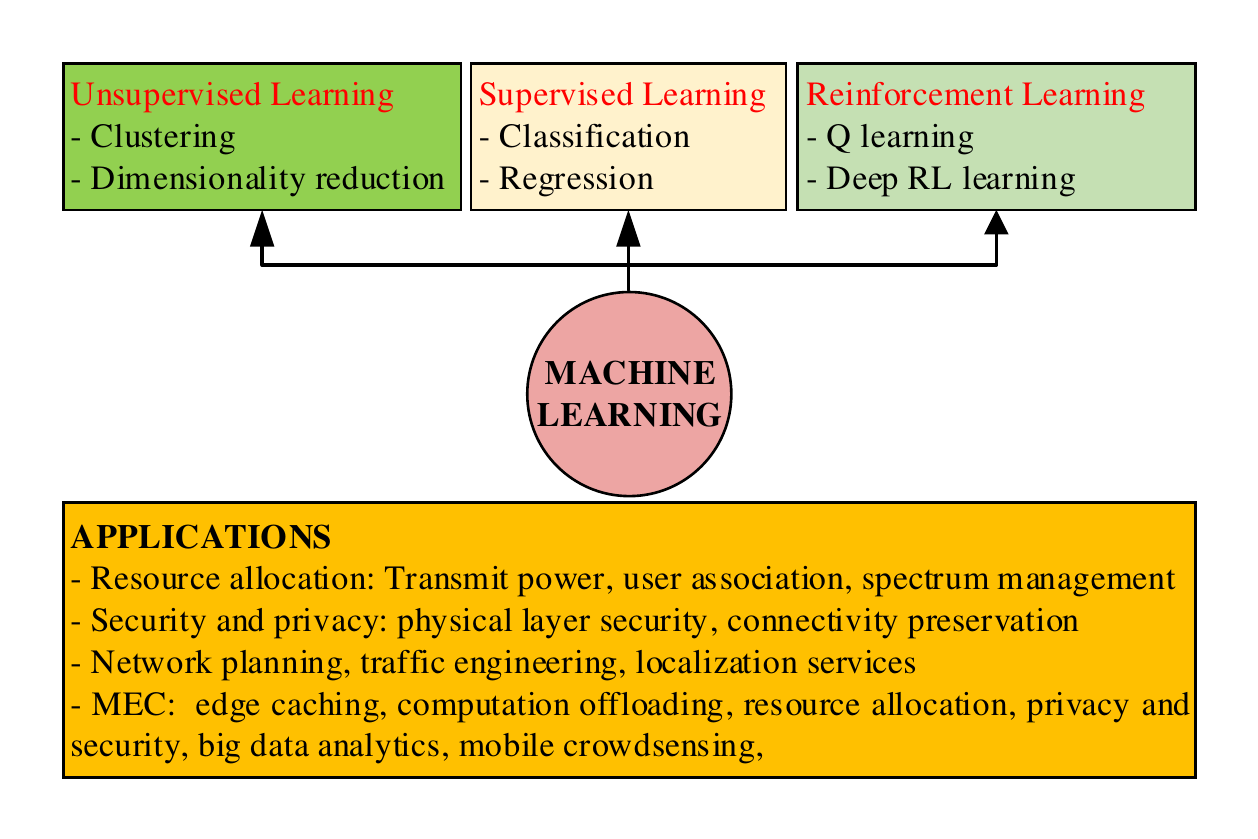}
	\caption{Classification and applications of ML in mobile and wireless networking.}
	\label{Fig:MachineLearning_and_ItsApplications}
\end{figure*}

ML was born from pattern recognition and the notion that computers are able to learn without being explicitly programmed. Recently, ML has been applied in a myriad of applications, for example, virtual personal assistants, video surveillance, social media services, email Spam and malware filtering, search engine result refining, and product recommendation. There are several reasons why ML algorithms are increasingly being used: 1) ML enables systems that can automatically adapt and customize themselves to individual users, 2) ML can discover new knowledge from large databases, 3) ML can mimic human and replace certain monotonous tasks, which requires some intelligence, 4) ML can develop systems that are difficult and expensive to construct manually because they require specific detailed skills or knowledge tuned to a specific task, and finally 5) there is a vast  increase in computational power, growing progress in available algorithms and theory developed by researchers, and increasing support from industries. Generally, ML is divided into three core types: {\color{black}supervised} learning, unsupervised learning, and reinforcement learning (RL). In supervised learning, the training data includes both inputs and labels (outputs) and the goal is to estimate the unknown model that maps known inputs to known labels whereas in the unsupervised learning, the training data does not include the labels and thus the goal is to learn a more efficient representation of a set of unknown inputs \cite{lapan2018deep}. In RL, the agent is not told what actions to do, but instead continuously interacts with environment and tries to follow a good policy that yields the best return \cite{lapan2018deep}. DL has been introduced as a breakthrough technique and a huge step forward in ML, which can achieve higher-level representations based on simpler ones. DL works in a similar way as the human brain processes information and learns and is heavily based on statistics and applied mathematics \cite{Goodfellow2016DeepLearning}. The classification and applications of ML in mobile and wireless networking, also in MEC and other edge computing paradigms, are illustrated in Fig~\ref{Fig:MachineLearning_and_ItsApplications}. {\color{black}Recently, the ITU Telecommunication Standardization Sector proposed a unified architecture for ML in future networks, where MEC is expected to play crucial roles as source, collector, pre-processor, model, policy, distributor, and sink \cite{ITUT_2019_UnifiedArchitecture}. For example, MEC can collect data from end users, then perform data preprocessing, and execute an ML model to extract necessary information before sending the output to the central cloud for further training.} Moreover, some surveys and tutorials on ML \cite{Simeone2018AVeryBriefIntroductiontoML}, DL \cite{Mao2018DeepLearning}, (deep) RL \cite{Mnih2015Humanlevel, Arulkumaran2017DeepRL}, as well as their applications in communications and networking \cite{Nguyen2008aSurveyOfTechniques, Luong2019Applications} have come out, and readers can refer to the aforementioned literature for more details.

Due to the rapid evolution of wireless communications and networks, it is believed that artificial intelligence in general and ML in particular will play vital roles in beyond 5G and 6G \cite{Chen2019Machine}. In general, ML can provide the following advantages:
\begin{itemize}
	\item First, the most natural advantage of ML is the ability to learn from big data to improve the network operation and performance, which can be done without any hand-crafting feature. The importance of learning arises naturally in wireless networks since 1) mobile data is massive, 2) mobile data increases at exponential rates, 3) mobile data is non-stationary (i.e., the time duration for data validity can be relatively short), 4) mobile data quality is not guaranteed (i.e., data collected can be low-quality and noisy), and 5) mobile data is heterogeneous (i.e., data can be generated from many sources, such as mobile users and IoT devices, and in different types) \cite{Alsheikh2016MobileBigData}. Examples of ML for channel estimation and signal detection \cite{Ye2018PowerOfDeepLearning}, localization based on channels state information \cite{Wang2017CSIBased}, and channel decoding \cite{Nachmani2018DeepLearning} show promising applications of ML for wireless communications. 
	\item Second, the design and optimization of wireless networks are sufficiently challenging without known channel and mobility models. Conventional optimization techniques are usually performed in an offline, heuristic, or iterative manner, which cannot guarantee the performance optimality or is not suitable for dynamic and time-varying systems. ML is a promising tool such that the network operation can be optimized over time, thus continuously improving the network performance. For example, ML showed a noticeable improvement in uplink data rate by managing uplink interference in cellular networks \cite{Deb2015LearningBasedULInterference}. 
	\item {\color{black}Third, joint 4C optimization in 5G and beyond is immensely complicated due to large state and action spaces, heterogeneous network devices, and various QoS requirements.} In such case, ML is capable of providing online and/or fully-distributed algorithms. Moreover, model-free wireless networks introduce various issues of channel modeling, problem formulation, and closed-form solution, which, however, can be efficiently solved by ML. 
	\item Next, ML should be deployed at the IoT device level and on large-scale distributed networks without violating user data privacy. In 2017, Google introduced an additional ML approach, called ``federated learning" that enables individual devices collaboratively learn a shared prediction model while keeping their own data locally, thus improving the training efficiency and data privacy. As the network will be highly dense and heterogeneous, federated learning is expected to be a major tool of beyond 5G. Motivated by the application of federated learning in Google board in Android \cite{FederatedLearning}, there have been a wide range of applications and problems in wireless networks that can adopt federated learning. For instance, in vehicular networks \cite{Samarakoon2018DistributedFL}, federated learning enables each vehicle user to estimate the tail distribution of network-wide queue length from locally available training data, which can achieve very high accuracy and reduce the amount of exchanged data by 79\%. 
	\item Last, since edge computing will play an important role in providing low-latency actions and the majority of intelligent applications will be deployed at the network edge, the emergence of edge learning is unavoidable. On the one hand, exploiting edge learning to extract useful information from a massive amount of mobile data can extend the capability of small IoT devices and enable the deployment of compute-intensive and low-latency application at the edge \cite{Khelifi2019BringingDL, Li2018LearningIoT}. On the other hand, edge learning can circumvent drawbacks of cloud AI and on-device AI through the tradeoff between the learning model complexity and the training time \cite{Zhu2018Intelligent}. 
\end{itemize}

\subsection{Machine Learning for Multi-Access Edge Computing}
Optimizing MEC face several challenges of caching placement, allocation of radio and computing resources, assignment of computation tasks, and joint 4C optimization. The existing literature has studied a number of problems in MEC systems, including \emph{computation offloading} \cite{Yang2018DeepRLbasedResourceAllocation, VanLe2018ADRLbasedOffloading, M_Min_TVT_2019, Yu2017ComputationOffloading, Xu2017OnlineLearning, Chen2019OptimizedCO, Wang2019TrafficAndComputation, Chen2019Machine}, \emph{caching} \cite{Zhu2018DeepRLforMEC, He20018SoftwareDN, Hou2019AQLearning, Tan2018MobilityAware}, \emph{joint 4C optimization} \cite{Li2018DeepRLbasedCO, Huang2019DeepRlforOnlineOffloading, Wei2018JointOptimization, He2018IntegratedNetworking, Tan2018MobilityAware, He2018Trustbased}, \emph{security and privacy} \cite{Gheisari2019ECA, Abeshu2018DeepLearning, Xiao2018SecurityInMEC, Tu2018SecurityInFogComputing, Du2018BigDataPrivacy, Du2018DifferentialPrivacy, Yang2018MachineLearning, Min2019LearningBased}, \emph{big data analytics} \cite{Meng2018QoEDriven, Du2018BigDataPrivacy, Du2018DifferentialPrivacy, Alsheikh2016MobileBigData}, and \emph{mobile crowd sensing} \cite{Zhou2018RobustMCS}. In what follows, we summarize the sate-of-the-art related to applications of ML approaches in these aspects.

\subsubsection{Edge Caching} 
Studies on mobile edge caching have focused on three main issues that are where to cache, what to cache, and how to cache \cite{Wang2017aSurvey, Wang2014CacheITA}. In terms of caching places, the state-of-the-art showed that the requested content can be cached at macro-eNBs, small-eNBs, and/or end users, where the storage resource of nearby mobile devices is exploited for content caching and D2D communication is used for content retrieving \cite{Liu2016CachingAtTheEdge}. {\color{black}To decide what to cache, one popular metric is the content popularity, which is defined as the ratio of the number of requests for a particular content to the total number of requests from all users within a specific region during a period of time}. The survey paper \cite{Wang2017aSurvey} showed that there are five main algorithms: content replacement policies such as the least frequently used (LFU) and least recently used (LRU), user preference based policies, learning based policies, non-cooperative caching, and cooperative caching. 

{\color{black}As the content popularity is time-varying and cannot be known in advance, many studies have focused on ML based caching strategies}. Most of the existing works focus on applications of deep RL (DRL) for proactive caching since DRL is able to learn caching policies automatically without any predefined network model and explicit assumption. The authors in \cite{Zhu2018DeepRLforMEC} explored the key challenges of edge caching and reviewed the state-of-the-art related to learning-based caching policies and algorithms. They showed that mobile edge caching schemes can be classified into two main approaches: \textit{popularity-prediction-based approach}, where the popularity estimation and caching policy are learned separately, and \textit{RL based approach}, where these two terms are learned simultaneously. Other studies on (deep) RL based caching algorithms can be found in \cite{He20018SoftwareDN, Hou2019AQLearning, Tan2018MobilityAware}. {\color{black}All of the experimental results from the above literature demonstrate the effectiveness of learning algorithms in terms of average cache hit ratio, training accuracy, and energy cost compared to baselines approaches, e.g., LFU and LRU}. It is a widely held axiom that besides the historical data, the correlation between social and geographic data of mobile users can be utilized to provide more accurate content popularity prediction. Thus, the authors in \cite{Chang2018LearnToCache} proposed using big data analytics techniques to advance edge caching designs and proved the effectiveness of these techniques via two case studies of eNB caching and device caching. However, big data analytics, particularly ML/DL mechanisms, has several challenging issues for implementation \cite{Thar2018DeepMEC}: huge computation resources required to process the high-dimensional big data, lack of an appropriate prediction model for various types of DL models, optimization of DL parameters, e.g., the depth of deep neural networks and learning rate. 

\subsubsection{Computation Offloading}
{\color{black}Due to the importance of computation offloading from the user perspective, recent years have seen many research works pertaining to computation offloading.} In \cite{VanLe2018ADRLbasedOffloading}, the authors formulated the computation offloading decision problem of a user in ad-hoc mobile clouds as an MDP. More specifically, both channel gains between the user and cloudlets and the user's and cloudlets' queue states are considered in the system state, the action is the task distribution decision (i.e., how many tasks to process locally and how many tasks to offload to each cloudlet), and the reward function is defined to maximize the user utility and minimize the cost of required payment, energy consumption, delay and, task loss probability. Simulation results showed that the DQN based offloading decision algorithm performed well under various task arrival rates. {\color{black}In \cite{M_Min_TVT_2019}, the combination of a ``hotbooting" Q-learning\footnote{The hotbooting Q-learning technique exploits experiences in similar scenarios to initialize the Q-function value so as to save the exploration time at the beginning of learning.} and DL was adopted to find the computation offloading decision and offloading rate in IoT with energy harvesting.} Another study on computation offloading in IoT with energy harvesting can be found in \cite{Min2019LearningBased}. The work in \cite{Yu2017ComputationOffloading} formulated the offloading decision problem as a multi-label classification problem and then utilized the deep supervised learning to minimize the computation and offloading overhead. Simulation results demonstrated that the proposed scheme can reduce the system cost in average by 49.24\%, 23.87\%, 15.69\%, and 11.18\% compared to the no offloading, random offloading, total offloading, and multi-label linear classifier-based offloading schemes, respectively, and can achieve a higher offloading accuracy. The joint optimization of bandwidth allocation and offloading decision was solved by the DQN-based algorithm in \cite{Huang2018DeepReinforcementLearning} and by the distributed DL-based algorithm in \cite{Huang2018DistributedDeepLearning}. The literature \cite{Xu2017OnlineLearning} jointly studied the offloading and autoscaling policy (i.e., the number of MEC servers is activated) in energy harvesting MEC systems, which was learned by a post-decision RL algorithm. In \cite{Chen2019OptimizedCO}, the DQN was deployed to learn the offloading decision and energy allocation of a representative mobile user in ultra-dense sliced RAN. The state is characterized by the {\color{black}energy level}, computation task queue, {\color{black}user association}, and channel gain quality, and the immediate reward is the weighted sum of satisfaction of the task execution delay and computation task drops, the task queuing delay, the penalty of failing to execute a computation task, and the payment of accessing the MEC service. The work in \cite{Wang2019TrafficAndComputation} utilized RL to jointly consider traffic and computation offloading for industrial applications in fog computing.


\subsubsection{Joint Optimization} 
Due to the facts that 1) the joint 4C optimization is needed for improving the network performance and 2) conventional approaches cannot efficiently solve the optimization problems with large action and state spaces, recent studies on MEC have addressed various problems pertaining to joint 4C optimization. For example, the authors in \cite{Tan2018MobilityAware} investigated two deep Q-learning models for mobile edge caching and computing in vehicular networks. To reduce the computational complexity of the original problem and circumvent the high mobility constraint of vehicles, the authors further proposed deploying two DQN models at two distinct timescales. In particular, each epoch is divided into several time slots and then the large timescale deep Q-learning model is executed at every epoch while the small timescale model is performed at every time slot. We note that the concept of multi-timescale control has been applied for some existing research works, e.g., cross-layer optimization \cite{Pham2015Multi_Timescale, Pham2017NetworkUtility}. {\color{black}The authors in \cite{Li2018DeepRLbasedCO} investigated two learning models, classical Q-learning and DQN method, for joint optimization of offloading decision and computation resource allocation in single-server MEC systems. Since DRL with discretized states suffers from the curse of dimensionality and slow convergence when a high quantization accuracy is required, a continuous control with DRL based framework of computation offloading and resource allocation in wireless powered MEC systems was studied in \cite{Huang2019DeepRlforOnlineOffloading}. As shown in \cite[Fig. 3]{Huang2019DeepRlforOnlineOffloading}, the proposed algorithm is composed of two alternating phases: i) offloading action generation to quantize the relaxed offloading decision as a set of binary actions, and ii) offloading policy update to select the best offloading action among quantized ones.}

More recently, there have been some works that study the joint optimization of computation, caching, and communication. The work in \cite{Wei2018JointOptimization} studied the joint optimization of resource allocation in hierarchical networks of fog-enabled IoT with edge caching and computing capability. In \cite{He2018IntegratedNetworking}, the authors proposed an integrated framework of networking, caching, and computing for connected vehicle networks and showed that the proposed DRL based algorithm is superior to the existing static scheme and those without virtualization, MEC offloading, or edge caching. Besides the integration of edge computing, in-network caching, and D2D communication, the literature \cite{He2018Trustbased} also took into consideration the social relationships among mobile users so as to improve the reliability and efficiency of resource sharing and delivery in mobile social networks.

\subsubsection{Security and Privacy} 
The following reasons explain why security and privacy are the greatest challenges \cite{Roman2018MobileEdgeComputing}. First, since there are many enabling technologies of MEC, it is necessary to not only protect individual enabling technology, but also orchestrate the diverse security algorithms. {\color{black}Second, the distributed nature of MEC causes many new network situations (e.g., heterogeneous computing capabilities and collaboration between edge devices), which call for new security mechanisms}. Third, it is possible that a large-scale edge computing system can be severely affected by the security threats of just a network component. Finally, there are many scenarios and aspects that can be influenced by privacy and security threats, e.g., private data generated by in-car sensors and critical emergency systems. In edge computing paradigms, there are numerous security and privacy threats, for example, wireless jamming, denial of service, man-in-the-middle, spoofing attacks, privacy leakage, virtual machine manipulation, and injection of information \cite{Roman2018MobileEdgeComputing, Xiao2018SecurityInMEC}. 

Recently, ML-based security and privacy in MEC have been studied from various perspectives. The use of DL for cyber-attack detection in edge networks was considered in \cite{Abeshu2018DeepLearning}, where the experiments demonstrate that the DL based model is better than that with shallow model in terms of learning accuracy, detection rate, and false alarm rate. The authors in \cite{Xiao2018SecurityInMEC} proposed different RL based edge caching security mechanisms of anti-jamming mobile offloading, physical authentication, and friendly jamming. Taking the randomness and variation of wireless channels between mobile users and fog nodes, the literature \cite{Tu2018SecurityInFogComputing} studied Q-learning based physical layer security in fog computing to improve the impersonation detection attack and the accuracy of receivers by learning from the dynamic environment. The work in \cite{Yang2018MachineLearning} investigated a new ML based privacy-preserving multifunctional data aggregation framework in order to overcome drawbacks of existing methods, which are high computation overhead, communication efficiency, and single aggregation function calculation. In \cite{Min2019LearningBased}, privacy-aware computation offloading in MEC-enabled IoT was studied, where the post-decision learning is used in conjunction with the standard DQN to accelerate the learning speed.

\subsubsection{Big Data Analytics} 
As aforementioned, there are three main challenges of mobile big data (MBD) analytics: \textit{large-scale and high-speed mobile networks} which reflect MBD volume and velocity, \textit{portability} which causes MBD volatility, and \textit{crowdsensing} which introduces MBD veracity and variety. Big data analytics enable the design of many smart applications, such as smart city, smart building, and smart manufacturing \cite{Patel2017OnUsing}. Intelligence at the edge is expected to play a major role in data analytics applications. In \cite{Alsheikh2016MobileBigData}, DL is considered as an attractive solution for MBD analytics by leveraging several advantages: 1) DL scores highly accurate results, 2) DL can automatically generate intrinsic features from MBD, 3) DL does not require labeled samples as the input training data, and 4) multimodal DL allows the learning from heterogeneous data sources. {\color{black} MEC is highly suitable for big data processing. However, there are several challenges \cite{Ndikumana2019JointCommunication, Zhang2018OptimalTask}: 1) how to distribute big data to distributed resource-finite servers, 2) collaborative MEC for resource sharing and optimization is needed, 3) 4C resources are tightly coupled, and 4) privacy is a critical issue due to the lack of a central management entity}. 


Some recent studies have utilized ML to address various problems pertaining to MEC big data. The work in \cite{Du2018DifferentialPrivacy, Du2018BigDataPrivacy} divided big data processing into three steps: data collection, aggregation, mining and analysis. Moreover, the authors proposed two privacy-preserving methods, namely output perturbation (OPP) and objective perturbation methods (OJP). In particular, training data privacy can be achieved by adding randomization noise to aggregated query results in the OPP method and to the objective function in the OJP method. Experimental results showed the high accuracy and data utility of OPP and OJP algorithms. In \cite{Meng2018QoEDriven}, the authors tried to provide users with better QoE in pervasive edge computing environments. The authors first deployed a Tensor-Fast convolutional neural network (TF-CNN) algorithm to guarantee accuracy and increase training speed with big data and next managed high-dimensional big data by using different accurate data transmission rates. It was shown that the proposed TF-CNN algorithm can achieve a higher QoE performance than the state-of-the-art training model.

\begin{table*}[t]
	\caption{Summary of key MEC problems that can be solved by machine learning techniques.}
	\label{Table:Summary_ML_based_Problems}
	\centering
	\begin{tabular}{|p{1.5cm}|p{2.3cm}|p{1.5cm}|p{11.0cm}|}
		\hline 
		\textbf{Applications} & \textbf{Existing Works} & \textbf{Proposed Framework} & \textbf{Challenges}\\ 
		\hline
		\multirow{8}{1.7cm}{Edge caching} & DRL-based caching strategies & \cite{Zhu2018DeepRLforMEC, He20018SoftwareDN, Hou2019AQLearning, Tan2018MobilityAware}  & 
		- Combination of transfer learning and DRL to exploit knowledge from other domains, e.g., content distribution in mobile social networks can be used to learn caching strategies in D2D. \newline
		- Tradeoff between exploration and exploitation due to edge dynamics.\newline
		- Competition and collaboration between caching nodes (e.g. eNB or device caching). 
		\\ 
		\cline{2-4}
		& DL-based caching & \cite{Thar2018DeepMEC} & 
		- Determining the suitable model among various types of deep learning models.\newline
		- Configuring hyperparameter settings.
		\\ 
		\cline{2-4}
		& Big data analytics based caching & \cite{Chang2018LearnToCache} & 
		- Utilization of different features of the previously requested data.\newline
		- Time-varying and spatio-temporal user behaviors.
		\\ 
		\hline
		\multirow{1}{1.7cm}{Computation offloading} & DRL-based computation offloading & \cite{Yang2018DeepRLbasedResourceAllocation, VanLe2018ADRLbasedOffloading, M_Min_TVT_2019, Yu2017ComputationOffloading, Xu2017OnlineLearning, Chen2019OptimizedCO, Wang2019TrafficAndComputation, Chen2019Machine}  & 
		- Dependence on statistical information of channel quality and task arrival rates.\newline
		- Time-varying user behaviors and unknown MEC network model.
		\\ 
		\hline
		\multirow{5}{1.7cm}{Joint resource optimization} & ML for caching, computation, communication, and control & \cite{Huang2019DeepRlforOnlineOffloading, Wei2018JointOptimization, He2018IntegratedNetworking, Tan2018MobilityAware, Li2018DeepRLbasedCO, He2018Trustbased, Huang2018DeepReinforcementLearning, Huang2018DistributedDeepLearning} & 
		- The complexity of joint optimization problems, and immense action and state spaces due to the combination of couple of different resource types.\newline
		- Real-time learning training model for time-varying and dynamic MEC systems.\newline
		- High overhead of signaling transmission and information exchange for Generation of the network state and action spaces, especially in ultra-dense networks.  
		\\ 
		\hline
		\multirow{7}{1.7cm}{Privacy and Security} & DRL-based privacy and security & \cite{Abeshu2018DeepLearning} & 
		- Lack of massive and high-quality training, validation, and test datasets, which is caused by heterogeneity of wireless networks, mobile devices, and edge nodes.\newline
		- Limited storage and computation for training DL models.  
		\\ 
		\cline{2-4}
		& DL-based privacy and security & \cite{Xiao2018SecurityInMEC, Tu2018SecurityInFogComputing, Min2019LearningBased} & 
		- Inaccurate and delayed state information, e.g., CSI and energy state information.\newline
		- Reward function evaluation that is usually estimated according to the security/privacy gain and the protection cost (e.g., computation and communication delay, and energy cost).\newline
		- Bad security policies at the beginning of learning (the basis of the trial and error methods), which can be effectively addressed by transfer learning techniques.\newline
		- Tight coupling between privacy and performance gain, thus requiring the optimization of privacy-aware computation offloading and resource allocation schemes. 
		\\ 
		\hline
		\multirow{1}{1.7cm}{Big data analytics} & ML-based big data processing & \cite{Meng2018QoEDriven, Du2018BigDataPrivacy, Du2018DifferentialPrivacy, Alsheikh2016MobileBigData} & 
		- Storage and computation burdens due to the curse of big data dimensionality.\newline
		- Tradeoff between the resource-limited MEC servers and the large-scale DL models.
		\\ 
		\hline
		\multirow{1}{1.7cm}{Mobile crowdsensing} & DL based MCS  & \cite{Zhou2018RobustMCS} &
		- Lack of privacy and security protection schemes for crowdsensing data.\newline
		- High computation overhead for collecting training data, i.e., the DL model requires a large amount of data to retrain the learning model due to edge dynamics.\newline
		- Lack of efficient DL approaches to be deployed at the lower-tier devices and to detect contaminated and/or fake data.
		\\ 
		\hline
	\end{tabular}
\end{table*}

\subsubsection{Mobile Crowdsensing} 
While mobile crowdsensing (MCS) has been widely studied in the literature, there are only a handful of studies on edge computing empowered MCS. There are several benefits of MEC in the context of MCS as follows \cite{Marjanovic2018EdgeComputing}. First, MEC enables the parallelization and partitioning of the centralized and large-scale problem, where MEC servers are responsible for controlling the sensing process on mobile devices located within their deployment area and manage MCS tasks within the same area. Second, the immense computational complexity of the central cloud that is caused by a large number of mobile users participating in MCS tasks with frequent context changes can be greatly reduced because of the distributed deployment of MEC. Third, MEC can reduce the latency of data and information propagation, that is suitable for real-time MCS services. Next, intensive computations can be offloaded from both mobile users and cloud servers to the edge and then being processed therein. Finally, MEC can reduce privacy threats since privacy-sensitive data can be distributed and handled across MEC servers. Recently, the work in \cite{Zhou2018RobustMCS} proposed a framework that integrates DL and MEC for robust MCS services. {\color{black}In particular, the proposed framework can be implemented by firstly designing an auction mechanism for participant recruitment, then using DL for data validation, and finally implementing data processing at the network edge.} In \cite{Zhou2018RobustMCS}, the authors also discussed several open research problems, including how to leverage DL to detect privacy and security threats, how to reduce computational overhead in vastly and rapidly changing environments, and how to implement DL in mobile users for energy and cost efficiency. {\color{black}A hierarchical computing architecture for task allocation was proposed in \cite{Zhou2019Privacy}, where the cloud layer does learning of participants' reputation and the edge layer communicates with participants for data collection and optimization.}

\subsection{Challenges and Future Works}
\label{Subsec:Challenges_ML}
Clearly, ML techniques will be an important tool for various problems in wireless networks and at the network edge so as to optimize edge caching, computation, enhance big data analytics, and improve security and data privacy. A summary of key problems solved by ML techniques in MEC is presented in Table.~\ref{Table:Summary_ML_based_Problems} along with major challenges. Despite many ML-based studies on MEC, there are still several key open problems that could be investigated in the future. 

\textit{Machine learning based frameworks of ultra-dense MEC systems}: It is widely expected that both wireless and wired backhaul solutions will coexist in future wireless networks. The simulation results in \cite{Pham2018Mobile} showed that the bandwidth allocation between wireless access and wireless backhaul plays a major role in the achievable performance. In this case, ML-based approaches can be deployed at the macro-eNB to predict the appropriate bandwidth partitioning factor based on user CSI and task characteristics. Moreover, a critical issue in ultra-dense MEC system is user association and its joint optimization with other aspects such as computation offloading and resource allocation. However, the joint problem of user association, offloading decision, and resource allocation are typically NP-hard non-convex, which are further exacerbated in time-varying and dynamic environments. In such networks, DRL can be used to provide fast and near-optimal solutions. 

\textit{Distributed and collaborative ML implementation in hierarchical and heterogeneous MEC}: The central implementation of ML-based algorithms faces serious challenges, such as learning complexity, storage and computation resources, and non-suitability for pervasive computing applications and large-scale systems. A potential solution is distributed ML, where the computation of a learning algorithm is divided into smaller parts and then these computations are allocated to distributed MEC servers. However, a number of questions need to be exhaustively answered when distributed ML is used: which computation parts can be divided, how to divide the computation to subtasks, how to synchronize the output among different MEC servers, and how to integrate the outputs from subparts into the output of the master model? Distributed ML becomes particularly important when a learning agent (e.g., MEC server) cannot observe the global state and action, and is merely aware of its local state, reward, and action. 

Actually, there is a tradeoff between the computation capability and learning efficiency when ML-based mechanisms are centralizedly implemented at resource-limited MEC servers. Thus, it is hard to efficiently implement a ML-based algorithm at MEC server with a very large number of users and an enormous amount of training data. Due to the fact that an artificial neural network (ANN) is composed of many layers (e.g., input, hidden, and output layers) \cite{Chen2019Machine}, the ANN model and the hierarchical MEC architecture are supposed to fit together, where an immediate layer of the entire ANN model can be offloaded to and performed by MEC layers (e.g., MEC at macro-eNBs and at small-eNBs) and the output of the edge learning is then transferred to higher-tier clouds for further processing. The collaborative learning offers considerable benefits from the reduction of training data size, the exploitation of ubiquitous computing, and the preservation of user data privacy. Moreover, DL approaches can be deployed at the MEC servers to detect contaminated and/or fake data, thus improving the data quality. For instance, Li \textit{et al.} in \cite{Li2018LearningIoT} considered a two-layer DL model for video recognition with IoT devices. Due to resource-limited MEC compared, the authors proposed determining the maximum number of computation tasks that can be handled at the edge layer. 

\textit{Federated learning and applications for MEC}: The conventional ML approaches require that the training server collects all data generated by individual devices and learns the training model centralizedly. Massive users can generate voluminous data, however, users may not be willing to share their data with the training server and other users. For example, such shared information can be any type of data collected by in-car sensing devices including cameras, radar, ultrasonic, and location-identifying sensors \cite{Coppola2016ConnectedCar}. Moreover, data privacy is applied to not only vehicle owners, but also the vehicle occupants. For instance, the image/video of customer captured by the in-car camera can be shared with the centralized cloud for the caching prediction model. The standard ML is not a suitable way to preserve data privacy. Federated learning leaves the training data distributed across individual users, thus enabling them to collaboratively learn a shared model while keeping their own data locally. Moreover, federated learning is able to address major drawbacks of distributed learning \cite{Konecny2016FederatedOptimization}, which are 1) lack of time and training data, 2) low performance due to heterogeneous user capabilities and network states, 3) unbalanced number of training data samples, and 4) nonindependent and identically distributed data among users.

Federated learning is expected to be a sharp tool for various problems in MEC. Take the computation offloading problem as an example, where massive users are trying to offload their computations to an MEC server for remote execution. Conventionally, to determine the offloading decision, users need to report their information such as channel gain, current battery level, and computation characteristics, to the MEC server \cite{Huang2019DeepRlforOnlineOffloading, Chen2019OptimizedCO}; however, such information can be revealed by eavesdroppers and can be used illegally to predict the user location. Applying federated learning, each user needs to download the master model from the MEC server and then learns the offloading decision based on its local information only, and the MEC server is merely responsible for updating the master model according to updates from individual users. In such way, federated learning can preserve data privacy and provide distributed offloading decisions, thus being suitable for large-scale MEC systems. Recently, the authors in \cite{Samarakoon2018DistributedFL} applied federated learning to estimate the tail distribution of the queues in URLLC vehicle communications and the works in \cite{Wang2019AdaptiveFL} proposed a new adaptive federated learning protocol in heterogeneous MEC systems. 

\section{Miscellaneous Researches}
\label{Sec:MiscellaneousResearches}
In this section, we first focus on recent open source activities. Then, we look at studies denoted to the testbed and implementation of MEC systems.

{\color{black}
\subsection{Open Source Activities}
\label{Subsec:OpenSourceActivities}

The ETSI ISG has created a new group, namely Deployment and Ecosystem Development working group (WG DECODE) to accelerate the adoption and implementation of MEC services in the industry\footnote{The announcement was issued at \url{www.etsi.org/newsroom/press-releases}.}. The group is expected to play a leading role in pursuing research activities defined in Phase 2 specifications. 

To achieve its objectives, the WG DECODE first exposes MEC descriptions based APIs to increase the adoption of MEC specifications and develop a strong MEC ecosystem. The set of open APIs (e.g., bandwidth management service API and radio network information API) are publicly available at \url{https://forge.etsi.org/rep/mec}. Moreover, the WG DECODE promotes the initiation of open source initiatives and facilitates the implementation of open source solutions for MEC applications. For instance, the Open Edge Computing Initiative\footnote{\url{https://www.openedgecomputing.org/}} was introduced in Jun. 2015 by Carnegie Mellon University and industry partners (e.g., Intel, Vodafone, and T-Mobile). Recently, the Open Edge and HPC Initiative\footnote{http://www.open-edge-hpc-initiative.org/} was launched in Nov. 2018 by Atos, E4, Forschungszentrum Jülich, Fraunhofer FOKUS, Huawei, Mellanox, and SUSE. But the availability of many platforms can cause edge market fragmentation, thus it leads to the interoperability problems and limits the industry collaboration. To circumvent these issues, the Linux Foundation started LF Edge in Jan 2019 to establish an open and interoperable framework, which currently includes five projects: Akraino Edge Stack, EdgeX Foundry, Open Glossary of Edge Computing, Home Edge, and Edge Virtualization Engine\footnote{https://www.lfedge.org/}. Due to the importance of edge computing, we believe that there will be many more groups and frameworks. More importantly, harmonizing open source platforms for MEC necessitates closer cooperation between ETSI and other edge organizations/standards like Open Edge Computing, LF Edge, OpenFog, and OpenStack in the future. 
}

\subsection{Testbed and Implementation}
\subsubsection{Single-Board Computer based Edge Cloud}
There are many ways to create an edge server; however, the implementation of single-board computers as edge clouds has been considered as an efficient and cost-effective solution. The increase in popularity of single-board computers (SBCs) (e.g., Raspberry Pi (RPi), Asus Tinker Board S, and Arduino Mega 2560) is due to their low cost, low energy, enough resource for various applications in not only education, but also in industry, hobbyists, prototype builders, and gamers \cite{Johnston2017TheRaspberryPi, Kempen2017MEC_ConPaaS}. The availability of SBCs has introduced a new concept, \emph{disposable computing}, such that SBCs are deployed as edge servers at any location where the edge service is not available or the current edge server is discarded and needs to be replaced by a new one. Another advantage is its potential use in emergency applications and security crises. For example, SBCs, built as edge servers, can be used for rescue missions in the area, where the underlying infrastructure has been destroyed by natural disasters, e.g., earthquakes and windstorms.

Elkhatib \textit{et al.} \cite{Elkhatib2017OnUsingMicro} considered the concept of ``micro-cloud" and examined the suitability and performance tradeoffs of RPi-based micro-clouds using four metrics: serving latency, hosting capability, the cost of memory writing/reading, and booting time.  {\color{black}Experimental evaluations in \cite{Elkhatib2017OnUsingMicro} demonstrated that RPi clouds can serve a large number of users with low latency and booting time, and can further reduce the cost compared with that of Amazon EC2. In \cite{Jaiswal2018AnIoT_Cloud}, the authors proposed an IoT-edge cloud framework for a smart healthcare information system using SBCs. The authors in \cite{Huang2017ApplicationAware} implemented an MEC framework with the OpenAirInterface\footnote{\url{ http://www.openairinterface.org/}} and evaluated their prototype framework with a streaming face detection application. Other studies have been conducted to realize SBCs for various applications: fast and accurate object analysis for AR applications \cite{Liu2017FastAndAccurate}, real-time image-based object tracking from live videos \cite{Zhao2018ECRT}, social sensing applications \cite{Zhang2018ARealTime}, and latency-aware video analytics \cite{Yi2017LAVEA}.}

\subsubsection{Lightweight Platforms for Edge Computing}
As MEC and D2D communication are both applications of the offloading concept \cite{Mach2017Mobile, Boabang2017NetworkAssisted, Girmay2019JointChannel}, the authors in \cite{Singh2016MobileEdgeFog, Wen2018EnergyEfficient} proposed different MEC architecture to further improve the network performance compared with the standard MEC. A D2D-based MEC architecture was proposed in \cite{Singh2016MobileEdgeFog}, where each relay gateway can act as a local cloud. Further, D2D communication is used to establish direct connections between a relay gateway and users so as to provide edge services and between two neighbor relay gateways to balance the traffic and computation demands among them. The work in \cite{Wen2018EnergyEfficient} introduced a concept of ``MEC D2D". Concretely, \textit{D2D MEC} enables the direct link between users and the MEC server, \textit{neighboring D2D} helps users to connect with the other server if they are not satisfied with the local MEC sever, \textit{cooperative relay} can extend the MEC service, \textit{conventional MEC} provides service to users via the collocated eNB, and \textit{remote cloud} let all users with Internet access use cloud services.

Wang \textit{et al.} \cite{Wang2018ALightweightEdgeComputing} proposed a lightweight edge computing platform that is based on SBCs, lightweight virtual switching, and lightweight container virtualization. Taking into account both the QoS requirements of edge services and the deployment cost and status of the underlying hardware, a lightweight platform for service deployment at the network edge was considered in \cite{Lertsinsrubtavee2017PiCasso}. To evaluate performance of the proposed platform, the authors developed RPis as edge servers and identified a set of the system parameters, such as, the number of services to be deployed and the number of supported users per service. The work in \cite{Korner2018OpenCarrierInterface} proposed an open carrier interface to offer a fair pay-on-use business model and to provide edge services in a distributed and autonomous manner.

\subsubsection{Middleware for Edge Computing} The very first context-adaptive middleware, named CloudAware, for computation offloading was proposed in \cite{Orsini2016CloudAware}. CloudAware is able to predict arbitrary context attributes, thus supporting a wide range of applications with dynamics of the underlying network. The evaluation showed that compared with local computing only, CloudAware can reduce the execution time by 276\% while maintaining the same level of offloading success rate. More recently, there have been a number of other studies on messaging middleware for edge computing applications. The middleware investigated in \cite{Rausch2018EMMA} optimized diverse user QoS requirements and orchestrated connections between users and brokers, \cite{Benson2018Ride} leveraged SDN to monitor network conditions for resilient data exchange of mission-critical applications, and the messaging middleware proposed in \cite{Carrega2017AMiddleware} enabled the development and deployment of emerging applications in distributed and heterogeneous edge computing systems. 

In \cite{Li2018MobileEdge}, the author proposed a middlebox approach to implement the MEC paradigm in 4G LTE networks. Some critical issues are needed to implement the proposed approach without the need to modify the underlying infrastructure: 1) how to intercept and forward the data packets, 2) how to serve the data packets by the MEC servers, 3) how to redirect data traffic to the MEC servers and to the centralized clouds, and 4) how to identify the tunnel for specific users? To solve these issues, the authors in \cite{Li2018MobileEdge} proposed implementing the MEC middlebox between the LTE eNB and the core network, and utilizing some novel design principles, for example, tunnel stateful tracking and traffic redirection. 

\section{Conclusion and Discussion}
\label{Sec:Conclusion}
This paper covers both fundamentals of MEC and a review of up-to-date research on ``integration of MEC with the forthcoming 5G technologies". In each section, we have presented a brief background, motivations, and overview in combining the corresponding individual technology in MEC systems. Moreover, we have outlined and discussed the lessons learned, open challenges, and future directions. 
{\color{black} A number of lessons have been learned from this survey paper:
\begin{itemize}
	\item There have been enormous efforts from academia and industry to realize MEC as the key enabler for applications and services (e.g., V2X, Tactile Internet, AR/VR, and big data) in the 5G and beyond network. MEC provides a great number of opportunities and potentials; however, some challenges exist and need to be further studied and tackled, e.g., distributed resource management, reliability and mobility, network integration and application portability, the coexistence of heterogeneous (i.e., H2H and MEC) traffic, data privacy, and security.
	
	\item There are three main types of MEC use cases: consumer-oriented services, operator and third-party services, and network performance and QoE improvements. To support these categorizations, the integration of MEC with the key enabling technologies in the 5G and beyond network is essential. Moreover, to enable a seamless integration of MEC into the 5G network architecture, the 3GPP has introduced several new functional enablers, namely user plane (re)selection, data network interface, local routing and traffic steering, session and service continuity, network capability expose, and QoS and charging. 
	
	\item By integrating with other 5G technologies, MEC systems can support massive IoT (NOMA), maintain the system self-sustainability and self-sufficiency (ET and WPT), improve the network performance, adaptability, and scalability (ML), improve the connectivity and coverage of terrestrial cellular networks (UAV), and help service/infrastructure providers make the economics of MEC services (collocation with C-RAN).
	
	\item To accelerate the adoption of MEC services, the ETSI ISG has defined and exposed a set of open APIs, and further participated in open source activities. Moreover, there have been many efforts and solutions for MEC testbeds and implementation. 
\end{itemize}

For the sake of achieving the seamless integration of MEC in the 5G and beyond network, a number of potential works have been given before. Here, we outline some open problems and challenges which need to be further studied and tackled.
\begin{itemize}
	
	\item \textit{Higher-Level Integration}: Although existing research integrates MEC with several enabling technologies, in fact, they are completely independent of each other. Therefore, it is possible to combine more than one of these technologies into a single MEC system. For example, IoT devices first harvest energy from a power source and then follow the NOMA principle to offload their computation tasks to a flying BS equipped with computing capability, where a DRL model is trained to determine the UAV's trajectory and adapt to the underlying dynamic network. 
	
	\item \textit{Coexistence of Multiple MEC Designs}: This issue becomes crucial when a number of proposals for the same problem of MEC systems are simultaneously proposed, e.g., offloading decision and resource allocation. There has been no answer for how different proposals can be integrated into a unique framework. One possible solution to overcome this issue is that different proposals are classified to find their common viewpoints and then a standard solution should be investigated to support MEC systems with these viewpoints.
	
	\item\textit{ More Opportunities and Challenges from 6G}: While the 5G standards are not well established yet, there have been some speculative studies for 6G wireless systems to circumvent limitations of the 5G network. For example, a wireless system must support ultra reliability, low latency, high data rate simultaneously, which cannot be fulfilled in the 5G system \cite{Saad2019AVision_6G}. It is expected that 6G will include new use cases like haptic communications for eXtended Reality (XR) services, massive IoT for smart city applications, automation and manufacturing. To support these new services, various promising technologies have been speculated and discussed recently, including pervasive and collective AI, radar-enabled communications, metamaterials and intelligent structures, cell-free networks, visible light communication, quantum computing and communications, and tiny cells with THz spectrum \cite{Saad2019AVision_6G, Tariq2019ASpeculative_6G}. It is inevitable that besides many more use cases and scenarios, new 6G technologies and application requirements also introduce hurdles in MEC and tremendous efforts need to be paid in the future. 
	
	\item \textit{More challenges and opportunities with distributed learning and FL}: To cope with stringent security requirements, data privacy concerns, massive connectivity, and network heterogeneity, enabling learning techniques (e.g., distributed and FL) in mobile edge networks is of crucial importance. Despite their considerable advantages, there are still many challenges and issues. In recent review articles \cite{Lim2019Federated, Li2019Federated}, several challenges and issues of deploying FL in mobile edge networks are discussed, which include, participant selection, tradeoff between privacy protection level and system performance, beyond supervised learning, interference management, communication security, incentive mechanism designs, and asynchronous FL approaches. Moreover, promising research directions, e.g., convergence guarantees for the non-convex loss function, heterogeneity diagnostics, and mobile crowdsensing for FL are outlined. In summary, providing solutions to these problems and enabling more applications of FL in MEC systems require interdisciplinary efforts from a variety of research communities.
	
\end{itemize}
}

We strongly believe that this survey can help the readers to deeply understand MEC and its interactions with the enabling technologies in 5G and beyond. We also hope that this survey will stimulate further 5G and MEC research activities.

\end{document}